\documentclass[9.5pt,journal,compsoc,english]{IEEEtran}

\usepackage{float}
\usepackage[T1]{fontenc}
\usepackage[latin9]{inputenc}
\usepackage{verbatim}
\usepackage{amsthm}
\usepackage{amsmath}
\usepackage{amssymb}
\usepackage{esint}
\usepackage{varwidth}
\usepackage{url}
\usepackage{constants}
\usepackage{graphicx}
\usepackage{float}
\usepackage{hyperref}
\usepackage{bm}
\usepackage{algorithm,algorithmic}
\usepackage{enumitem}
\usepackage{tikz}
\usepackage{setspace}
\usepackage{epstopdf}
\usepackage{subfig}
\usepackage{float}
\usepackage{etoolbox}
\usepackage{caption}
\usepackage{comment}

\usepackage{color}
\usepackage{soul}
\providecommand{\keywords}[1]{\textbf{\textit{Keywords---}} #1}

\newtoggle{OVERLEAF}

\newcommand{\edit}[1]{{\color{black} #1}}
\newcommand{\editr}[1]{{\color{black} #1}}

\makeatletter

\theoremstyle{plain}
\newtheorem{thm}{Theorem}

  \theoremstyle{plain}
  \newtheorem{lem}{Lemma}
  
  \theoremstyle{plain}

 \theoremstyle{plain}

   \theoremstyle{plain}
  \newtheorem{prop}{Proposition}

\usepackage{babel}

\def \I {\mathcal{I}}

\def \SVD{\operatorname{SVD}}

\def \SU {\mathcal{S}}

\def \alg {\operatorname{SNIPE}}
\def \Q {\boldsymbol{Q}}
\def \E {\mathbb{E}}
\def \h {\widehat}
\def \U {\mathcal{U}}

\def \Y {\mathcal{Y}}
\def \R {\mathbb{R}}
\def \GR {\text{G}}

\def \X {\boldsymbol{X}}
\def \Y{\boldsymbol{Y}}
\def \y{\boldsymbol{y}}
\newcommand{\B}[1]{\boldsymbol{#1}}
\def \P{\boldsymbol{P}}
\def \rate{\eta}
\def \Qsub{\mathcal{Q}}
\def \Ysub{\mathcal{S}}
\def \rank {\operatorname{rank}}
\def \p{p} 
\def \l{\left}
\def \r{\right}
\def \G{\B{G}}
\def \g{\B{g}}

\def \wh{\widehat}
\def \wt{\widetilde}

\def \J {\mathcal{J}}
\hypersetup{
   colorlinks=true,%
   citecolor=black,%
   filecolor=black,%
   linkcolor=black,%
   urlcolor=black
}

\def \alg {\operatorname{MOSES}}

\makeatother

\usepackage{babel}
\begin{document}
{\setstretch{1.0}
\title{MOSES: A Streaming Algorithm for\\ Linear Dimensionality Reduction}
\author{Armin Eftekhari, Raphael A.\ Hauser, Andreas Grammenos\thanks{
 AE is with the Institute of Electrical Engineering at the \'{E}cole Polytechnique F\'{e}d\'{e}rale de Lausanne. RAH is with the Mathematical Institute at the University of Oxford. AG is with the Department of Computer Science at the University of Cambridge. RAH and AG are also affiliated with the Alan Turing Institute in London. Emails: \tt{armin.eftekhari@epfl.ch, hauser@maths.ox.ac.uk, ag926@cl.cam.ac.uk}.
}}}

\maketitle

\begin{abstract}

This paper introduces Memory-limited Online Subspace Estimation Scheme (MOSES) for both estimating the principal components of \edit{streaming} data and reducing its dimension. More specifically, \edit{in various applications such as sensor networks,} the  data vectors are presented sequentially to a user who has limited storage and processing time available.  \edit{Applied to such problems}, MOSES \edit{can provide a running} estimate of leading principal components of the data that has arrived so far and also reduce its dimension. 

\edit{MOSES  generalises the popular incremental Singular Vale Decomposition (iSVD)  to handle thin blocks of data, rather than just vectors}. This minor generalisation \edit{in part allows us} to complement MOSES with a comprehensive statistical analysis, \edit{ thus providing the first theoretically-sound variant of  iSVD, which has been lacking despite the empirical success of this method}. This generalisation also enables us to concretely interpret MOSES  as an approximate solver for the underlying non-convex optimisation program. We find that MOSES \edit{consistently surpasses the state of the art} in our numerical experiments with both synthetic and real-world datasets\edit{, while being computationally inexpensive.}

\end{abstract}


\section{Introduction \label{sec:intro0}}
Linear models are  pervasive in data and computational sciences and, \edit{in particular}, {Principal Component Analysis} (PCA) is an  indispensable tool for detecting linear structure in collected data \cite{van2012subspace,ardekani1999activation,krim1996two,tong1998multichannel,vidal2016generalized,hauser2018pca,hauser2018pca2}. 
{Principal components} are the  directions that preserve most of the   ``energy'' of a dataset and can be used for linear dimensionality reduction,  among other things. In turn, successful dimensionality reduction is at the heart of \edit{statistical learning and serves to tackle the ``curse of dimensionality''~\cite{hastie2013elements}.}

In this work, we are  interested in both computing the principal components \emph{and} reducing the dimension of data that is presented sequentially to a user. Due to hardware limitations, the user can only store small amounts of data, which in turn \edit{would} severely limit the available processing time for each incoming data vector.

For example, consider a network of battery-powered and cheap sensors that must  relay their measurements to a central node on a daily basis. Each sensor has a small storage and does not have the power to relay all the raw data to the central node. One solution is then for each sensor to reduce the dimension of its data to make transmission to the central node possible. 
Even if each sensor had unlimited storage, the frequent daily updates scheduled by the central node would force each sensor to reduce the dimension of its data ``on the go'' before transmitting it to the central node. 
A number of similar problems are listed in~\cite{balzano2010online}. 
%

Motivated by such scenarios, we are interested in developing a \emph{streaming}  algorithm for linear dimensionality reduction, namely, an algorithm with minimal storage and computational requirements. As more and more data vectors arrive, this algorithm would keep a running estimate of the principal components of the data \emph{and} project the available data onto this estimate to reduce its dimension. \edit{As discussed in Section \ref{sec:intro}, this is equivalent to designing a streaming algorithm for truncated Singular Value Decomposition (SVD).} 

Indeed, incremental SVD (iSVD) is a successful streaming algorithm that updates its estimate of the truncated SVD of the data matrix with every new incoming  vector \cite{bunch1978rank,brand2006fast,brand2002incremental,comon1990tracking,li2004incremental}. 
\edit{However,} to the best of our knowledge and despite its popularity and empirical success, iSVD lacks comprehensive statistical guarantees. In fact, \cite{balsubramani2013fast} only very recently provided stochastic analysis for two of the variants of iSVD in \cite{krasulina1969method,oja1983subspace}. \edit{More specifically,} in \cite{balsubramani2013fast} the authors studied how well the output of iSVD approximates  the leading principal component of data, in expectation. Crucially,~\cite{balsubramani2013fast} does \emph{not} offer any guarantees for dimensionality reduction; see Section~\ref{sec:lit review} \edit{for more details on iSVD} and  review of the prior art. 

\paragraph*{\textbf{Contributions}} In this paper, \edit{to address the shortcomings of iSVD,} we propose Memory-limited Online Subspace Estimation Scheme ($\alg$) for streaming  dimensionality reduction. $\alg$  generalises iSVD to update its estimate with every incoming  \emph{thin} block of data, rather than with every incoming vector. This small \edit{generalisation} is in part what enables us to complement $\alg$ with a comprehensive statistical analysis, \edit{thus providing (to the best of our knowledge) the first theoretically-sound variant of iSVD.}

Indeed, Theorem~\ref{thm:stochastic result} below  considers the important case where the incoming data vectors are drawn from a zero-mean normal distribution. This stochastic setup is a powerful generalisation of the popular \emph{spiked covariance} model common in statistical signal processing~\cite{johnstone2001distribution}. Theorem \ref{thm:stochastic result} states that $\alg$ nearly matches the performance of ``offline'' truncated SVD (which has unlimited memory and computing resources), provided that the corresponding covariance matrix is well-conditioned and has a small residual. 

Moreover, we concretely interpret $\alg$ as an approximate solver for the underlying non-convex optimisation program. We also find that $\alg$ consistently surpasses the state of the art in our numerical experiments with both synthetic and real-world datasets, \edit{while  being computationally inexpensive.}


\section{Introducing MOSES \label{sec:intro}}

Consider a sequence of  vectors $\{y_t\}_{t=1}^T\subset \mathbb{R}^n$, presented to us sequentially, and let 
\begin{equation}
\label{eq:conc of yts}
\Y_{T} := \l[ 
\begin{array}{cccc}
y_1 & y_2 & \cdots & y_T
\end{array}
\r] \in\R^{n\times T},
\end{equation} 
for short. 
We conveniently assume throughout that $\Y_T$ is \emph{centred}, namely, the entries of each row of $\Y_T$ sum up to zero. 
\edit{For an integer $r\le \rank(\Y_T)$, let us partition the SVD of $\Y_T$ into two orthogonal components as 
%
\begin{align}
\Y_T & \overset{\text{SVD}}{=} \B{S}_T \B{\Gamma}_T \Q_T^*
\qquad ( \B{S}_T \in \text{O}(n), \, \B{Q}_T \in \text{O}(T) )
 \nonumber\\
& = \B{S}_{T,r} \B{\Gamma}_{T,r} \Q_{T,r}^*+ \B{S}_{T,r^+} \B{\Gamma}_{T,r^+} \Q_{T,r^+}^* \nonumber\\
& =: \Y_{T,r}+\Y_{T,r^+},
\label{eq:decomposition of Y}
\end{align}
where $\text{O}(n)$ is the orthogonal group, containing all $n\times n$ matrices with orthonormal columns, and $\B{\Gamma}_T\in \R^{n\times T}$ above contains the singular values in nonincreasing order. Moreover, $\B{S}_{T,r} \in \R^{n\times r}$ contains leading $r$ principal components of $\Y_T$, \edit{namely,  the singular vectors corresponding to  largest $r$ singular values of $\Y_T$  \cite{eckart, mirsky}.} 

 Given     $\B{S}_{T,r}$,  we can reduce the dimension of data from $n$ to $r$ by projecting $\Y_T$ onto the column span of $\B{S}_{T,r}$, that is,
\begin{align}
\B{S}_{T,r}^*\cdot \Y_T 
& =  \B{\Gamma}_{T,r} \B{Q}_{T,r}^* \in\mathbb{R}^{r\times T}.
\qquad \mbox{(see \eqref{eq:decomposition of Y})}
\label{eq:projected data matrix}
\end{align}
Above, the projected data matrix $\B{S}_{T,r}^* \Y_T\in\R^{r\times T}$ again has $T$ data vectors (namely, columns) but these  vectors are embedded in (often much smaller) $\mathbb{R}^r$ rather than $\mathbb{R}^n$.}  
\edit{Note that
$
\Y_{T,r}  
$
in \eqref{eq:decomposition of Y} is a rank-$r$ truncation of $\Y_T$, which we denote with $\Y_{T,r}=\SVD_r(\Y_T)$.} That is, $\Y_{T,r}$ is a best rank-$r$ approximation of $\Y_T$ with the corresponding residual 
\begin{align}
\| \Y_T - \Y_{T,r} \|_F^2 & = \min_{\operatorname{rank}(\B{X})=r} \| \Y_T - \B{X}\|_F^2 \nonumber\\
& = \| \Y_{T,r^+}\|_F^2  = \sum_{i\ge r+1} \sigma_i^2(\Y_T)\nonumber\\
& =: \rho_r^2(\Y_T),
\label{eq:residual}
\end{align}
where $\sigma_1(\Y_T)\ge \sigma_2(\Y_T)\ge \cdots $ are the singular values of $\Y_T$. 
\edit{We also observe that} 
\begin{align}
\Y_{T,r}& =\SVD_r(\Y_T) \nonumber\\
&  = \underset{\operatorname{PCs}}{\underbrace{\B{S}_{T,r}}} \cdot \underset{\operatorname{projected\; data}}{\underbrace{\B{\Gamma}_{T,r} \B{Q}_{T,r}^*}}. 
\qquad \mbox{(see (\ref{eq:decomposition of Y},\ref{eq:projected data matrix}))}
\label{eq:low-rank est kept}
\end{align}
That is, rank-$r$ truncation of $\Y_T$  encapsulates both  leading $r$ principal components of $\Y_T$, namely $\B{S}_{T,r}$,  and the projected data matrix $\B{S}_{T,r}^*\B{Y}_T = \B{\Gamma}_{T,r} \B{Q}_{T,r}^*$. In other words, computing a rank-$r$ truncation of the data matrix  both yields  its principal components and reduces the dimension of data at once.
\vspace{-14pt}
\begin{center}
\fbox{\begin{minipage}[t]{1\columnwidth}%
We are in this work interested in developing a {streaming} algorithm to compute $\Y_{T,r}=\SVD_r(\Y_T)$, \edit{namely,} a rank-$r$ truncation of the data matrix $\Y_T$. More specifically, to compute $\Y_{T,r}$, we are only allowed one pass through the columns of $\Y_T$ and have \edit{access to} a limited amount of storage, namely, $O(n)$ bits. 
\end{minipage}}
\end{center}
\noindent For a block size $b$, our strategy is to iteratively group every $b$ incoming vectors into an $n\times b$ block and then update a rank-$r$ estimate of the data received so far. We assume throughout that $r\le b\le T$ and in fact often take the block size as  $b=O(r)$, where $O$ is the standard Big-O notation.  It is convenient to  assume that the number of blocks  $K:=T/b$ is an integer.   
We call this simple algorithm $\alg$ for Memory-limited Online Subspace Estimation Scheme\editr{, presented in an accessible fashion in  Algorithm  \ref{alg:MOSES}.} 

The output of $\alg$ after $K$ iterations is 
$$
\wh{\Y}_{Kb,r}=\wh{\Y}_{T,r},
$$
which contains both  an estimate of  leading $r$ principal components of $\Y_T$ and \edit{also} the projection of $\Y_T$ onto this estimate, as discussed below. 
A \editr{computationally} efficient implementation of $\alg$ is given in Algorithm \ref{alg:MOSES implement}, which explicitly maintains both the estimates of principal components and the projected data. As also discussed below, the storage and computational requirements of Algorithm~\ref{alg:MOSES implement} are nearly \edit{minimal}.

\paragraph*{\textbf{Discussion}}
$\alg$  maintains a rank-$r$ estimate of the data received so far, and  updates its estimate in every iteration to account for the new incoming block of data. 
More specifically, note that the final output of $\alg$, namely $\widehat{\Y}_{T,r}\in\mathbb{R}^{n\times T}$, is at most rank-$r$, and let 
\begin{equation*}
\wh{\Y}_{T,r} \overset{\text{tSVD}}{=} \wh{\B{S}}_{T,r} \wh{\B{\Gamma}}_{T,r} \wh{\B{Q}}_{T,r}^*
\end{equation*}
\begin{equation}
\l( \wh{\B{S}}_{T,r}\in \text{St}(n,r), \, \wh{\B{Q}}_{T,r}\in \text{St}(T,r) \r)
\end{equation}
\edit{be its thin SVD. Above, $\text{St}(n,r)$ is the Stiefel manifold, containing all $n\times r$ matrices with orthonormal columns, and the diagonal matrix $\wh{\B{\Gamma}}_{T,r}\in \R^{r\times r}$ contains the singular values in nonincreasing order.}  
Then, $\wh{\B{S}}_{T,r}\in\mathbb{R}^{n\times r}$ is  $\alg$'s estimate of \edit{leading} principal components of the data matrix $\Y_{T}$, and  
$$
\wh{\B{S}}_{T,r}^* \wh{\Y}_{T,r}  = \wh{\B{\Gamma}}_{T,r} \wh{\B{Q}}_{T,r}
\in \mathbb{R}^{r\times T}
$$ 
is the projection of $\wh{\Y}_{T,r}$  onto this estimate. That is, $\wh{\B{S}}_{T,r}^* \wh{\Y}_{T,r}$ is the $\alg$'s estimate of the projected data matrix.


\paragraph*{\textbf{Origins}}
{iSVD} is a streaming algorithm that updates its estimate of (truncated) SVD of the data matrix with every new incoming vector \cite{bunch1978rank,brand2006fast,brand2002incremental,comon1990tracking,li2004incremental}.  
\edit{Despite its popularity and empirical success, iSVD lacks  comprehensive statistical guarantees, as detailed in Section \ref{sec:lit review}.} 
 
$\alg$  generalises iSVD to update its estimate with every incoming block of data, rather than  every data vector. As detailed later in Section \ref{sec:lit review}, \edit{this minor extension
in part  enables us to complement $\alg$ with a comprehensive statistical analysis, summarised  in Theorem~\ref{thm:stochastic result} below. In this sense, $\alg$ might be interpreted as a variant of iSVD that is both successful in practice and theoretically grounded.}

\edit{Working with data blocks} also allows us to concretely interpret $\alg$ as an approximate solver for the underlying non-convex program, as detailed in Section 3 of the supplementary material.

\paragraph*{\textbf{Storage and computational requirements}}
The efficient implementation of $\alg$ in Algorithm~\ref{alg:MOSES implement} is based on the ideas from iSVD and it is straightforward to verify that Algorithms~\ref{alg:MOSES}~and~\ref{alg:MOSES implement} are indeed equivalent; at  iteration $k$,  the relation between the  output of Algorithm~\ref{alg:MOSES} ($\wh{\Y}_{kb,r}$) and the output of Algorithm~\ref{alg:MOSES implement} ($\wh{\B{S}}_{kb,r},\wh{\B{\Gamma}}_{kb,r},\wh{\B{Q}}_{kb,r}$) is 
$$
\wh{\Y}_{kb,r}\overset{\text{tSVD}}{=}\wh{\B{S}}_{kb,r}\wh{\B{\Gamma}}_{kb,r}\wh{\B{Q}}_{kb,r}^*.
$$ 
More specifically,  $\wh{\B{S}}_{kb,r} \in\text{St}(n,r)$ is  $\alg$'s estimate of  leading $r$ principal components of $\Y_{kb}\in \mathbb{R}^{n\times kb}$, where we recall that $\Y_{kb}$ is the data received \edit{up to iteration $k$}.  Moreover, 
$$\wh{\B{S}}_{kb,r}^*  \wh{\Y}_{kb,r}= \wh{\B{\Gamma}}_{kb,r}\wh{\B{Q}}_{kb,r}^*\in \mathbb{R}^{r\times kb}
$$ is the projection of $\wh{\Y}_{kb,r}$ onto this estimate, namely,  $\wh{\B{S}}_{kb,r}^*  \wh{\Y}_{kb,r}$ is  $\alg$'s estimate of the projected data matrix so far. In words, the efficient implementation of $\alg$ in Algorithm \ref{alg:MOSES implement} explicitly maintains estimates of both  principal components and the projected data, at every iteration. 

\edit{As detailed in Section 2 of the supplementary material, Algorithm \ref{alg:MOSES implement}  requires $O(r(n+kr))$  bits of memory at iteration $k$.  This is optimal, as it is impossible to store a rank-$r$  matrix of size $n\times kb$ with fewer  bits when $b=O(r)$. On the other hand, Algorithm~\ref{alg:MOSES implement} performs $O(r^2(n+kb))=O(r^2(n+kr))$ flops in iteration $k$. The dependence of both storage and computational complexity on $k$ is because $\alg$  maintains both an estimate of  principal components  ($\wh{\B{S}}_{kb,r}$) and  an estimate of the projected data ($\B{\Gamma}_{kb,r}\B{Q}_{kb,r}^*$). To maximise the efficiency, one might optionally ``flush out'' the projected data after every  $n/b$ iterations, as described in the last step in Algorithm \ref{alg:MOSES implement}.}

%

{\setstretch{1.0}
\begin{center}%
\begin{algorithm}[H]
{
\caption{$\alg$: A streaming algorithm for linear dimensionality reduction \editr{(accessible version)}}\label{alg:MOSES}
\vspace{3mm}
\textbf{Input:} Sequence of vectors $\{y_t\}_{t\ge 1}\subset\mathbb{R}^n$, rank $r$, and block size $b\ge r$.
\vspace{3mm}

\textbf{Output:} Sequence \edit{$\{\wh{\Y}_{kb,r}\}_{k\ge 1}$}, where $\wh{\Y}_{kb,r}\in\mathbb{R}^{n\times kb}$ for every $k\ge 1$.
\vspace{3mm}

\textbf{Body:} 
\\
\begin{enumerate}[leftmargin=*]
\item  Set $\wh{\Y}_{0,r}\leftarrow \{\}$.
\item  For $k\ge 1$, repeat 
\begin{enumerate}
\item  Form $\y_k\in\mathbb{R}^{n\times b}$ by concatenating $\{y_t\}_{t=(k-1)b+1}^{kb}$.  
\item Set $\widehat{\Y}_{kb,r}=\text{SVD}_{r}([
\widehat{\Y}_{(k-1)b,r} \,\,\, \y_{k}])$, where $\text{SVD}_r(\cdot)$ returns a rank-$r$ truncated SVD of its argument.
\end{enumerate}
\end{enumerate}
}
\end{algorithm}
\end{center}
}

\section{Performance of MOSES \label{sec:results}}

In this section, we study the performance of $\alg$ in a stochastic setup. Consider the probability space $(\R^n,\mathcal{B},\mu)$, where $\mathcal{B}$ is the Borel $\sigma$-algebra and $\mu$ is an \emph{unknown} probability measure with zero mean, namely, $\int_{\R^n} y \, \mu(dy) = 0$.  \edit{Informally,} we are interested in finding an $r$-dimensional subspace $\U$ that   \editr{captures most of the mass of} $\mu$. That is, with $y$ drawn from this probability space, we are interested in finding an $r$-dimensional subspace $\U$ that 
minimises the \emph{population risk}. \editr{To be specific, we wish to solve} 
\begin{align}
& \min_{\U\in\GR(n,r)} \underset{y\sim \mu}{\E}\l\| y - \P_{\U} y\r\|_2^2 \nonumber\\
 & = \min_{\U\in\GR(n,r)} \int_{\R^n} \l\| y- \P_{\U} y\r\|_F^2 \, \mu(dy) 
 =: \rho_r^2(\mu),
\label{eq:direct}
\end{align} 
\edit{where the Grassmanian  $\text{G}(n,r)$ is the set of all $r$-dimensional subspaces in $\R^n$}\editr{, and  $\rho_r(\mu)$ denotes the corresponding {residual}.}

Since $\mu$ is unknown, we cannot directly solve Program~\eqref{eq:direct}, but  suppose that instead we have access to the \emph{training samples} $\{y_t\}_{t=1}^T \subset \R^{n}$, drawn independently from \editr{the probability measure $\mu$}. Let us form $\Y_T\in\R^{n\times T}$ by concatenating these vectors, \editr{similar to } \eqref{eq:conc of yts}. In lieu of Program~\eqref{eq:direct}, \editr{we minimise the \emph{empirical risk}, namely,  we  solve the  optimisation program}
\begin{align}
 & \min_{\U\in\GR(n,r)} \frac{1}{T}\sum_{t=1}^T \l\| y_t - \P_{\U}y_t \r\|_2^2 \nonumber\\
 & = 
 \min_{\U\in\GR(n,r)} \frac{1}{T} \l\|\Y_T - \P_{\U}\Y_T \r\|_F^2 .
 \qquad \text{(see \eqref{eq:conc of yts})}
\end{align}
\editr{Let $\SU_{T,r}\in\GR(n,r)$ be a minimiser of the above program, with the orthonormal basis  $\B{S}_{T,r}\in\text{St}(n,r)$.} \edit{By the Eckart-Young-Mirsky Theorem \cite{eckart,mirsky}, $\B{S}_{T,r}$ consists of  leading $r$ principal components of $\Y_T$, namely, it contains leading $r$ left singular vectors of $\Y_T$.} \editr{Therefore, 
\begin{align} 
&  \min_{\U\in\GR(n,r)} \frac{1}{T}\sum_{t=1}^T \l\| y_t - \P_{\U}y_t \r\|_2^2 
 \nonumber\\
 & = \frac{1}{T}\l\|\Y_T - \P_{\SU_{T,r}}\Y_T \r\|_F^2
 \nonumber\\
 & = \frac{1}{T} \l\|\Y_T -\Y_{T,r} \r\|_F^2
 \qquad \l(\Y_{T,r}=\text{SVD}_r(\Y_T)\r) \nonumber\\
& =: \frac{\rho_r^2(\Y_T) }{T}.
\qquad \text{(see \eqref{eq:residual})}
 \label{eq:empirical}
\end{align}}Given its principal components, we can then reduce the dimension of the data matrix $\Y_T\in\R^{n\times T}$ from $n$ to $r$ by computing $\B{S}_{T,r}^*\Y_T\in \R^{r\times T}$. Note also that \editr{the subspace}  $\SU_{T,r}$ is a possibly sub-optimal choice in Program~\eqref{eq:direct}, namely,
\begin{equation}
\underset{y\sim \mu}{\E} \|y-\P_{\SU_{T,r}}y\|_2^2 \ge \rho_r^2(\mu).
\qquad \text{(see \eqref{eq:direct})}
\end{equation}
But one would  hope that \edit{$\SU_{T,r}$} still nearly minimises Program \eqref{eq:direct}, in the sense that 
\begin{equation}
 \underset{y\sim \mu}{\E} \| y- \P_{\SU_{T,r}}y\|_2^2 \approx \rho_r^2(\mu),
\label{eq:indirect 1}
\end{equation}
with high probability over the choice of training data $\{y_t\}_{t=1}^T$. That is, one would hope that the \emph{generalisation error} of Program \eqref{eq:empirical} is small. Above, $\underset{y\sim \mu}{\E}$ stands for expectation over $y$, so that the left-hand side of \eqref{eq:indirect 1} is still a random variable because of its dependence on \edit{$\SU_{T,r}$ and, in turn, \editr{on }the training data.} 

If the training data $\{y_t\}_{t=1}^T$ is presented to us sequentially and little storage is available, we cannot hope to directly solve Program \eqref{eq:empirical}. 
In this streaming scenario, we may apply $\alg$ to obtain the (rank-$r$) output $\wh{\Y}_{T,r}$. We then set  
 \begin{equation}
 \wh{\SU}_{T,r} = \mbox{span}(\wh{\Y}_{T,r}),
 \label{eq:span of output of MOSES}
\end{equation} 
with orthonormal basis \edit{$\wh{\B{S}}_{T,r}\in \text{St}(n,r)$}. Note that \edit{$\wh{\B{S}}_{T,r}$} is $\alg$' estimate of  leading $r$ principal components of the data matrix $\Y_T$ and is possibly suboptimal in the sense that 
\begin{equation}
\| \Y_T-\wh{ \Y}_{T,r} \|_F  \ge 
\rho_r(\Y_T). \qquad \mbox{(see \eqref{eq:empirical})}
\end{equation}
\editr{Again,} we would still hope that the output $\widehat{\Y}_{T,r}$ of $\alg$ is a nearly optimal choice in Program~\eqref{eq:empirical}, in the sense that
\begin{equation}
\| \Y_T-\wh{ \Y}_{T,r} \|_F \approx \rho_r(\Y_T),
\label{eq:indirect 1.5}
\end{equation}
with high probability over the choice of $\{y_t\}_{t=1}^T$. 
Moreover,  \editr{similar to}  \eqref{eq:indirect 1}, $\wh{\SU}_{T,r}$ is again a possibly \editr{sub-optimal} choice for Program \eqref{eq:direct}, and yet we  hope that 
\begin{equation}
\E_y\| y- \P_{\wh{\SU}_{T,r}}y\|_2^2 \approx  \rho_r^2(\mu),
\label{eq:ideal}
\end{equation}
with high probability over the choice of $\{y_t\}_{t=1}^T$.

To summarise, the key questions are whether (\ref{eq:indirect 1},\ref{eq:indirect 1.5},\ref{eq:ideal}) hold. \editr{We} answer these questions for the important case where $\mu$ is a zero-mean Gaussian probability measure with covariance matrix $\B{\Xi}\in\R^{n\times n}$. 
For this choice of $\mu$ in \eqref{eq:direct}, it is not difficult to verify that 
\begin{equation}
\rho_r^2(\mu)  = \sum_{i=r+1}^n \lambda_i(\B{\Xi}),  
\label{eq:res of gaussian dist}
\end{equation}
where $\lambda_1(\B{\Xi}) \ge \lambda_2(\B{\Xi})\ge \cdots $ are the eigenvalues of the covariance matrix $\B{\Xi}$. From now on, \editr{we use}
$$
\rho_r = \rho_r(\mu), \qquad \lambda_i=\lambda_i(\B{\Xi}),
\qquad i\in [1:n].
$$  
For our choice of $\mu$ above, 
one can use standard tools 
to show that  \eqref{eq:indirect 1} holds when $T$ is sufficiently large, \editr{see} Section B of the supplementary material  \cite{eftekhari2016learning,hauser2017quantifying,vershynin2012close}. 
\begin{prop}\label{prop:first app}
Suppose that $\{y_t\}_{t=1}^T\subset\R^n$  are drawn independently from a zero-mean Gaussian \edit{distribution} with \edit{the} covariance matrix $\B{\Xi}\in\R^{n\times n}$ and form $\Y_T\in\R^{n\times T}$ by concatenating these vectors, as in (\ref{eq:conc of yts}). Suppose also that $\SU_{T,r}\in\GR(n,r)$ is the span of leading $r$ principal components of $\Y_T$. For $1\le \alpha \le \sqrt{T/\log T}$, it then holds that 
\begin{equation}
\frac{\rho_r^2(\Y_T)}{T}  \lesssim {{ \alpha \rho_r^2}}, 
\label{eq:true residual prop}
\end{equation}
\begin{equation}
\underset{y\sim \mu}{\E} \|y - \P_{\SU_{T,r}} y\|_2^2 
\lesssim
{{ \alpha \rho_r^2}} +
{{\alpha (n-r)\lambda_1 \sqrt{\frac{ {\log T} }{T}}}},
\label{eq:gen err easy}
\end{equation}
except with a probability of at most $T^{-C\alpha^2}$, \editr{see (\ref{eq:residual})}. \editr{Here,} $C$ is a universal constant, \edit{the value of which may change in every appearance.} Above, $\lesssim$ suppresses \editr{some of the} universal constants for a more tidy presentation. 
\end{prop}
In words, \eqref{eq:gen err easy} states that the generalisation error of Program \eqref{eq:empirical} is  small, namely,  \editr{offline truncated SVD successfully estimates the unknown subspace from training data.} 
\editr{Indeed}, \eqref{eq:indirect 1} holds when $\alpha = O(1)$ and $ T$ is sufficiently large.
As the dimension $r$ of the subspace fit to the data approaches the ambient dimension $n$, \editr{the right-hand of }
\eqref{eq:gen err easy} vanishes, \editr{see \eqref{eq:direct}.}

\editr{In the streaming setup,} Theorem \ref{thm:stochastic result} below states that  $\alg$ approximately solves Program~\eqref{eq:empirical}, namely, $\alg$ approximately  estimates  leading principal components of $\Y_T$ \emph{and} reduces the dimension of data from $n$ to $r$ with only $O(r(n+T))$ bits of memory, rather than $O(nT)$ bits required for solving Program~\eqref{eq:empirical} with offline truncated SVD. Moreover, $\alg$ approximately solves Program \eqref{eq:direct}. In other words, $\alg$ satisfies \emph{both} (\ref{eq:indirect 1.5},\ref{eq:ideal}). 
These statements are made concrete below and proved in Section C of the supplementary material. 
\begin{thm}\label{thm:stochastic result}
\textbf{\emph{(Performance of MOSES)}}
Suppose that $\{y_t\}_{t=1}^T\subset\R^n$  are drawn independently from a zero-mean Gaussian \edit{distribution with the} covariance matrix $\B{\Xi}\in\R^{n\times n}$  \edit{and form $\Y_T$ by concatenating these vectors, as in (\ref{eq:conc of yts}).} Let us define 
{\footnotesize\begin{equation}
\kappa_r^2 := \frac{\lambda_1}{\lambda_r},
\qquad 
\rho_r^2 = \sum_{i=r+1}^n \lambda_i, \qquad 
\rate_r := \kappa_r +  \sqrt{\frac{2\alpha \rho_r^2}{p^{\frac{1}{3}}\lambda_r}},
\label{eq:spect of guassian thrm}
\end{equation}}where $\lambda_1\ge \lambda_2 \ge \cdots$ are the eigenvalues of $\B{\Xi}$. Let $\wh{\SU}_{T,r} = \operatorname{span}(\wh{\Y}_{T,r})$ be the span of the output of MOSES, as in (\ref{eq:span of output of MOSES}).
Then, for tuning parameters $1\le \alpha\le \sqrt{T/\log T}$ and $p>1$,  \editr{the output $\wh{\Y}_{T,r}$  of MOSES satisfies}  
{\footnotesize
\begin{align}
& \frac{\| \Y_T - \wh{\Y}_{T,r} \|_F^2}{T}  \nonumber\\
& \lesssim \frac{\alpha \p^{\frac{1}{3}} 4^{p\eta_r^2} }{(\p^{\frac{1}{3}}-1)^2} \cdot \min\l (   \kappa_r^2 \rho_r^2,  r\lambda_1+\rho_r^2\r)\cdot 
\l(\frac{T}{\p\eta^2_r b} \r)^{p\eta_r^2-1},
\label{eq:near residual of moses thm}
\end{align}}\editr{explained in words after this theorem.}
\editr{Moreover, the output subspace $\wh{\SU}_{T,r}$ of MOSES satisfies} 
{\footnotesize
\begin{align}
&  \underset{y\sim \mu}{\E} \|y - \P_{\wh{\SU}_{T,r}} y\|_2^2 \nonumber\\
& \lesssim   \frac{\alpha\p^{ \frac{1}{3}}4^{p\eta_r^2} }{(\p^{\frac{1}{3}}-1)^2}  \cdot \min\l (   \kappa_r^2 \rho_r^2,  r\lambda_1+  \rho_r^2\r) 
 \l(\frac{T}{\p\eta^2_r b} \r)^{p\eta_r^2-1}
  \nonumber\\
& +
\alpha (n-r)\lambda_1 \sqrt{\frac{ {\log T} }{T}} ,
\label{eq:final bnd}
\end{align}
}\editr{explained after this theorem.} 
\edit{\editr{Both (\ref{eq:near residual of moses thm},\ref{eq:final bnd})}  hold} except with a probability of at most $T^{-C\alpha^2}+ e^{-C\alpha r}$ and provided that 
\begin{equation}
b
\ge \frac{\alpha \p^{\frac{1}{3}} r}{({\p^{\frac{1}{6}}}-1)^2},
\qquad b\ge C\alpha r, 
\qquad T \ge \p\eta_r^2 b.
\end{equation}
\end{thm}
The requirement $T\ge p\eta_r^2 b$  above is only for cleaner bounds.
 A general expression for arbitrary $T$ is given in the proof. 

 \paragraph*{\textbf{Discussion of Theorem \ref{thm:stochastic result}}}
On the one hand, \eqref{eq:near residual of moses thm} states that \editr{$\alg$  successfully reduces the dimension of streaming data, namely, \eqref{eq:indirect 1.5} holds under certain conditions:} \editr{Loosely-speaking,}  \eqref{eq:near residual of moses thm} states that  $\|\Y_T-\wh{\Y}_{T,r}\|_F^2$ scales with $\rho_r^2 T^{p\eta_r^2}/b^{p\eta^2_r -1}$, \editr{where $\wh{\Y}_{T,r}$ is the output of MOSES.} \editr{In contrast with~\eqref{eq:true residual prop}, we have}
\begin{align}
& \|\Y_T - \wh{\Y}_{T,r}\|_F^2  \propto   \l(\frac{T}{b}\r)^{p\eta_r^2-1}   \rho_r^2(\Y_T) & \nonumber\\
& = \l(\frac{T}{b}\r)^{p\eta_r^2-1} \|\Y_T - \Y_{T,r}\|_F^2,
\qquad \text{(see \eqref{eq:empirical})}
\label{eq:gist} 
\end{align}
after ignoring the less important terms. 
In words,  applying \edit{(offline)} truncated SVD to $\Y_T$ outperforms the \edit{(streaming)} $\alg$ by  a polynomial factor in $T/b$.

This polynomial factor can be negligible when the covariance matrix $\B{\Xi}$ of the Gaussian data distribution is well-conditioned ($\kappa_r =O(1)$) and has a small residual  ($\rho_r^2 =O(\lambda_r)$), in which case we will have $\eta_r =O(1)$, see \eqref{eq:spect of guassian thrm}. \editr{Also setting} $p=O(1)$, \eqref{eq:gist} then reads as 
\begin{align}
\|\Y_T - \wh{\Y}_{T,r}\|_F^2 \propto \l(\frac{T}{b}\r)^{O(1)}\rho_r^2(\Y_T),
\label{eq:intuition}
\end{align}
\editr{namely, (streaming) MOSES is comparable to offline truncated SVD.}
In particular, when $\rank(\B{\Xi}) \le r$, we have by \eqref{eq:res of gaussian dist} that $\rho_r=0$. Consequently,  \eqref{eq:gist} reads as $\wh{\Y}_{T,r}=\Y_{T,r}=\Y_T$. \editr{That is,}  the outputs of offline truncated SVD and $\alg$ coincide \editr{in this case}. 


The dependence \edit{in Theorem \ref{thm:stochastic result}} on the condition number $\kappa_r$ and the residual $\rho_r$ is very likely \emph{not} an artifact of the proof techniques; see \eqref{eq:spect of guassian thrm}. Indeed, when $\kappa_r\gg 1$, certain directions are less often observed in the incoming data vectors $\{
y_t\}_{t=1}^T$, which tilts the estimate of $\alg$ towards the dominant principal components corresponding to the very large singular values. Moreover, if $\rho_r \gg 1$, there are too many significant principal components, \edit{while} $\alg$ can at most ``remember'' $r$ of them from its previous iteration. In this scenario, \edit{fitting an $r$-dimensional subspace to data} is not a good idea in the first place \edit{because even} the residual $\rho_r(\Y_T)$ of the offline truncated SVD will be large, and we should perhaps increase the dimension $r$ of the subspace fitted to the incoming data.

As \edit{the block size} $b$ increases, \editr{the} performance of $\alg$ approaches that of the offline truncated SVD. In particular, when $b=T$, $\alg$ \editr{reduces to offline truncated SVD, processing all of the data at once.} This trend is somewhat imperfectly reflected in \eqref{eq:near residual of moses thm}. 

On the other hand, Theorem \ref{thm:stochastic result} and specifically \eqref{eq:final bnd} state that
\editr{MOSES approximately estimates the (unknown) underlying subspace, namely,}  \eqref{eq:ideal} holds under certain conditions. Indeed, for sufficiently large $T$, \eqref{eq:final bnd} \editr{loosely-speaking} reads as 
\begin{align}
& \underset{y\sim\mu}{\E} \|y-\P_{\wh{\SU}_{T,r}}y\|_2^2  \propto \l( \frac{T}{b} \r)^{p\eta_r^2-1} \rho_r^2 \nonumber\\
&  = \l( \frac{T}{b} \r)^{p\eta_r^2-1} \min_{\U\in\GR(n,r)} \underset{y\sim \mu}{\E}\l\| y - \P_{\U} y\r\|_2^2,
\end{align}
see Program \eqref{eq:direct}. 
That is, the output of $\alg$ is sub-optimal for Program \eqref{eq:direct} by a polynomial factor in $T$, which is negligible if the covariance matrix $\B{\Xi}$ of the data distribution $\mu$ is well-conditioned and has a small residual, \edit{as discussed earlier.}

\edit{As the closing remark, Section 3 of the supplementary material applies Theorem \ref{thm:stochastic result} to the popular spiked covariance model as a special case, outlines the proof technique, and discusses extension to non-Gaussian stochastic models.}

\section{Prior Art \label{sec:lit review}}

In this paper, we presented $\alg$ for streaming (linear) dimensionality reduction, an algorithm with (nearly) minimal storage and computational requirements. \editr{We can} think of $\alg$ as an online ``subspace tracking'' algorithm that identifies the linear structure of data as it arrives. Once the data has fully arrived, both principal components and the projected data are \editr{already} made available by $\alg$\edit{, ready for any additional processing.} 


\edit{\editr{The well-known iSVD is a special case of MOSES}. At iteration $t$ and  given the (truncated) SVD of $\Y_{t-1}$, iSVD aims to compute the  (truncated) SVD of $\Y_t = [\Y_{t-1}\; y_t] \in\R^{n\times t}$, where $y_t\in\R^{n}$ is the newly arrived data vector and $\Y_{t-1}$ is the matrix formed by concatenating the previous data vectors}
\cite{bunch1978rank,brand2006fast,brand2002incremental,comon1990tracking,li2004incremental}. $\alg$  generalises iSVD to handle data blocks; see Algorithm~\ref{alg:MOSES}. 
This \edit{minor generalisation} in part enables us to complement $\alg$ \editr{(and iSVD)} with  comprehensive statistical analysis in Theorem~\ref{thm:stochastic result},  
which has been lacking despite the popularity and empirical success of iSVD.

\edit{In fact},  \cite{balsubramani2013fast} only very recently provided stochastic analysis for two of the variants of iSVD in \cite{krasulina1969method,oja1983subspace}. The results in \cite{balsubramani2013fast} hold in expectation and for the special case of $r=1$, the first leading principal component. Crucially, these results measure the angle $\angle[\SU_{T,r},\wh{\SU}_{T,r}]$ between the true leading principal components of the data matrix and those estimated by iSVD.

Such results are \editr{therefore} inconclusive because \edit{they are silent about the  dimensionality reduction task \editr{of iSVD}. Indeed,} iSVD \edit{can} estimate both left and right leading singular vectors of the data matrix, namely, iSVD \edit{can estimate} both the leading principal components of the data matrix $\wh{\B{S}}_{T,r}$ \emph{and} reduce the dimension of data by computing $\wh{\B{S}}_{T,r}^* \wh{\Y}_{T,r}\in \R^{r\times T}$, where $\wh{\B{S}}_{T,r}$ and $\wh{\Y}_{T,r}$ are the final outputs of iSVD.  \editr{Unlike} \cite{balsubramani2013fast},  Theorem \ref{thm:stochastic result} and specifically \eqref{eq:near residual of moses thm} assesses the quality of both of these tasks and establishes that, under certain conditions, $\alg$ performs nearly as well as \edit{the} offline \editr{truncated} SVD.  

$\operatorname{GROUSE}$  \cite{balzano2013grouse} is another algorithm for streaming PCA\edit{, for data with possibly missing entries. GROUSE} can be interpreted as  projected stochastic gradient descent on the Grassmannian manifold. GROUSE is effectively identical to iSVD when the incoming data is low-rank \cite{balzano2013grouse}.  In \cite{zhang2016global} and \editr{on the basis of} \cite{balsubramani2013fast}, the authors offer theoretical guarantees for GROUSE that again does {not} account for \edit{the quality of dimensionality reduction}. Their results hold  without any missing data, in expectation, and in a setup similar to the spiked covariance model. An alternative to GROUSE is SNIPE that has  stronger theoretical guarantees in \edit{the} case of missing data~\cite{eftekhari2016snipe,eftekhari2016expect}. In Section~\ref{sec:numerics}, we will numerically compare $\alg$ with GROUSE.

\edit{The closely-related method of Frequent Directions (FD) replaces the hard thresholding of the singular values in iSVD with soft thresholding \cite{ghashami2016frequent,desai2016improved}. Later, robust FD \cite{luo2017robust}  improved the performance of FD and addressed some of its numerical issues. 
On the algorithmic side, FD keeps an estimate of $\B{S}_{t,r} \B{\Gamma}_{t,r}$, whereas $\alg$ also calculates the projected data, namely, it keeps an estimate of $\{\B{S}_{t,r},\B{\Gamma}_{t,r},\B{Q}_{t,r}\}_t$; see Algorithm~\ref{alg:MOSES implement}. In terms of guarantees, the available guarantees for FD control the error incurred in estimating the principal components, whereas Theorem \ref{thm:stochastic result} for $\alg$ also controls the error incurred in \editr{the} dimensionality reduction step. Indeed, 
\begin{align}
& \l\| \B{S}_{T,r} \B{\Gamma}_{T,r}^2 \B{S}_{T,r}^* - \wh{\B{S}}_{T,r} \wh{\B{\Gamma}}_{T,r}^2 \wh{\B{S}}_{T,r}^*   \r\|_F \nonumber\\
& \le  \| \B{Y}_{T,r} - \wh{\B{Y}}_{T,r}   \|_F \cdot \l( \|\B{Y}_{T,r}\| + \|\wh{\B{Y}}_{T,r}\|  \r ),
\end{align}
namely, Theorem~\ref{thm:stochastic result}  is a more general result than those exemplified by Theorem~1.1 in \cite{ghashami2016frequent} which, except for symmetric matrices, do not convey any information about the row-span. Another key difference is that Theorem~\ref{thm:stochastic result} is a stochastic result versus  the deterministic Theorem~1.1 in \cite{ghashami2016frequent}  and similar results. Indeed, an intermediate step to prove Theorem~\ref{thm:stochastic result} is the deterministic Lemma~2 in the supplementary material. An important feature of this work is to translate Lemma~2 into a stochastic result in learning theory, of interest to the machine learning and statistical signal processing communities. In Section~\ref{sec:numerics}, we will numerically compare MOSES with FD. 
}

\edit{
As detailed in Section~4 of the supplementary material, $\alg$ can be adapted to the dynamic case, where the distribution of data changes over time. This is achieved by using a ``forgetting factor'' in Step b of Algorithm~\ref{alg:MOSES}. Such an extension is crucial, as there are  pathological examples where (static) $\alg$ and iSVD both fail to follow the changes in the distribution of data~\cite{desai2016improved}. This important research direction is left for future work.
}

One might also view $\alg$ as a stochastic algorithm for PCA. Indeed, note that Program~\eqref{eq:direct} is equivalent to 
\begin{align}
\begin{cases}
\max \,\,\, \underset{y\sim\mu}{\E} \| \B{U} \B{U}^* y \|_F^2 \\
\B{U}^* \B{U} = \B{I}_r
\end{cases}
 & = 
\begin{cases}
\max \,\,\, \underset{y\sim\mu}{\E} \langle \B{U} \B{U}^*, y y^* \rangle  \\
\B{U}^* \B{U} = \B{I}_r
\end{cases} \nonumber\\
&=
\begin{cases}
\max \,\,\, \underset{y\sim\mu}{\E} \langle \B{U} \B{U}^*, y y^* \rangle  \\
\B{U}^* \B{U} \preccurlyeq \B{I}_r,
\end{cases}
\label{eq:prg review}
\end{align}
where the maximisation is over \edit{the} matrix $\B{U}\in\R^{n\times r}$. Above,  
$\B{U}^*\B{U}\preccurlyeq \B{I}_r$ is the unit ball with respect to the spectral norm and  $\B{A} \preccurlyeq \B{B}$ means that $\B{B}-\B{A}$ is a positive semi-definite matrix. The last identity above holds because a convex function is always maximised on the boundary of the feasible set. \edit{With} the Schur complement, we can equivalently write the last program above as 
\begin{align}
 \begin{cases}
\max \,\,\, \underset{y\sim\mu}{\E} \langle \B{U} \B{U}^*, y y^* \rangle  \\
\l[
\begin{array}{cc}
\B{I}_n & \B{U} \\
\B{U}^* & \B{I}_r
\end{array}
\r]
\succcurlyeq \B{0}
\end{cases} 
& = 
\begin{cases}
\max \,\,\, \langle \B{U} \B{U}^*, \B{\Xi} \rangle  \\
\l[
\begin{array}{cc}
\B{I}_n & \B{U} \\
\B{U}^* & \B{I}_r
\end{array}
\r]
\succcurlyeq \B{0},
\end{cases}
\label{eq:sdp formulation}
\end{align}
where $\B{\Xi}=\E [yy^*] \in\R^{n\times n}$ is the covariance matrix of the data distribution $\mu$. 
Program \eqref{eq:sdp formulation} has a convex (in fact, quadratic) objective function  that is \emph{maximised} on a convex (conic) feasible set. 
 We cannot hope to directly compute the gradient of the objective function above \edit{(namely, $2\B{\Xi}\B{U}$)} because the distribution of $y$ and hence its covariance matrix $\B{\Xi}$ are unknown. Given an iterate $\h{\B{S}}_{t}$, one might instead draw a random vector $y_{t+1}$ from the probability measure $\mu$ and move along the direction $2y_{t+1}y_{t+1}^*\h{\B{S}}_t$, motivated by the observation that $\E [2y_{t+1}y_{t+1}^*\h{\B{S}}_t ]= 2\B{\Xi}\widehat{\B{S}}_t$. This is then followed by back projection onto the feasible set of Program \eqref{eq:prg review}. That is, 
\begin{equation}
\h{\B{S}}_{t+1} = \mathcal{P}\l(\B{S}_t + 2\alpha_{t+1} y_{t+1}y_{t+1}^* \h{\B{S}}_t  \r),
\end{equation}
for an appropriate step size $\alpha_{t+1}$. Above, $\mathcal{P}(\B{A})$ projects onto the unit spectral norm ball by setting to one all  singular values of $\B{A}$ that exceed one.

The stochastic projected gradient ascent  for PCA, described above, is itself closely related to the so-called \emph{power method} and is at the heart of \cite{mitliagkas2013memory,oja1985stochastic,sanger1989optimal,kim2005iterative,arora2012stochastic}, all lacking a statistical analysis similar  to Theorem \ref{thm:stochastic result}. One notable exception is the power method in \cite{mitliagkas2013memory} which in a sense applies \emph{mini-batch} stochastic projected gradient ascent   to solve Program \eqref{eq:sdp formulation}, with data blocks (namely, batches) of size $b=\Omega(n)$. There the authors offer statistical guarantees for the spiked covariance model,  defined in Section \ref{sec:results} of the supplementary material. As before, these guarantees are for the quality of estimated principal components and silent about the quality of projected data, which is addressed in Theorem \ref{thm:stochastic result}. Note also that, especially when the data dimension $n$ is large, one disadvantage of this approach is its large block size; it takes a long time \edit{(namely, $\Omega(n)$ iterations)} for the algorithm to update its estimate of the principal components,\edit{ a big disadvantage in the dynamic case.} In this setup, we may think of $\alg$ as a stochastic algorithm for PCA based on alternative minimisation rather than gradient ascent, as detailed in Section~3 of the supplementary material. Moreover, $\alg$ updates its estimate frequently, after receiving every $b=O(r)$ data vectors, and  also maintains the projected data. In Section \ref{sec:numerics}, we numerically compare $\alg$ with the power method in \cite{mitliagkas2013memory}. A few closely related works are~\cite{hardt2014noisy,de2014global,jain2016streaming,de2014global}.

In the context of online learning and \emph{regret minimisation}, \cite{warmuth2008randomized,arora2012stochastic} offer two algorithms, the former of which is not memory optimal and the latter does not have guarantees similar to Theorem \ref{thm:stochastic result}; see also \cite{boutsidis2015online}. A Bayesian approach to PCA is studied in \cite{roweis1998algorithms,tipping1999probabilistic} and the \emph{expectation maximisation} algorithm therein  could be implemented in an online fashion but without theoretical guarantees.

\edit{More generally, $\alg$ might be interpreted as a deterministic \emph{matrix sketching} algorithm. Common sketching algorithms either randomly sparsify a matrix, randomly combine its rows (columns), or randomly subsample its rows (columns) according to its \emph{leverage scores}  \cite{tropp2017fixed,chiu2013sublinear,pourkamali2016randomized,gittens2016revisiting,ghashami2016frequent,gilbert2012sketched}. In particular, FD was observed to outperform random sketching algorithms in practice \cite{ghashami2016frequent}.  The relation between streaming algorithms and distributed computing is also perhaps worth pointing out; see Figure~\ref{fig:cone tree} of the supplementary material and \cite{iwen2016distributed,sameh2005parallel}}. Lastly, when the data vectors have missing entries, a closely related problem is low-rank matrix completion~\cite{davenport2016overview,candes2009exact,eftekhari2018weighted,eftekhari2016mc}.

\section{Experiments \label{sec:numerics}}

In this section, we investigate  the numerical performance of $\alg$ and compare it against \edit{competing  algorithms, namely},   GROUSE~\cite{balzano2013grouse}, \edit{the method of frequent directions (FD)~\cite{desai2016improved,luo2017robust},} and the power method (PM)~\cite{mitliagkas2014streaming}, \editr{all detailed in Section~\ref{sec:lit review}.}
\edit{On both synthetic and real-world datasets,} we reveal one by one the data vectors $\{y_t\}_{t=1}^T\subset \mathbb{R}^n$  and, for every $t$, \edit{wish} to  compute a rank-$r$ truncated SVD of \edit{$[y_1,\cdots,y_t]$, the data  arrived so far.}

For the tests \edit{on synthetic datasets}, the vectors $\{y_t\}_{t=1}^T$ are drawn independently from a zero-mean Gaussian distribution with \edit{the} covariance matrix $\B{\Xi}=\B{S}\B{\Lambda}\B{S}^*$, where $\B{S}\in \text{O}(n)$ is a generic orthonormal basis obtained by orthogonalising a standard random Gaussian matrix. The entries of the diagonal matrix $\B{\Lambda}\in\mathbb{R}^{n\times n}$ (the eigenvalues of the covariance matrix $\B{\Xi}$) are selected according to the power law\edit{, namely,}  $\lambda_i =  i^{-\alpha}$, for a positive $\alpha$. To be more succinct, where possible we \edit{will} use {MATLAB}'s notation for specifying the value ranges in this section. 

To assess the performance of $\alg$, let $\Y_t=[y_1,\cdots,y_t]\in \mathbb{R}^{n\times t}$ be the data received by time $t$ and let $\widehat{\Y}^m_{t,r}$ be the output of $\alg$ at time $t$.\footnote{Note that $\alg$ updates its estimate after receiving each block of data, namely after every $b$ data vectors. For the sake of an easier comparison with other algorithms (with different block sizes), we properly ``interpolate'' the outputs of all algorithms over time.} Then the error incurred by $\alg$ is 
\begin{equation}
\frac{1}{t}\|\Y_t - \widehat{\Y}^m_{t,r}\|_F^2, 
\end{equation} 
see Theorem \ref{thm:stochastic result}.
Recall from \eqref{eq:residual} that the above error is always larger than the residual of $\Y_t$, namely
\begin{equation}
\|\Y_t - \widehat{\Y}^m_{t,r}\|_F^2 \ge \|\Y_t - {\Y}_{t,r}\|_F^2 = \rho_r^2(\Y_t),
\end{equation}
see \eqref{eq:residual}. Above, ${\Y}_{t,r}=\SVD_r(\Y_t)$ is a rank-$r$ truncated SVD of $\Y_t$ and $\rho_r^2(\Y_t)$ is the corresponding residual.

Later in this section, we compare $\alg$ against {GROUSE}~\cite{balzano2013grouse}, \edit{FD~\cite{desai2016improved}},  \edit{PM}~\cite{mitliagkas2013memory}, described in Section~\ref{sec:lit review}. \edit{In contrast with MOSES,} these algorithms only estimate the principal components of the data. 
More specifically, \edit{let $\widehat{\SU}_{t,r}^g \in \text{G}(n,r)$ be the span of the output of GROUSE, with the outputs of the other algorithms defined similarly.} These algorithms then incur the errors
\begin{equation*}
\frac{1}{t} \| \Y_t - \P_{\widehat{\SU}_{t,r}^g} \Y_t \|_F^2,
\qquad
\frac{1}{t} \| \Y_t - \P_{\widehat{\SU}_{t,r}^f} \Y_t \|_F^2,
\end{equation*}
\begin{equation}
\frac{1}{t} \| \Y_t - \P_{\widehat{\SU}_{t,r}^p} \Y_t \|_F^2,
\end{equation}
respectively. 
Above, $\P_{\mathcal{A}}\in \mathbb{R}^{n\times n}$ is the orthogonal projection onto the subspace $\mathcal{A}$. \edit{Even though robust FD \cite{luo2017robust} improves over FD in the quality of matrix sketching, since the subspaces produced by FD and robust FD coincide, there is no need here for computing a separate error for  robust FD.}
We now set out to do various tests and report the results. \edit{To ensure the reproducibility of our results, both the accompanying MATLAB code and the datasets used are publicly available}.\footnote{\href{https://github.com/andylamp/moses}{github.com/andylamp/moses}} 
\\

\begin{figure}[h]
\vspace{-12pt}
\begin{center}
\subfloat[Variable $n$, for $r=15$, $b=2r$ \label{fig:num_eval_mos_scaling_var_n}]
{
\iftoggle{OVERLEAF}{
    \includegraphics[width=.3\textwidth]{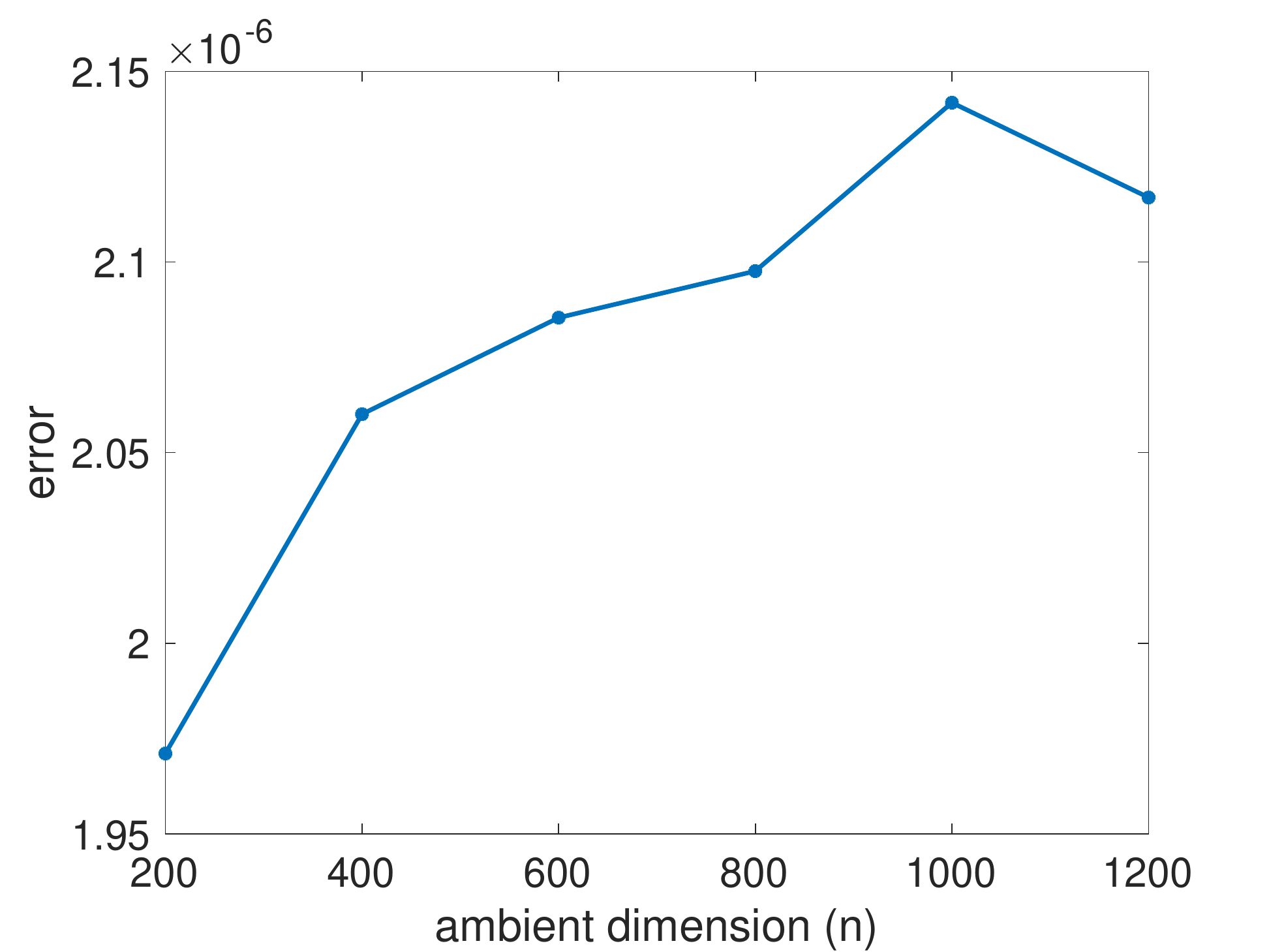}
}{
    \includegraphics[width=.3\textwidth]{moses_scaling_exp1_final_error_fixed_b_r_var_n_T_3k_notitle}
}
}
 \subfloat[Variable $b$, for $n=1200$, $r=15$ \label{fig:num_eval_mos_scaling_var_b}]{
 \iftoggle{OVERLEAF}{
        \includegraphics[width=.3\textwidth]{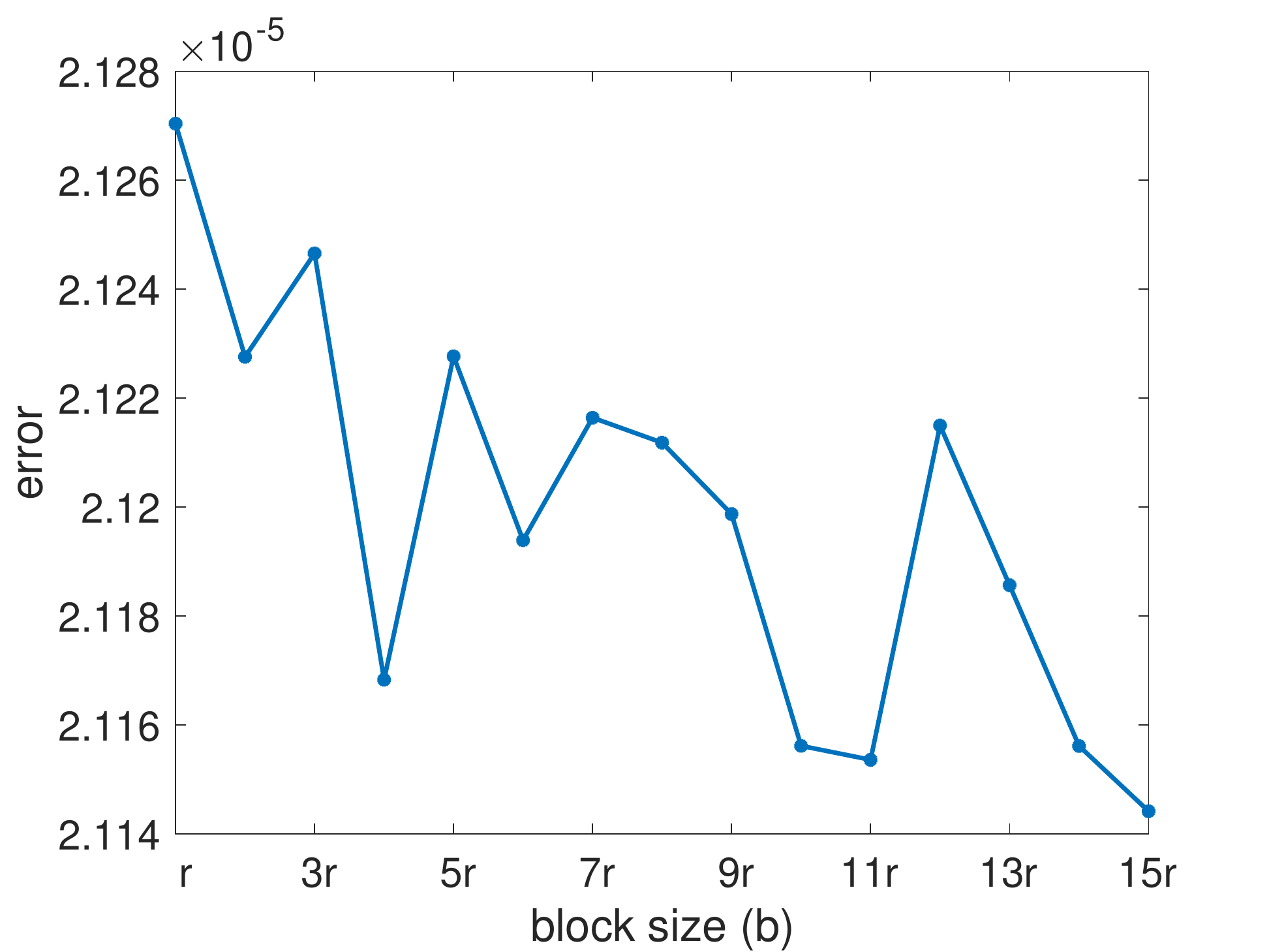}
        }{
        \includegraphics[width=.3\textwidth]{moses_scaling_exp2_final_error_fixed_n_r_var_b_T_3k_notitle}
        }
        }
        \subfloat[Variable $r$, for $n=1200$, $b=2r$ \label{fig:num_eval_mos_scaling_var_r}]{
\iftoggle{OVERLEAF}{
        \includegraphics[width=.3\textwidth]{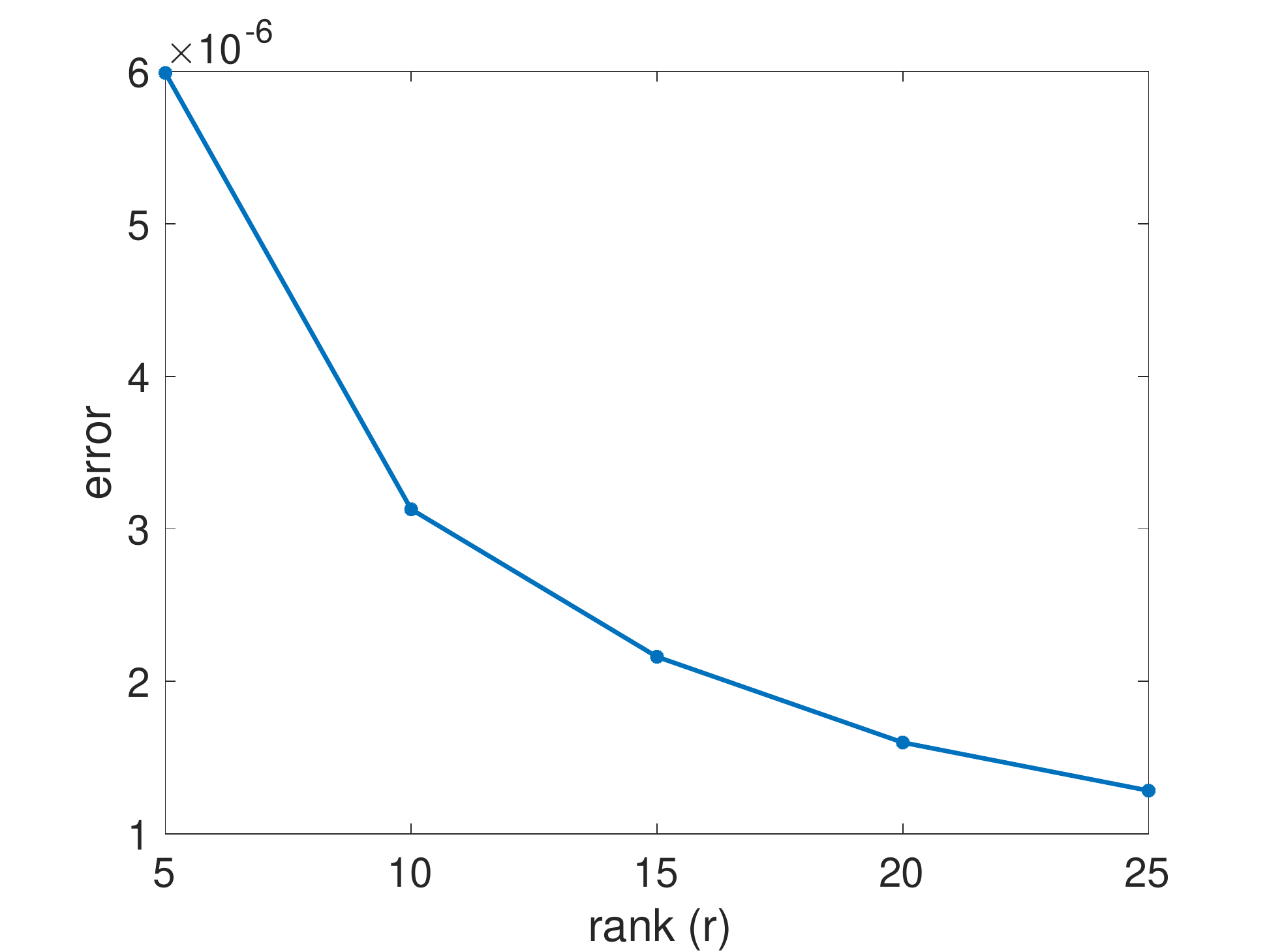}
        }{
        \includegraphics[width=.3\textwidth]{moses_scaling_exp3_final_error_fixed_n_b_var_r_T_3k_notitle}
        }
        }
    \vspace{-6pt}
    \caption{Performance of $\alg$ on synthetic datasets, see Section \ref{sec:numerics} for the details.  \label{fig:num_eval_mos_scaling_exp}}
    \end{center}
\end{figure}

\noindent \textbf{Ambient dimension:} On a synthetic dataset with $\alpha =1$ and $T=2000$, we first test $\alg$ by varying the ambient dimension as  $n\in\{200:200:1200\}$, and setting the rank and block size to $r=15$, $b=2r=30$. The average error over ten trials is reported in Figure~\ref{fig:num_eval_mos_scaling_var_n}. Note that the error is increasing in  $n$, \edit{which indeed agrees with Theorem \ref{thm:stochastic result}, as detailed in the supplementary material under the discussion of the spiked covariance model in Section~\ref{sec:results} therein. }

\noindent{\textbf{Block size:}} 
On a synthetic dataset with $\alpha =1$ and $T=2000$, we  test $\alg$ by setting the ambient dimension and rank to $n=1200$, $r=15$, and varying the block size as $b\in\{r:r:15r\}$. The average error over ten trials is reported in Figure~\ref{fig:num_eval_mos_scaling_var_b}. Note that  $\alg$ is robust against the choice of the block size and that, at the extreme case of $b=T$, error vanishes and $\alg$ reduces to offline truncated SVD. \edit{This is predicted by Theorem \ref{thm:stochastic result}, as seen in \eqref{eq:gist}. }

\noindent{\textbf{Rank:}} 
On a synthetic dataset with $\alpha =1$ and $T=2000$, we  test $\alg$ by setting the ambient dimension and block size to $n=1200$, $b=2r$, and varying the rank as $r\in\{5:5:25\}$. The average error over ten trials is reported in Figure~\ref{fig:num_eval_mos_scaling_var_r}. As  expected, the error is decreasing in the dimension $r$ of the subspace that we fit to the data and in fact, at the extreme case of $r=n$, there would be no error at all.  \edit{This observation is corroborated with Theorem \ref{thm:stochastic result} and, in particular,~\eqref{eq:gist}.}

\begin{figure}[htb]
\vspace{-24pt}
\begin{center}
\subfloat[$\alpha=0.01$ \label{fig:num_eval_mos_synth_alpha_0_01}]{
\iftoggle{OVERLEAF}{
    \includegraphics[clip, trim=1cm 7.5cm 1cm 8cm,width=0.4\textwidth]{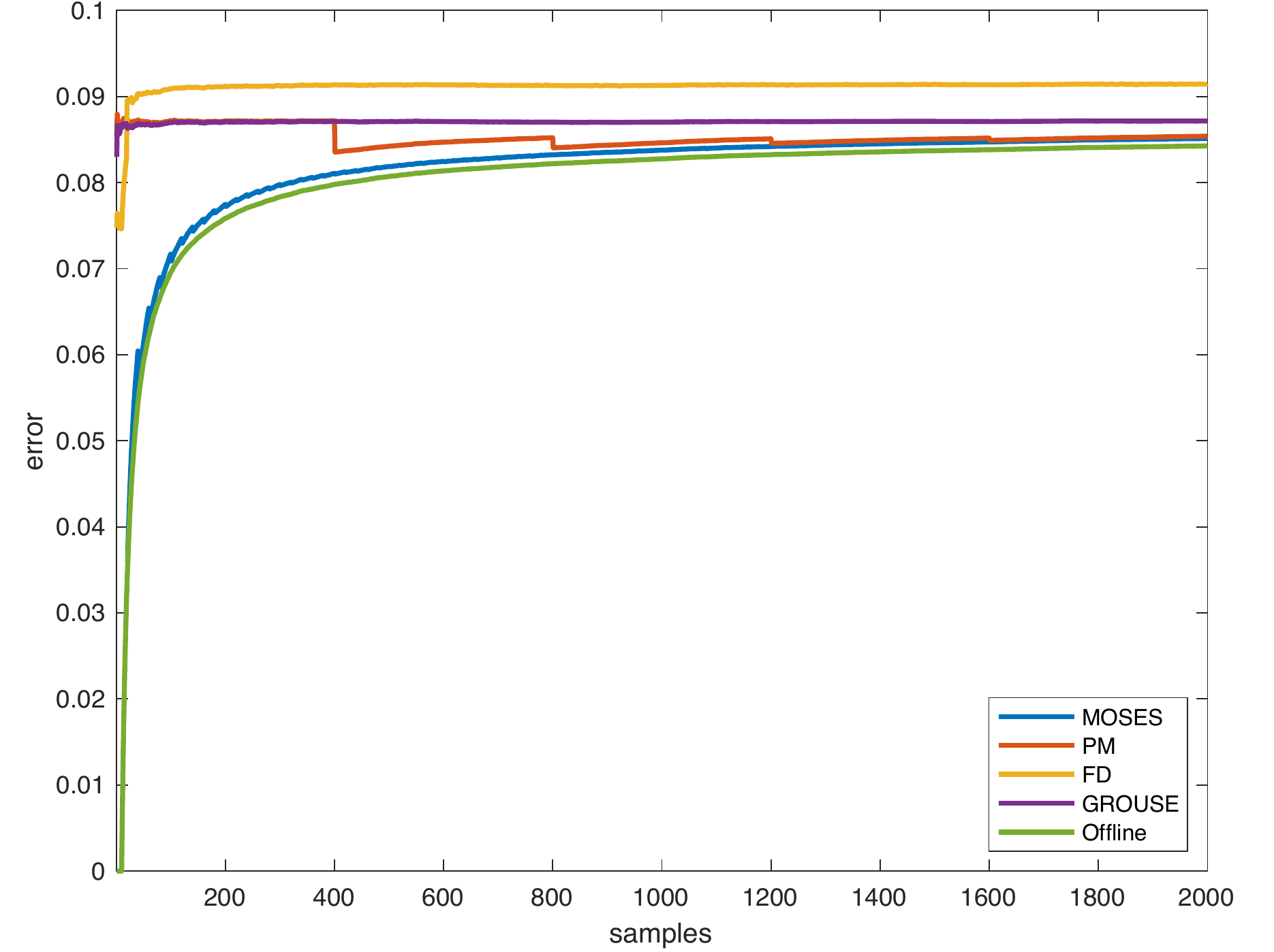}
}{
    \includegraphics[clip, trim=1cm 7.5cm 1cm 8cm,width=0.4\textwidth]{synthetic_froerror_noscree_n_200_r_10_alpha_0_01_nsim_10_notitle}
}
 }
\subfloat[$\alpha=0.1$ \label{fig:num_eval_mos_synth_alpha_0_1}]{
\iftoggle{OVERLEAF}{
    \includegraphics[clip, trim=1cm 7.5cm 1cm 8cm,width=0.4\textwidth]{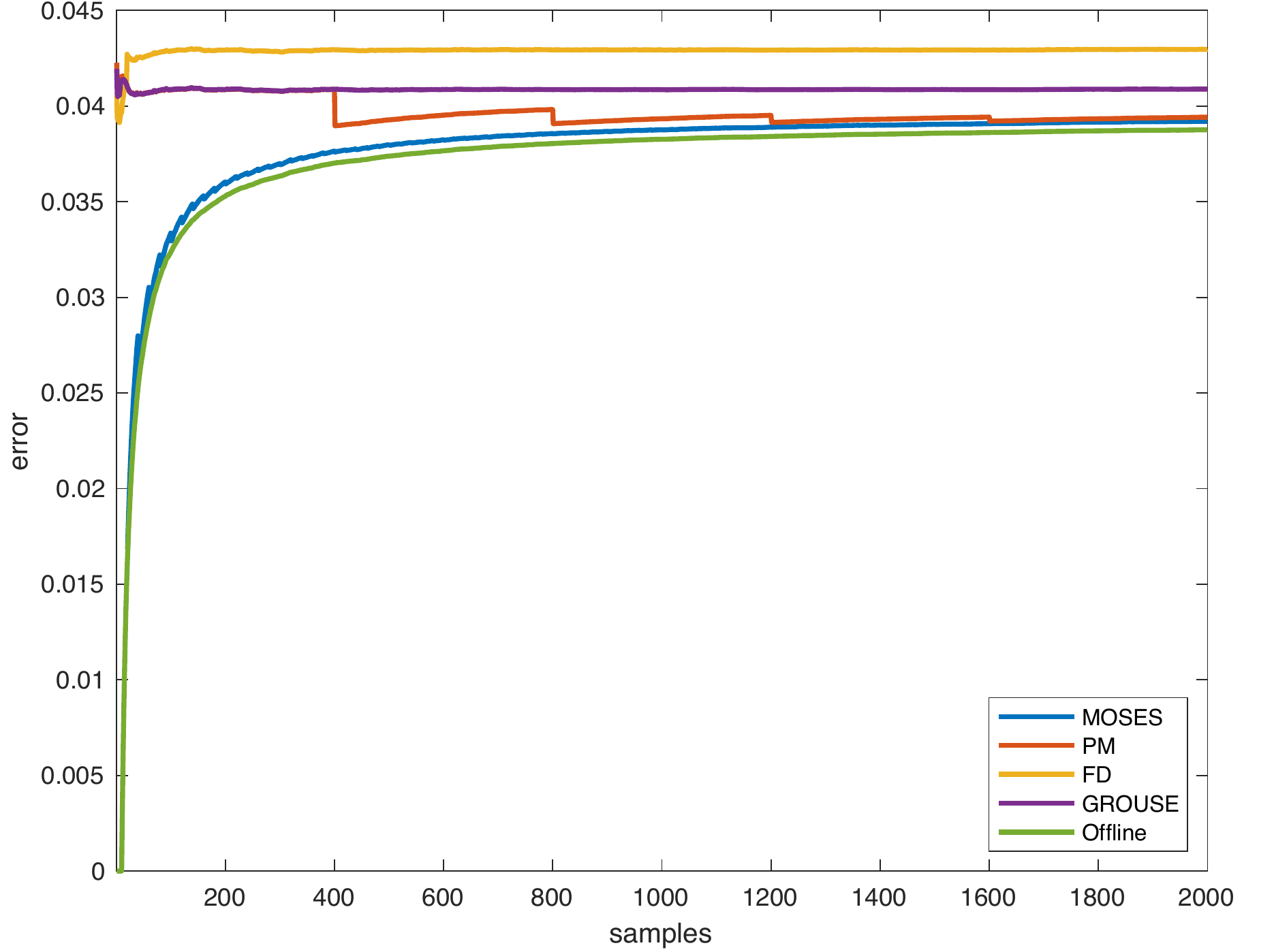}
}{
    \includegraphics[clip, trim=1cm 7.5cm 1cm 8cm,width=0.4\textwidth]{synthetic_froerror_noscree_n_200_r_10_alpha_0_1_nsim_10_notitle}
}
}\\[-2ex]
\subfloat[$\alpha=0.5$ \label{fig:num_eval_mos_synth_alpha_0_5}]{
\iftoggle{OVERLEAF}{
    \includegraphics[clip, trim=1cm 7.5cm 1cm 8cm,width=0.4\textwidth]{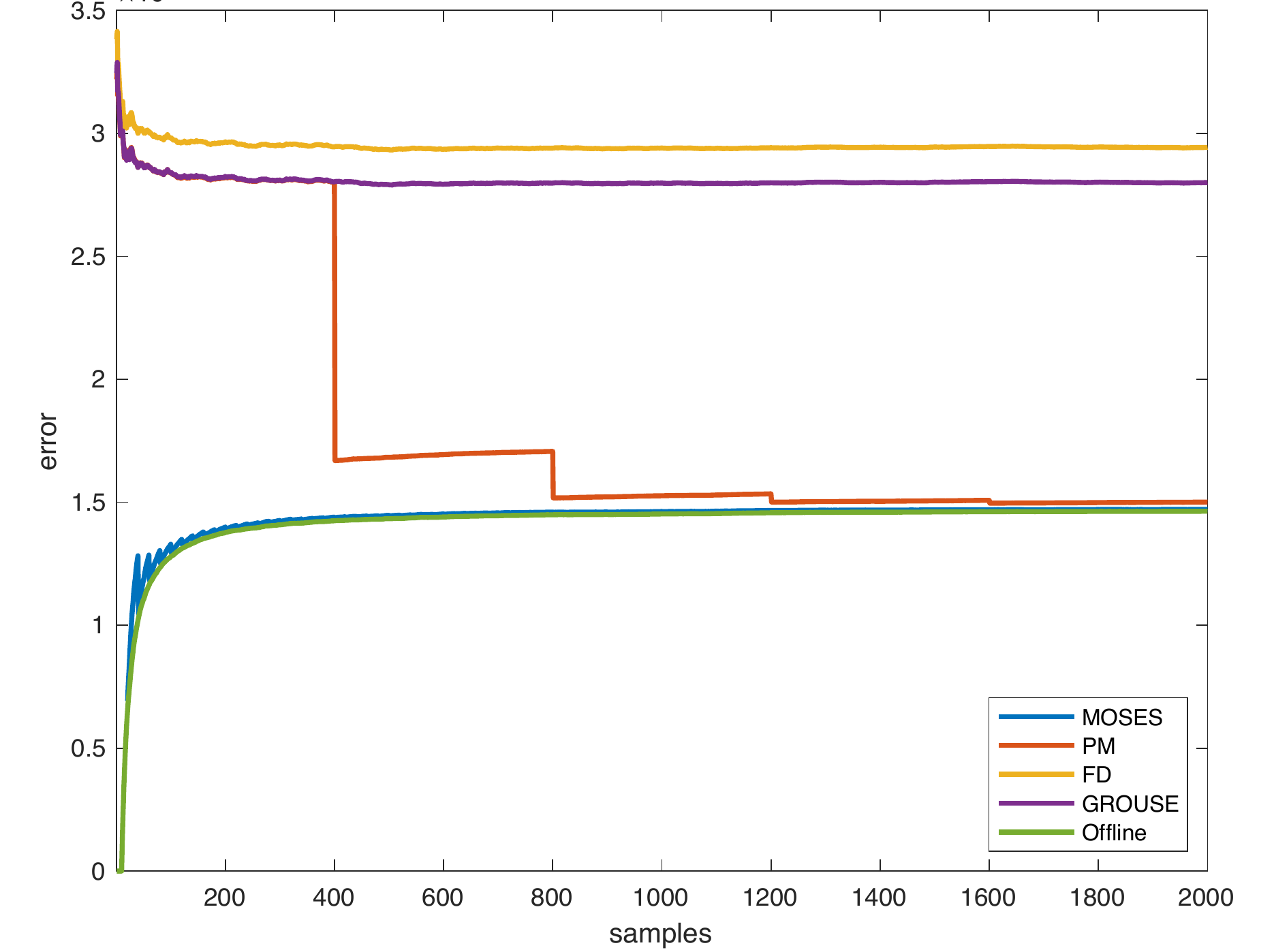}
}{
    \includegraphics[clip, trim=1cm 7.5cm 1cm 8cm,width=0.4\textwidth]{synthetic_froerror_noscree_n_200_r_10_alpha_0_5_nsim_10_notitle}
}
}
\subfloat[$\alpha=1$ \label{fig:num_eval_mos_synth_alpha_1}]
{        
\iftoggle{OVERLEAF}{
\includegraphics[clip, trim=1.5cm 7.5cm 1cm 8cm,width=0.4\textwidth]{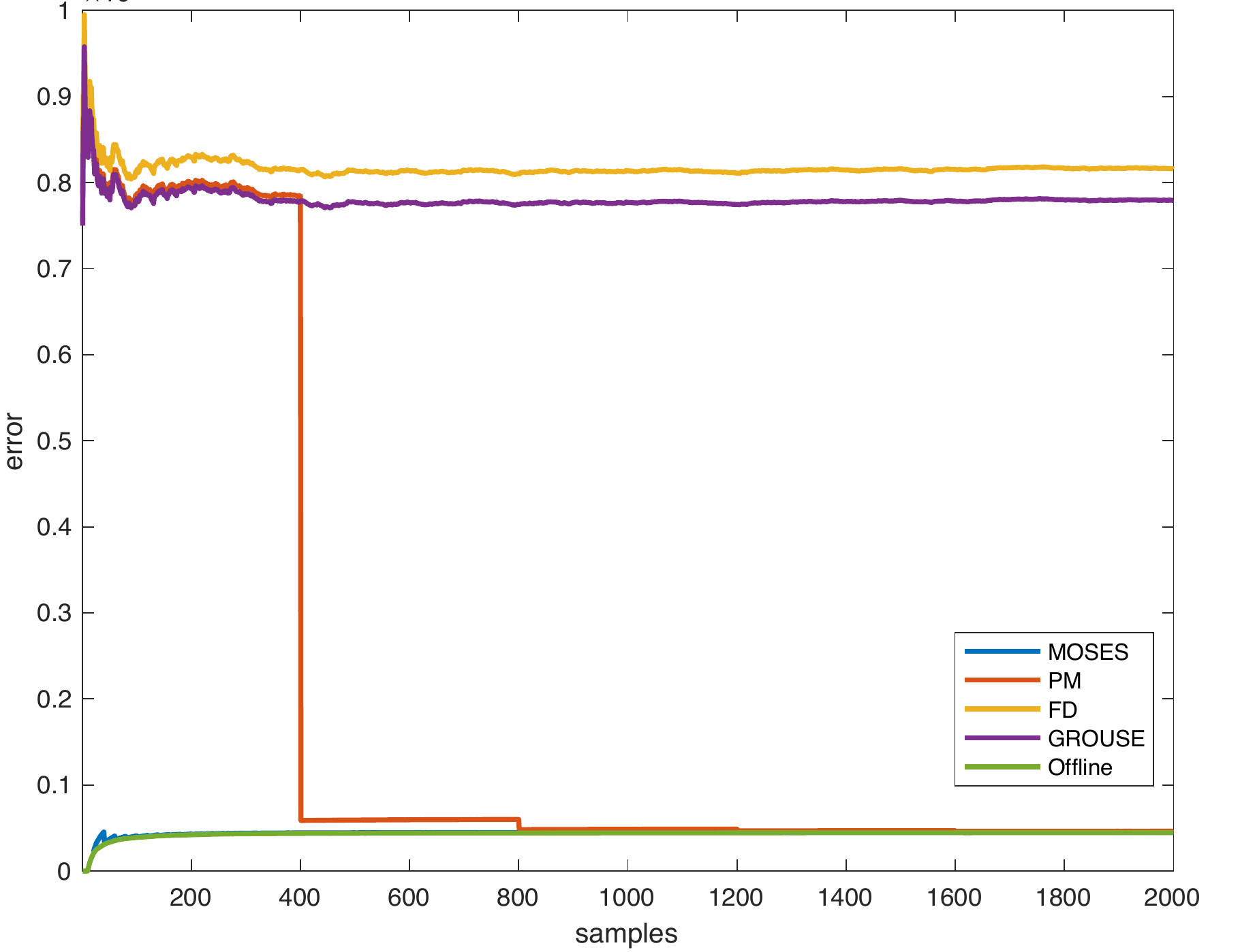}
}{
\includegraphics[clip, trim=1.5cm 7.5cm 1cm 8cm,width=0.4\textwidth]{synthetic_froerror_noscree_n_200_r_10_alpha_1_nsim_10_notitle}
}
}
    \vspace{-6pt}
    \caption{Comparisons on synthetic datasets, see Section \ref{sec:numerics} for the details. \label{fig:num_eval_mos_synth}}
\end{center}
\end{figure}

\noindent{\textbf{Comparisons on synthetic datasets:}} On synthetic datasets with $\alpha\in\{0.01,0.1,0.5,1\}$ and $T=2000$, we compare $\alg$ against  GROUSE, \edit{FD,  and PM}.\footnote{The MATLAB code for GROUSE is publicly available at \href{https://web.eecs.umich.edu/~girasole/grouse/}{web.eecs.umich.edu/{\raise.17ex\hbox{$\scriptstyle\sim$}}girasole/grouse}. \edit{The Python code for FD is available at \href{https://github.com/edoliberty/frequent-directions/}{github.com/edoliberty/frequent-directions/}}} More specifically, we set the ambient dimension to $n=200$ and the rank to $r=10$. For $\alg$, the block size was set to $b=2r$. \edit{For GROUSE, we set the step size to $2$. For FD and PM, the block size was set to  $2r=20$  and $2n=400$,} respectively, as these values seemed to produced the best results overall for these  algorithms.  Both GROUSE and PM were initialised randomly, as prescribed in \cite{balzano2013grouse,mitliagkas2014streaming}, \edit{while FD does not require any  initialisation}. The average errors of all three algorithms over ten trials versus time is shown in Figure~\ref{fig:num_eval_mos_synth}. 

Because of its large blocks size of $O(n)$~\cite{mitliagkas2014streaming}, PM updates its estimate of the principal components much slower than $\alg$, but the two algorithms converge to similar errors. The slow updates of \edit{PM is a major problem in a} dynamic scenario, where the distribution of data changes over time. We will also see later that $\alg$ is much faster than \edit{PM and performs better on the real datasets we tested.}

We \edit{next} evaluate all these \edit{four} algorithms on  
publicly-available \edit{sensor network data}; we use four different datasets that contain \emph{mote} (sensor node)  voltage, humidity, light, and 
temperature measurements over time~\cite{deshpande2004model}.

\begin{figure}[htb]
\begin{center}
    \vspace{-24pt}
    \subfloat[Voltage dataset \label{fig:num_eval_mos_real_volt}]{
    \iftoggle{OVERLEAF}{
    \includegraphics[clip, trim=1cm 7.5cm 1cm 8cm,width=0.4\textwidth]{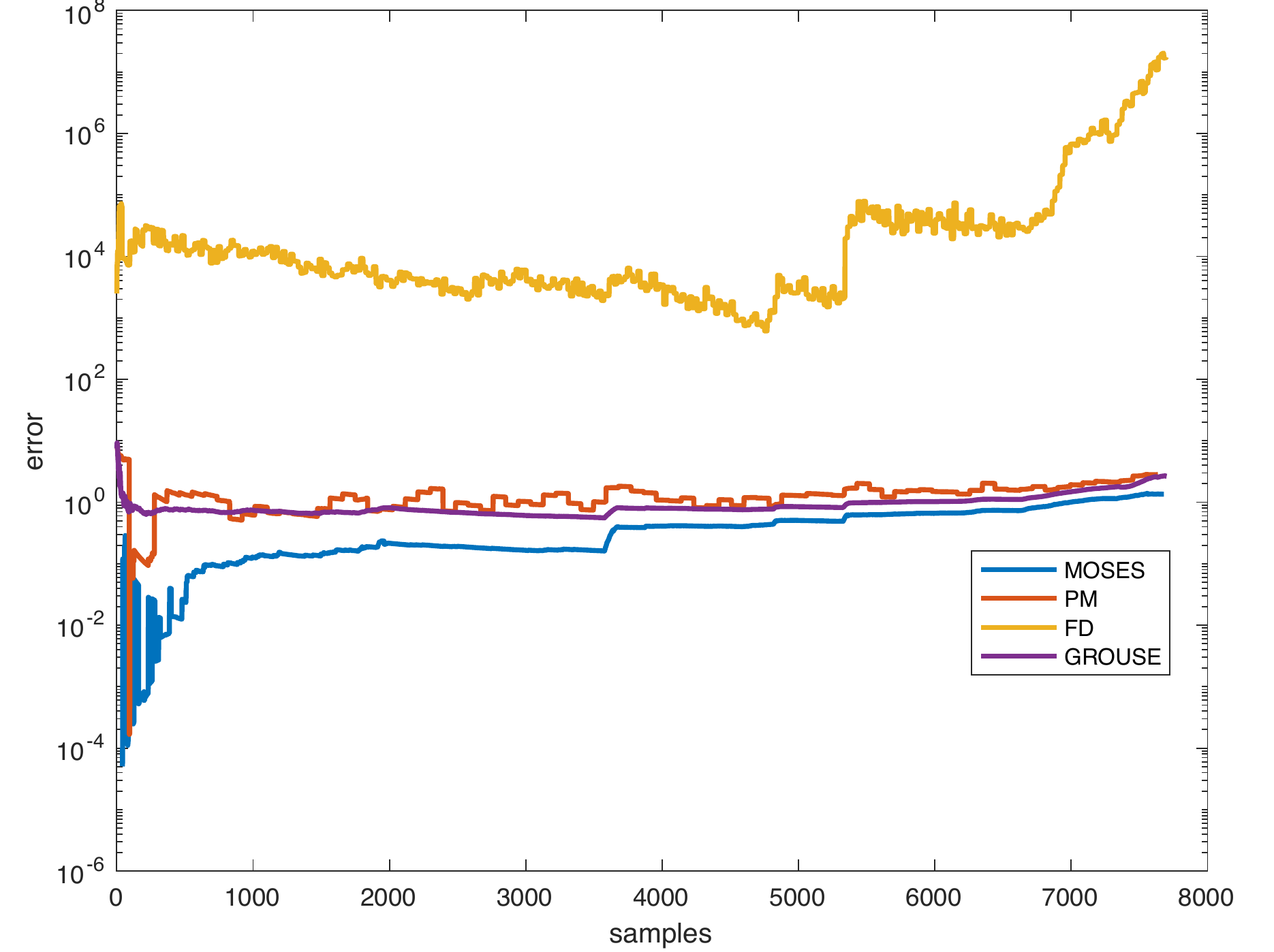}
    }{
    \includegraphics[clip, trim=1cm 7.5cm 1cm 8cm,width=0.4\textwidth]{real_froerror_n_46_r_20_Voltage_dataset_notitle}
    }
}
\subfloat[Humidity dataset \label{fig:num_eval_mos_real_humidity}]{
    \iftoggle{OVERLEAF}{
        \includegraphics[clip, trim=1cm 7.5cm 1cm 8cm,width=0.4\textwidth]{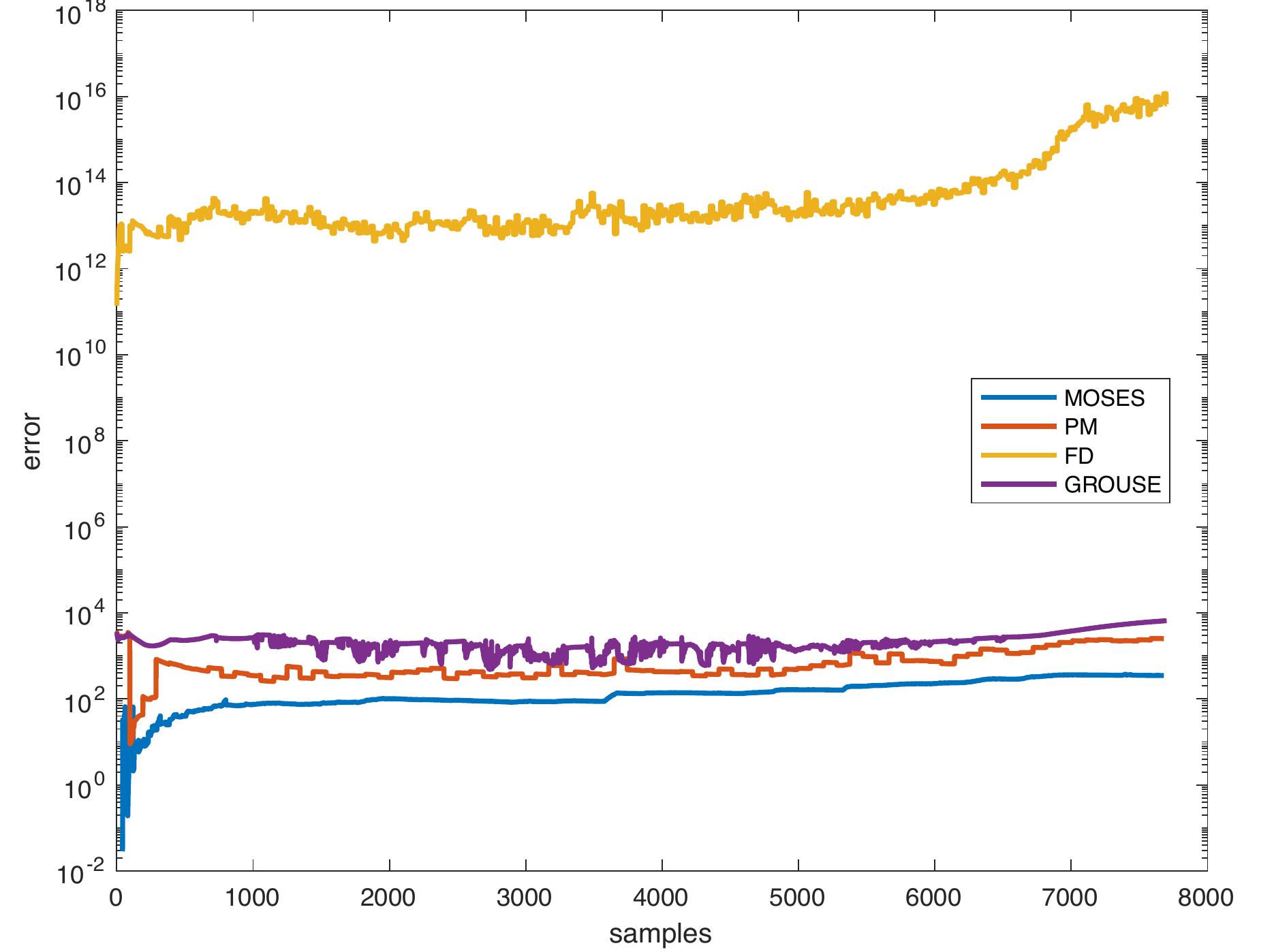}
    }{
        \includegraphics[clip, trim=1cm 7.5cm 1cm 8cm,width=0.4\textwidth]{real_froerror_n_48_r_20_Humidity_dataset_notitle}
    }
}\\[-2ex]
\subfloat[Light dataset \label{fig:num_eval_mos_real_light}]{
    \iftoggle{OVERLEAF}{
        \includegraphics[clip, trim=1cm 7.5cm 1cm 8cm,width=0.4\textwidth]{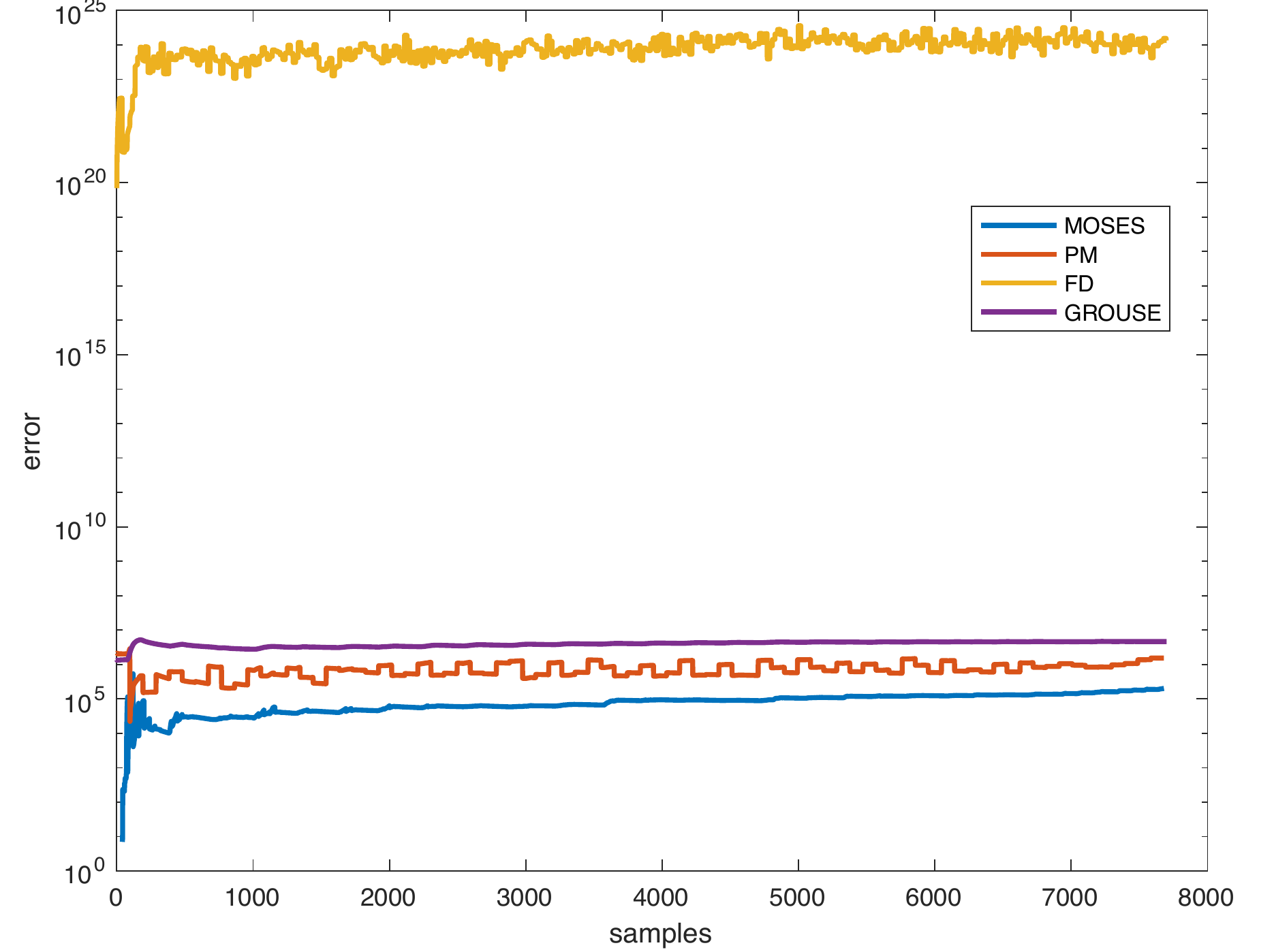}
    }{
        \includegraphics[clip, trim=1cm 7.5cm 1cm 8cm,width=0.4\textwidth]{real_froerror_n_48_r_20_Light_dataset_notitle}
    }
}
\subfloat[Temperature dataset \label{fig:num_eval_mos_real_temp}]{ 
    \iftoggle{OVERLEAF}{
       \includegraphics[clip, trim=1cm 7.5cm 1cm 8cm,width=0.4\textwidth]{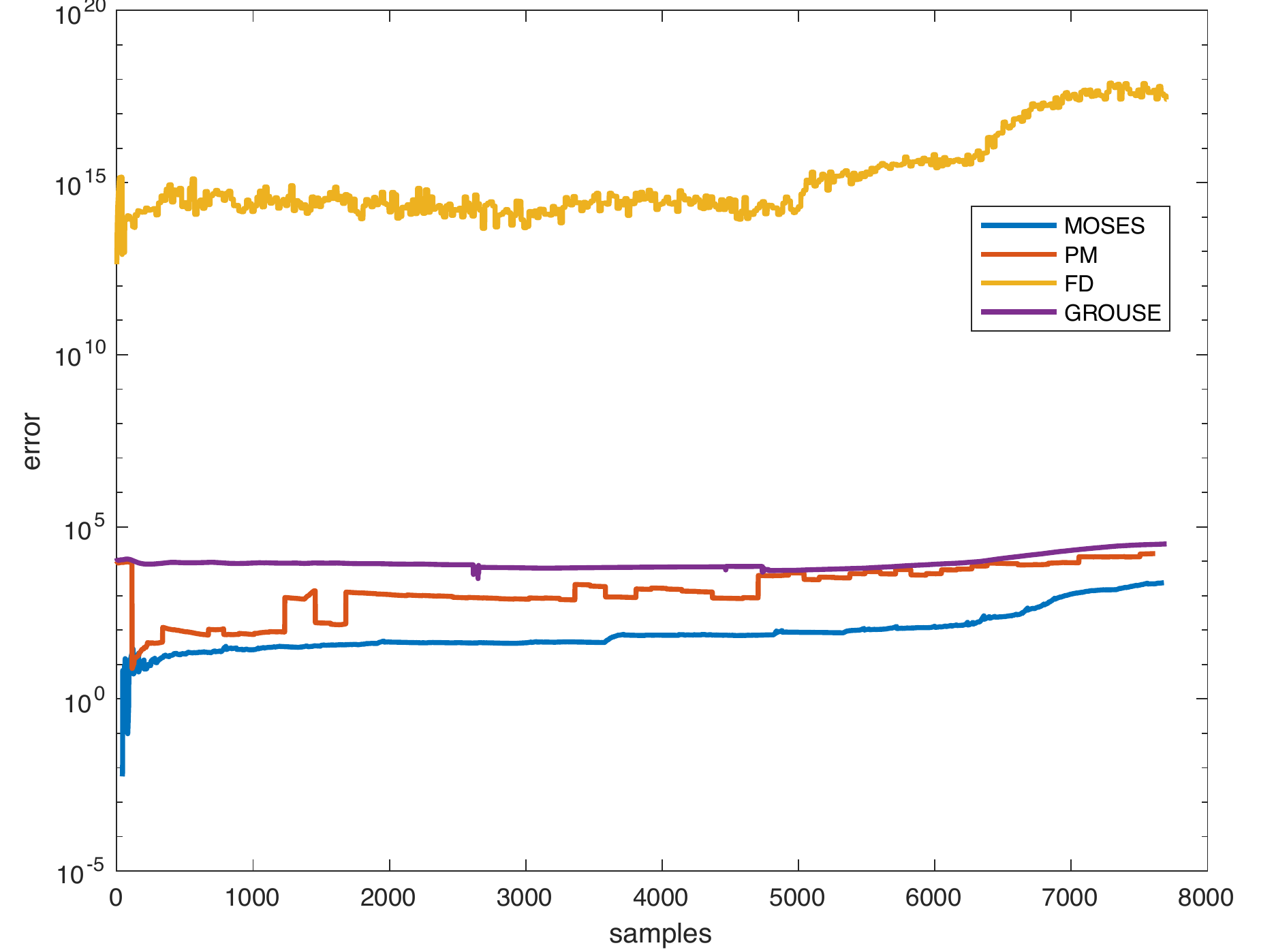}
    }{
       \includegraphics[clip, trim=1cm 7.5cm 1cm 8cm,width=0.4\textwidth]{real_froerror_n_56_r_20_Temperature_dataset_notitle}
    }
}
    \vspace{-6pt}
    \caption{Comparisons on real-world datasets, see Section \ref{sec:numerics} for the details. \label{fig:num_eval_mos_real}}
    \end{center}
\end{figure}

\noindent{\textbf{Mote voltage dataset:}} The first dataset we evaluate has an ambient dimension of $n=46$ and has $T=7712$ columns.
With $r=20$ and the rest of the  parameters as described in the synthetic comparison above, the errors over time for all  algorithms is shown in~Figure~\ref{fig:num_eval_mos_real_volt} in logarithmic scale. $\alg$ \edit{here outperforms GROUSE, FD, and PM}.

\noindent{\textbf{Mote humidity dataset:}} The second dataset evaluated has an ambient dimension of $n=48$ and has $T=7712$ columns. 
This dataset contains the 
humidity measurements of motes.
With $r=20$ and the rest of the  parameters as described in the synthetic comparison above, the errors over time for all  algorithms is shown in~\edit{Figure}~\ref{fig:num_eval_mos_real_humidity} in logarithmic scale. $\alg$ again outperforms the \edit{other  algorithms}.

\noindent{\textbf{Mote light dataset:}} The third dataset has an ambient dimension $n=48$ and has $T=7712$ columns. This dataset contains the light measurements of the motes.
With $r=20$ and the rest of  the parameters as described in the synthetic comparison above, the errors over time for all  algorithms is shown in~\autoref{fig:num_eval_mos_real_light} in logarithmic scale. As before, $\alg$  outperforms the  \edit{competing} algorithms.

\noindent{\textbf{Mote temperature dataset:}} The last real dataset we consider in this instance has an ambient dimension of $n=56$ and has 
$T=7712$ columns. This dataset contains the temperature measurements of the sensor motes and has mostly periodic value changes and infrequent 
spikes. With $r=20$ and the rest of  the parameters as described in the synthetic comparison above, the errors over time for all  algorithms is shown in~\edit{Figure}~\ref{fig:num_eval_mos_real_temp} in logarithmic scale. It is evident that $\alg$ outperforms the other \edit{competing} algorithms.

\editr{\noindent{\textbf{Storage:}}
We also performed memory complexity tests for all mote datasets above. Our experiments showed that often $\alg$ required the least amount 
of memory allocation against competing methods. 
Specifically, $\alg$ required $189.36$ Kb, 
$92.25$Kb, $127.02$Kb, and $215.97$Kb for the voltage, humidity, light, and 
temperature datasets, respectively. PM required $572.06$Kb, $638.80$Kb, 
$548.63$Kb, and $682.59$Kb, GROUSE required $2896.87$Kb, $2896.45$Kb, 
$2897.45$Kb, and $3769.42$Kb, and finally
FD required $173.86$Kb, $281.82$Kb, $194.32$ Kb, $655.78$Kb for the 
same datasets.
}

\begin{figure}[h]
  \begin{center}
\subfloat[Running time with $r=1$ \label{fig:comp_eval_mos_r1}]{
    \iftoggle{OVERLEAF}{
        \includegraphics[clip, trim=1cm 7.5cm 1cm 8cm,width=0.23\textwidth]{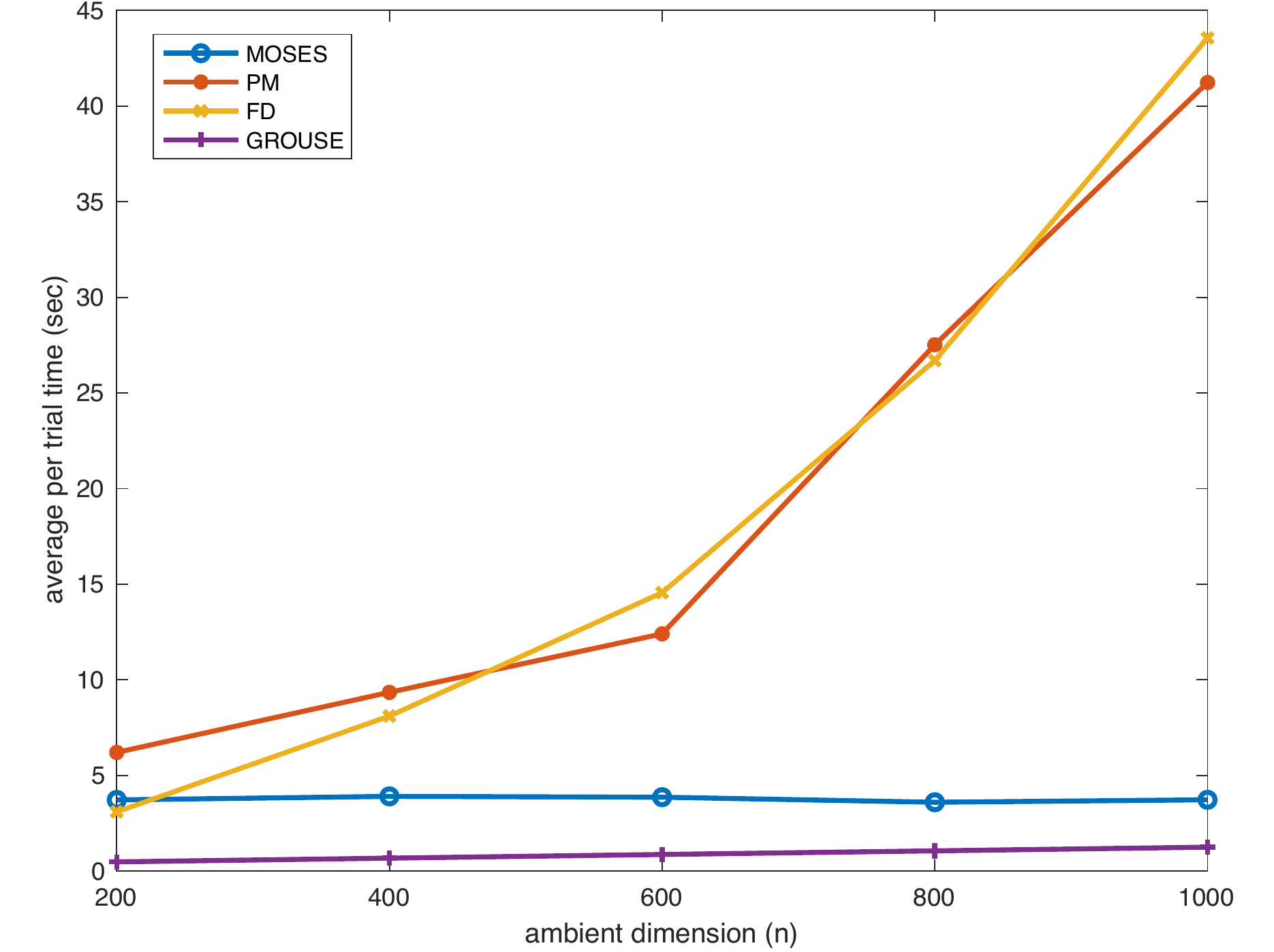}
    }{
        \includegraphics[clip, trim=1cm 7.5cm 1cm 8cm,width=0.23\textwidth]{speedtest_T_10k_kr_1_alpha_1_trials_5_notitle}
    }
}
\subfloat[Running time with $r=10$ \label{fig:comp_eval_mos_r10}]{
    \iftoggle{OVERLEAF}{
        \includegraphics[clip, trim=1cm 7.5cm 1cm 8cm,width=0.23\textwidth]{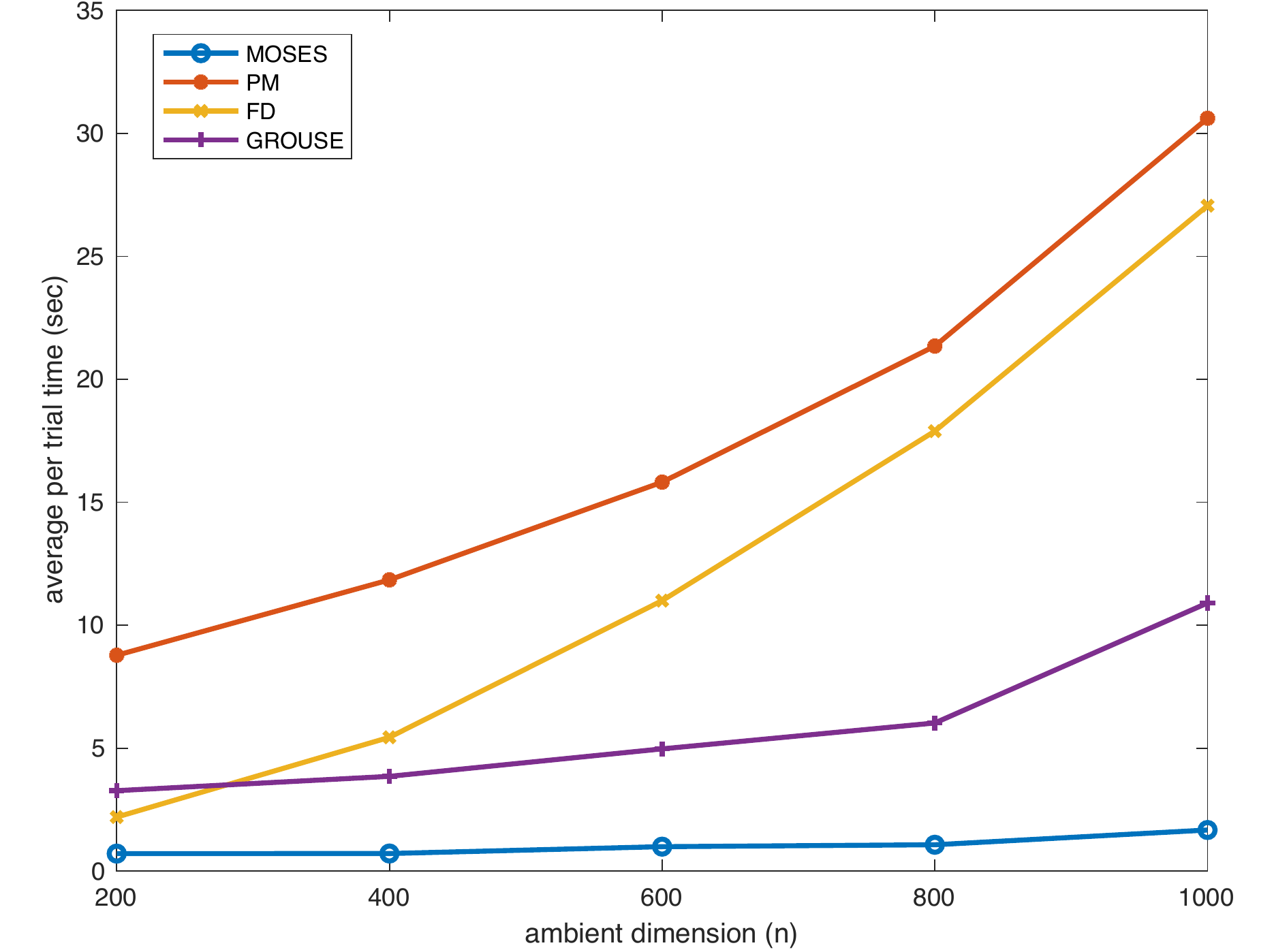}
    }{
        \includegraphics[clip, trim=1cm 7.5cm 1cm 8cm,width=0.23\textwidth]{speedtest_T_10k_kr_10_alpha_1_trials_5_notitle}
    }
}
\subfloat[Running time with $r=50$ \label{fig:comp_eval_mos_r50}]{
    \iftoggle{OVERLEAF}{
        \includegraphics[clip, trim=1cm 7.5cm 1cm 8cm,width=0.23\textwidth]{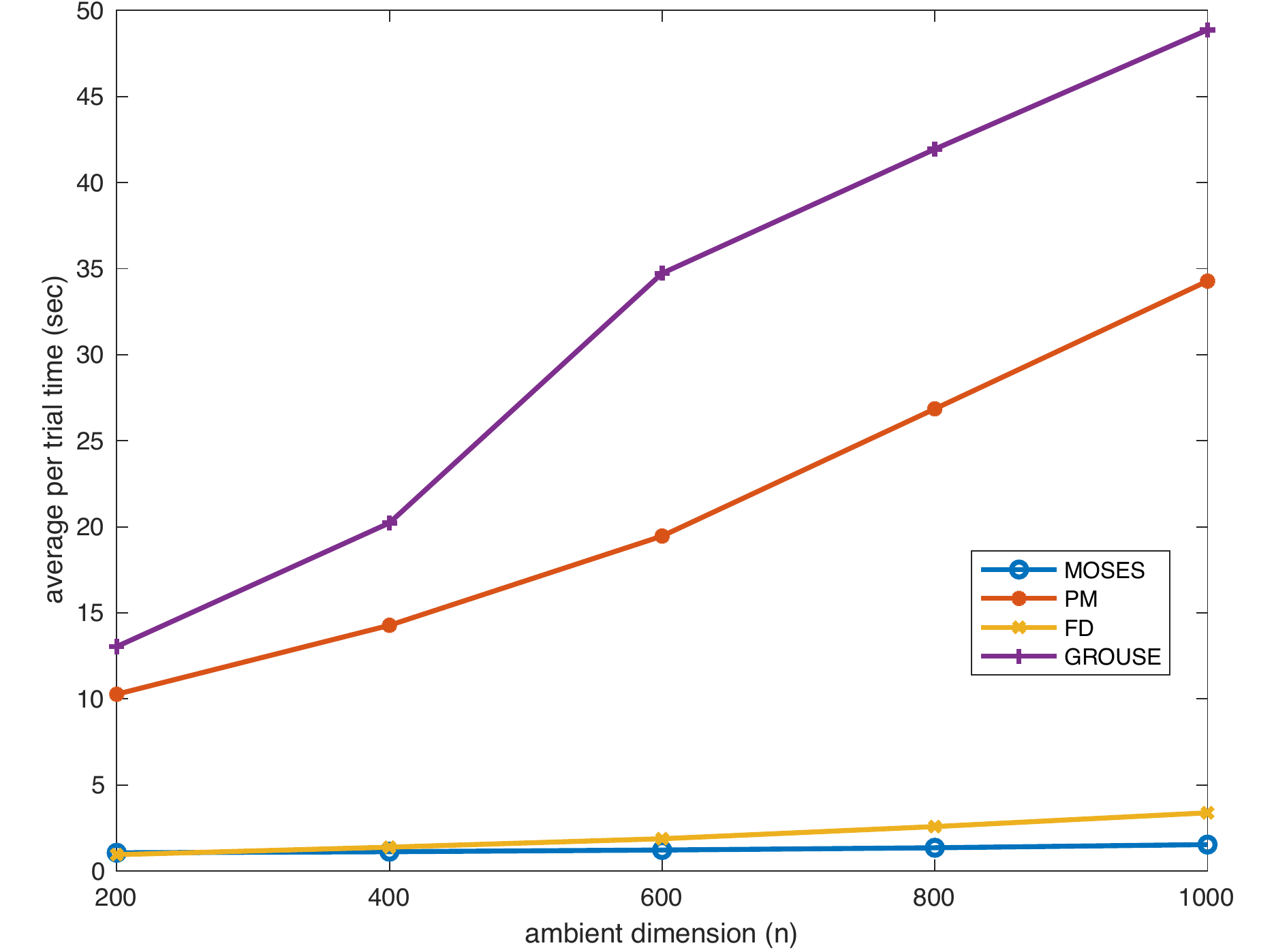}
    }{
        \includegraphics[clip, trim=1cm 7.5cm 1cm 8cm,width=0.23\textwidth]{speedtest_T_10k_kr_50_alpha_1_trials_5_notitle}
    }
}
\subfloat[Running time with $r=100$ \label{fig:comp_eval_mos_r100}]{
    \iftoggle{OVERLEAF}{
        \includegraphics[clip, trim=1cm 7.5cm 1cm 8cm,width=0.23\textwidth]{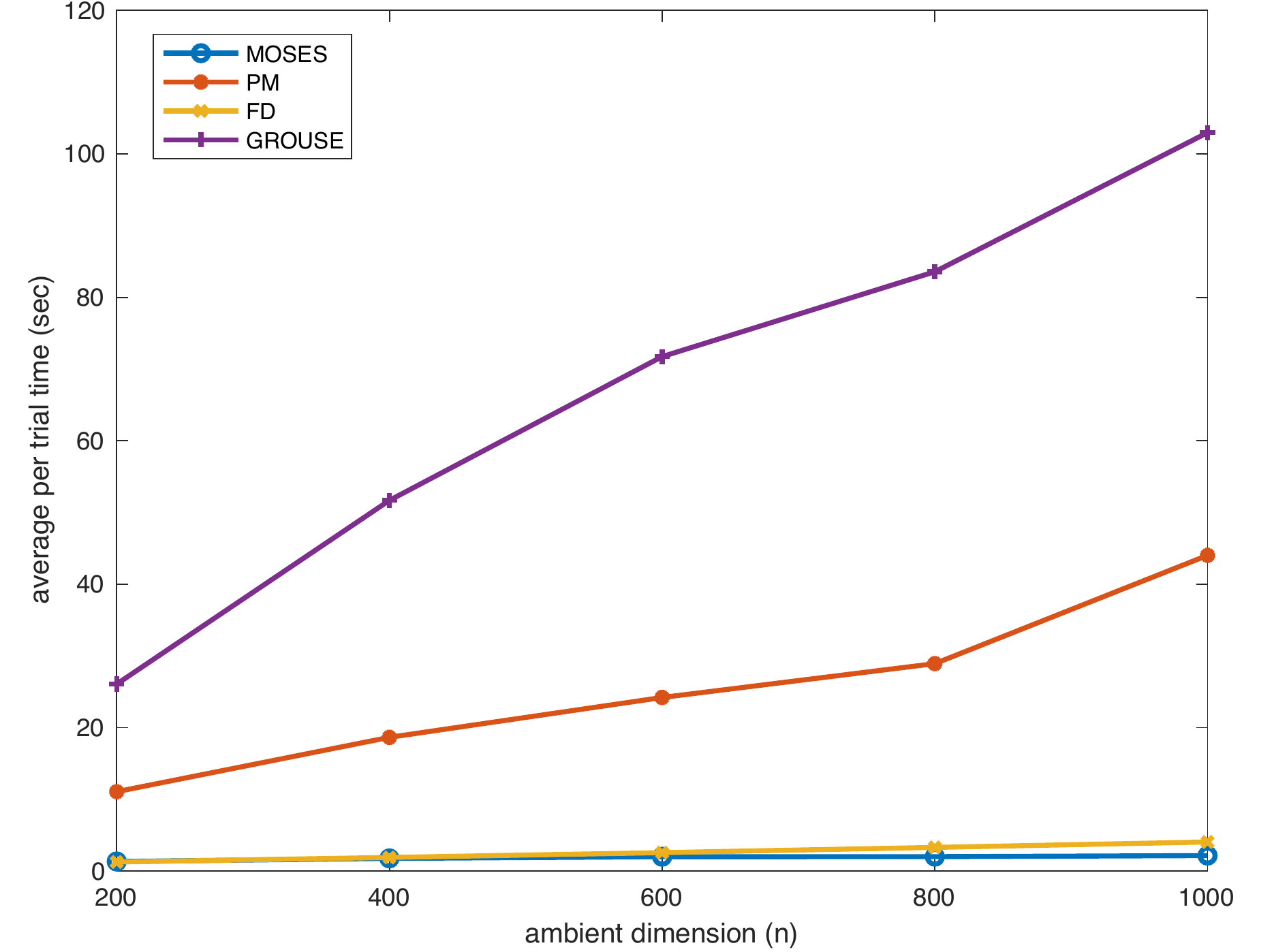}
    }{
        \includegraphics[clip, trim=1cm 7.5cm 1cm 8cm,width=0.23\textwidth]{speedtest_T_10k_kr_100_alpha_1_trials_5_notitle}
    }
}
\caption{Computational comlexity of all algorithms on synthetic datasets, see Section \ref{sec:numerics} for the details. \label{fig:comp_eval_mos}}
\end{center}
\end{figure}

\noindent{\textbf{Complexity on synthetic datasets:}}
Let us now turn our attention to the computational \edit{efficiency} of these \edit{four} algorithms. On synthetic datasets with $\alpha = 1$ and \edit{$T=10000$}, we compare the run-time of $\alg$ against \edit{GROUSE, FD, and PM, with the parameters set as described in the synthetic tests earlier}. This simulation was carried out \edit{using} {MATLAB} \edit{2018b} on a  2012 Mac Pro configured with Dual $6$-core Intel Xeon X5690 with $64$GB of {DDR3} {ECC} {RAM}. The average run-time of all three algorithms over five trials and for various choices of rank $r$ is shown in Figure~\ref{fig:comp_eval_mos}.  We note that the computational cost of $\alg$ remains consistently small throughout these simulations, especially for large ambient dimensions and ranks where GROUSE and PM perform poorly \edit{regardless of the desired recovery rank $r$ used; see \edit{Figure}~\ref{fig:comp_eval_mos_r50}. Interestingly enough, FD performs poorly when attempting a relatively low rank recovery ($r\in \{1,10\}$) and closely matches $\alg$ as $r$ increases, which can be attributed to the buffering size of FD.}

{\footnotesize
\bibliographystyle{unsrt}
\iftoggle{OVERLEAF}{
\bibliography{References/References}
}{
\bibliography{References}
}
}

\onecolumn


\begin{center}
\scalebox{0.9}{
\begin{minipage}{0.95\linewidth}
\begin{algorithm}[H]
{\footnotesize
\caption{$\alg$: A streaming algorithm for linear dimensionality reduction \editr{(efficient version)}}
\label{alg:MOSES implement} 
\vspace{1mm}

\textbf{Input:} Sequence of vectors $\{y_t\}_t\subset\mathbb{R}^n$ and block size $b$.
\vspace{1mm}

\textbf{Output:} Sequence $\{\wh{\B{S}}_{kb,r},\wh{\B{\Gamma}}_{kb,r},\wh{\B{Q}}_{kb,r}\}_k$.
\vspace{1mm}
 
\textbf{Body:} 
\vspace{1mm}

\begin{enumerate}
\item For $k=1$,
\begin{enumerate}
\item Form $\y_1\in\R^{n\times b}$ by concatenating $\{y_t\}_{t=1}^b$. 
\item Set 
$$
[\wh{\B{S}}_{b,r}, \wh{\B{\Gamma}}_{b,r},\wh{\B{Q}}_{b,r}] = \SVD_r (\y_1),
$$
where $\wh{\B{S}}_{b,r}\in\R^{n\times r}$ and $\wh{\B{Q}}_{b,r}\in\R^{b\times r}$ have orthonormal columns, and the diagonal matrix $\wh{\B{\Gamma}}_{b,r}\in\R^{r\times r}$ contains  leading  $r$  singular values. 
\end{enumerate}
\item For $k\ge 2$, repeat 
\begin{enumerate}
\item Form $\y_k\in\mathbb{R}^{n\times b}$ by concatenating $\{y_t\}_{t=(k-1)b+1}^{kb}$.  
\item Set $$\dot{\B{q}}_k=\wh{\B{S}}_{(k-1)b,r}^* \y_k \in\R^{r\times b},\qquad\wh{\B{z}}_k = \y_k - \wh{\B{S}}_{(k-1)b,r} \dot{\B{q}}_k\in\R^{n\times b}.$$
\item Let $[\wh{\B{s}}_k ,\B{v}_k] = \mbox{QR}(\wh{\B{z}}_k)$ be the QR decomposition of $\wh{\B{z}}_k$, where $\wh{\B{s}}_k\in\R^{n\times b}$ has orthonormal columns and $\B{v}_k\in\R^{b\times b}$. 
\item Let 
\begin{equation}
\l[{\B{u}}_k ,\wh{\B{\Gamma}}_{kb,r}, \wh{\B{q}}_k \r] =\SVD_r\l(
\l[
\begin{array}{cc}
\wh{\B{\Gamma}}_{(k-1)b,r} & \dot{\B{q}}_k \\
\B{0}_{b\times r} & \B{v}_k 
\end{array}
\r] \r),
\end{equation}
where ${\B{u}}_k,\wh{\B{q}}_k\in\R^{(r+b)\times r}$ have orthonormal columns and the diagonal matrix $\wh{\B{\Gamma}}_{kb,r}\in\R^{r \times r}$ contains  leading $r$ singular values in nonincreasing order.  
\item Let 
\begin{equation*}
\wh{\B{S}}_{kb,r} = \l[ 
\begin{array}{cc}
\wh{\B{S}}_{(k-1)b,r} & \wh{\B{s}}_k
\end{array}
\r] {\B{u}}_k.
\end{equation*}
\item \textbf{(optional)} If the number of rows of $\wh{\B{Q}}_{(k-1)b,r}$ exceeds $n$ and $\wh{\B{Q}}_{(k-1)b,r}$ is not needed any more, it is optional  in order to improve efficiency to set $\wh{\B{Q}}_{kb,r} = \wh{\B{q}}_k$.
\item  Otherwise, set 
\begin{equation}
\wh{\B{Q}}_{kb,r} = 
\l[
\begin{array}{cc}
\wh{\B{Q}}_{(k-1)b,r} & 0\\
0 & \B{I}_b
\end{array}
\r]
 \wh{\B{q}}_k.
\end{equation}
\end{enumerate}
\end{enumerate}
}
\vspace{-6pt}
\end{algorithm}
\end{minipage}
} 

\end{center}

\vspace{-12pt}

\clearpage
\noindent\textbf{{\huge Supplementary Material}}

\setcounter{section}{0}
\renewcommand\thesection{\alph{section}}
\section{Overview}
The supplementary material contains more details about MOSES, an optimisation interpretation of MOSES, the proofs of the main results, as well as the acknowledgements.

\section{Storage and Computational Requirements of MOSES \label{sec:intro}}

The efficient implementation of $\alg$ in Algorithm \ref{alg:MOSES implement} is based on the ideas from incremental SVD and it is straightforward to verify that Algorithms \ref{alg:MOSES} and \ref{alg:MOSES implement} are indeed equivalent; at  iteration $k$,  the relation between the  output of Algorithm \ref{alg:MOSES} ($\wh{\Y}_{kb,r}$) and the output of Algorithm \ref{alg:MOSES implement} ($\wh{\B{S}}_{kb,r},\wh{\B{\Gamma}}_{kb,r},\wh{\B{Q}}_{kb,r}$) is 
$$
\wh{\Y}_{kb,r}\overset{\text{SVD}}{=}\wh{\B{S}}_{kb,r}\wh{\B{\Gamma}}_{kb,r}\wh{\B{Q}}_{kb,r}^*,
$$ 
where the right-hand side above is the SVD of $\wh{\Y}_{kb,r}$. More specifically,  $\wh{\B{S}}_{kb,r} \in\mathbb{R}^{n\times r}$ has orthonormal columns and is the $\alg$'s estimate of  leading $r$ principal components of $\Y_{kb}\in \mathbb{R}^{n\times kb}$, where we recall that $\Y_{kb}$ is the data received so far.  Moreover, 
$$\wh{\B{S}}_{kb,r}^*  \wh{\Y}_{kb,r}= \wh{\B{\Gamma}}_{kb,r}\wh{\B{Q}}_{kb,r}^*\in \mathbb{R}^{r\times kb}
$$ is the projection of $\wh{\Y}_{kb,r}$ onto this estimate, namely $\wh{\B{S}}_{kb,r}^*  \wh{\Y}_{kb,r}$ is  $\alg$'s estimate of the projected data matrix so far. In words, the efficient implementation of $\alg$ in Algorithm \ref{alg:MOSES implement} explicitly maintains estimates of both  principal components and the projected data, at every iteration.

Let us now evaluate the storage and computational requirements of $\alg$. At the start of iteration $k$, Algorithm \ref{alg:MOSES implement}  stores the matrices
$$
\wh{\B{S}}_{(k-1)b,r}\in\mathbb{R}^{n\times r},
\qquad 
\wh{\B{\Gamma}}_{(k-1)b,r} \in \mathbb{R}^{r\times r},
\qquad 
\wh{\B{Q}}_{(k-1)b,r} \in\mathbb{R}^{(k-1)b\times r},
$$ 
and after that also receives and stores the  incoming block $\y_{k}\in\mathbb{R}^{n\times b}$.  This requires $O(r(n + (k-1)b+1))+O(bn)$ bits of memory, because $\widehat{\B{\Gamma}}_{(k-1)b,r}$ is diagonal. Assuming that $b=O(r)$,  Algorithm \ref{alg:MOSES implement} therefore requires $O(r(n+kr))$  bits of memory at iteration $k$.  Note that this is optimal, as it is impossible to store a rank-$r$  matrix of size $n\times kb$ with fewer  bits when $b=O(r)$. 

It is also easy to verify that Algorithm \ref{alg:MOSES implement} performs $O(r^2(n+kb))=O(r^2(n+kr))$ flops in iteration $k$. The dependence of both storage and computational complexity on $k$ is due to the fact that $\alg$  maintains both an estimate of  principal components in  $\wh{\B{S}}_{kb,r}$ and  an estimate of the projected data in $\B{\Gamma}_{kb,r}\B{Q}_{kb,r}^*$. To maximise the efficiency, one might optionally ``flush out'' the projected data after every  $n/b$ iterations, as described in the last step in Algorithm \ref{alg:MOSES implement}.

\section{Optimisation Viewpoint \label{sec:opt view}}
$\alg$ has a natural interpretation as an approximate solver for the non-convex optimisation program underlying PCA, which serves as its motivation. More specifically, recall that   leading $r$ principal components of $\Y_T$ are obtained by solving the non-convex program 
\begin{equation}
\min_{\U\in \GR(n,r)} \l\| \Y_T - \P_{\U} \Y_T\r\|_F^2,
\label{eq:pca 1}
\end{equation}
where the minimization is over the Grassmannian $\GR(n,r)$, the set of all $r$-dimensional subspaces in $\R^n$. Above, $\P_{\U}\in\R^{n\times n}$ is the orthogonal projection onto the subspace $\U$. 
By construction in Section \ref{sec:intro}, note that 
\begin{align}
\Y_T & = 
\l[ 
\begin{array}{cccc}
y_1 & y_2 & \cdots & y_T
\end{array}
\r] 
\qquad \mbox{(see \eqref{eq:conc of yts})}
\nonumber\\
& = \l[ 
\begin{array}{cccc}
\y_1 & \y_2 & \cdots & \y_K
\end{array}
\r] \in\R^{n\times T},
\label{eq:brk blcks}
\end{align}
where $\{\y_k\}_{k=1}^K$ are the incoming blocks of data. 
This allows us to rewrite Program \eqref{eq:pca 1} as 
\begin{align}
\min_{\U\in \GR(n,r)} \l\| \Y_T - \P_{\U} \Y_T\r\|_F^2 & 
= \min_{\U\in \GR(n,r)} \sum_{k=1}^K \l\| \y_{k} - \P_{\U} \y_{k}\r\|_F^2
\qquad \text{(see \eqref{eq:brk blcks})}
 \nonumber\\
& = 
\begin{cases}
\min \sum_{k=1}^K \l\| \y_{k} - \P_{\U_K}\cdots \P_{\U_k}  \y_{k}\r\|_F^2 & \\
\U_1 = \U_2 = \cdots =\U_K,
\label{eq:nested}
\end{cases}
\end{align}
where the last minimisation above is over all identical subspaces $\{\U_k\}_{k=1}^K\subset \GR(n,r)$.
Our strategy is to make a sequence of approximations to the program in the last line above. In the first approximation, we only keep the first summand in the last line of \eqref{eq:nested}. That is, our first approximation reads as 
\begin{align}
\begin{cases}
\min \sum_{k=1}^K \l\| \y_{k} - \P_{\U_K}\cdots \P_{\U_k}  \y_{k}\r\|_F^2 & \\
\U_1 = \U_2 = \cdots =\U_K
\end{cases}
& \ge 
\begin{cases}
\min  \l\| \y_{1} - \P_{\U_K}\cdots \P_{\U_1}  \y_{1}\r\|_F^2 & \\
\U_1 = \U_2 = \cdots =\U_K
\end{cases} \nonumber\\
& = \min_{\U\in\GR(n,r)} \l\| \y_1 - \P_{\U}\y_1 \r\|_F^2,
\end{align}
where the second line above follows by setting $\U =\U_1=\cdots =\U_K$. 
 Let $\wh{\SU}_{b,r}$ be a minimiser of the program in the last line above. Note that $\wh{\SU}_{b,r}$ simply spans    leading $r$ principal components of $\y_1$, akin to Program \eqref{eq:pca 1}. This indeed coincides with the output of $\alg$ in the first iteration, because  
 \begin{align}
 \wh{\Y}_{b,r} & = \SVD_r(\y_1) 
\qquad \mbox{(see Algorithm \ref{alg:MOSES})} 
 \nonumber\\
&  = \P_{\wh{\SU}_{b,r}}\y_1. \qquad 
\mbox{(similar to the second line of \eqref{eq:low-rank est kept})}
\label{eq:first iteration checks out}
 \end{align}
Next consider the next approximation in which we  keep two of the summands in the last line of  \eqref{eq:nested}, namely 
\begin{align}
\begin{cases}
\min \sum_{k=1}^K \l\| \y_{k} - \P_{\U_K}\cdots \P_{\U_k}  \y_{k}\r\|_F^2 & \\
\U_1 = \U_2 = \cdots =\U_K
\end{cases}
& \ge 
\begin{cases}
\min  \l\| \y_{1} - \P_{\U_K}\cdots \P_{\U_1}  \y_{1}\r\|_F^2 
+ \l\| \y_{2} - \P_{\U_K}\cdots \P_{\U_2}  \y_{2}\r\|_F^2 
& \\
\U_1 = \U_2 = \cdots =\U_K,
\end{cases}
\end{align}
and then we substitute $\U_1 = \wh{\SU}_{b,r}$ above to arrive at the new program 
\begin{align}
&
\begin{cases}
\min \,\,\, \| \y_{1} - \P_{\U_K}\cdots \P_{\U_2} \P_{\wh{\SU}_{b,r}}  \y_{1}\|_F^2 
+ \l\| \y_{2} - \P_{\U_K}\cdots \P_{\U_2}  \y_{2}\r\|_F^2 
& \\
\U_2 = \U_3  = \cdots =\U_K
\end{cases}
\nonumber\\
 & = 
\min_{\U \in \GR(n,r)} \,\,\, \| \y_{1} - \P_{\U} \P_{\wh{\SU}_{b,r}}  \y_{1}\|_F^2 
+ \l\| \y_{2} - \P_{\U} \y_{2}\r\|_F^2,
\end{align}
where the second program above follows by setting $\U=\U_2=\cdots =\U_K$. 
We can rewrite the above program as 
\begin{align}
& 
\min_{\U \in \GR(n,r)} \,\,\, \| \y_{1} - \P_{\U} \P_{\wh{\SU}_{b,r}}  \y_{1}\|_F^2 
+ \l\| \y_{2} - \P_{\U} \y_{2}\r\|_F^2  
\nonumber\\
& = \min_{\U\in \GR(n,r)} \,\,\, 
\l\| 
\l[
\begin{array}{cc}
\y_1 - \P_{\U}\P_{\wh{\SU}_{b,r}} \y_1 & 
\y_2 - \P_{\U} \y_2
\end{array}
\r]
\r\|_F^2
 \nonumber\\
& = \min_{\U\in\GR(n,r)} \,\,\,
\l\|
\l[
\begin{array}{cc}
\P_{\wh{\SU}_{b,r}^\perp} \y_1 & \B{0}_{n\times b}
\end{array}
\r]
+ \P_{\U^\perp}
\l[
\begin{array}{cc}
\P_{\wh{\SU}_{b,r}}\y_1 & \y_2
\end{array}
\r]
\r\|_F^2 \nonumber\\
& = 
\| \P_{\wh{\SU}_{b,r}^\perp} \y_1 \|_F^2 +  \min_{\U\in\GR(n,r)}  \,\,\, \l\| 
\P_{\U^\perp}
\l[
\begin{array}{cc}
\P_{\wh{\SU}_{b,r}}\y_1 & \y_2
\end{array}
\r]
\r\|_F^2 
\qquad \text{(see the text below)}
\nonumber\\
& = 
\| \P_{\wh{\SU}_{b,r}^\perp} \y_1 \|_F^2 +  \min_{\U\in\GR(n,r)}  \,\,\, \l\| 
\P_{\U^\perp}
\l[
\begin{array}{cc}
\wh{\Y}_{b,r} & \y_2
\end{array}
\r]
\r\|_F^2,
\qquad \mbox{(see \eqref{eq:first iteration checks out})}
\end{align}
and let $\wh{\SU}_{2b,r}$ be a minimiser of the last program above. 
Above, $\perp$ shows the orthogonal complement of a subspace. The second to last line above follows because   $\wh{\SU}_{2b,r}$  is always within the column span of $[\P_{\wh{\SU}_{b,r}}\y_1\,\,\, \y_2]$. 
Note also that $\wh{\SU}_{2b,r}$ is the span of  leading $r$ principal components of the matrix $[\wh{\Y}_{1,r}\,\,\, \y_2]$, similar to  Program~\eqref{eq:pca 1}.  
This again coincides with the output of $\alg$ in the second iteration, because
\begin{align}
\wh{\Y}_{2b,r} & = \SVD_r\l( 
\l[
\begin{array}{cc}
\wh{\Y}_{b,r} & \y_2
\end{array}
\r]
 \r)
 \qquad \mbox{(see Algorithm \ref{alg:MOSES})} \nonumber\\
 & = \P_{\wh{\SU}_{2b,r}}  \l[
\begin{array}{cc}
\wh{\Y}_{b,r} & \y_2
\end{array}
\r].
 \qquad 
\mbox{(similar to the second line of \eqref{eq:low-rank est kept})}
\end{align}
Continuing this  procedure precisely produces the iterates of $\alg$. Therefore we might interpret $\alg$ as an optimisation algorithm for  solving Program \eqref{eq:pca 1} by making a sequence of approximations.

\begin{center}
\begin{figure}[h]
\begin{center}

\iftoggle{OVERLEAF} {
    \includegraphics[width=0.7\textwidth]{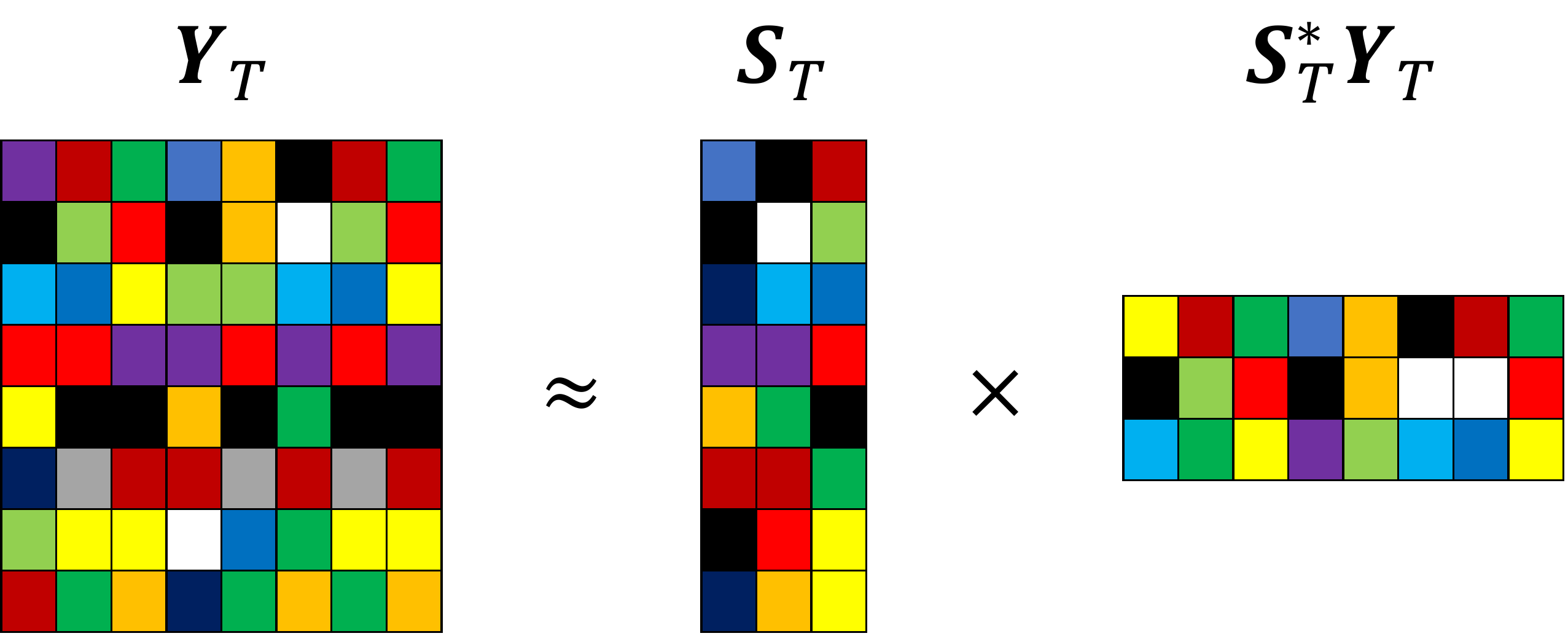}
} {
    \includegraphics[width=0.7\textwidth]{TruncatedSVD}
}
\caption{Given a data matrix $\Y_T\in\R^{n\times T}$, truncated SVD finds the best low-dimensional linear model to represent the data: For a typically small integer $r$, we compute $\Y_{T,r}=\text{SVD}_r(\Y_T)=\B{S}_{T,r}\cdot \B{S}_{T,r}^*\Y_T$, where $\B{S}_{T,r}\in\R^{n\times r}$ contains leading $r$ principal components of $\Y_T$ and $\B{S}_{T,r}^*\Y_T\in \R^{r\times T}$ is the projected data matrix with reduced dimension $r$ (instead of $n$). This paper presents $\alg$, a streaming algorithm for truncated SVD. Put differently, $\alg$ keeps both a running estimate of the principal components  {and} the projection of data, received so far, onto this estimate. \label{fig:demo}}
\end{center}
\end{figure}
\end{center}

\section{Spiked Covariance Model and Additional Remarks\label{sec:results}}

 A popular  model in the  statistics literature is the {spiked covariance} model, where the data vectors $\{y_t\}_{t=1}^T$ are drawn from a distribution with a covariance matrix $\B{\Xi}$. Under this model, $\B{\Xi}$ is a low-rank perturbation of the identity matrix \cite{johnstone2001distribution,vershynin2012close}, namely $\lambda_1(\B{\Xi}) = \cdots =\lambda_r(\B{\Xi})=\lambda$ and $\lambda_{r+1}(\B{\Xi})=\cdots =\lambda_n(\B{\Xi})=~1$. Proposition \ref{prop:first app} in this case reads as
\begin{equation}
\E \|y-\P_{{\SU}_{T,r}} y\|_2^2  \propto {{  (n-r)}} +
{{(n-r) \lambda \sqrt{\frac{ {\log T} }{T}}}},
\label{eq:sp cov 1}
\end{equation}
where ${\SU}_{T,r}$ spans leading $r$ principal components of the data matrix $\Y_T$. In contrast, Theorem \ref{thm:stochastic result} roughly speaking states that 
\begin{equation}
\E \|y-\P_{\wh{\SU}_{T,r}} y\|_2^2 \propto 
(n-r) \l( \frac{T\lambda}{bn}\r)^{\frac{n}{\lambda}} + 
{{(n-r) \lambda \sqrt{\frac{ {\log T} }{T}}}},
\label{eq:sp cov 2}
\end{equation}
where $\wh{\SU}_{T,r}$ spans the output of $\alg$. 
When $\lambda\gtrsim n \log (T/b) = n\log K$ in particular,  we find that the error bounds in (\ref{eq:sp cov 1},\ref{eq:sp cov 2}) are of the same order. That is, under the spiked covariance model, $\alg$ for streaming truncated SVD matches the performance of ``offline'' truncated SVD, provided that the underlying distribution has a sufficiently large spectral gap. In practice, (\ref{eq:sp cov 2}) is often a conservative bound. 

\paragraph{Proof strategy.}  \label{rem:proof} Starting with \eqref{eq:final bnd}, the proof of Theorem \ref{thm:stochastic result} in Section \ref{sec:theory} of the supplementary material breaks down the error associated with $\alg$ into two components as 
\begin{align}
\E_y\| y - \P_{\wh{\SU}_{T,r}}y \|_2 &  \le 
\frac{1}{T} \| \Y_T - \P_{\wh{\SU}_{T,r}}\Y_T\|_F^2 +  
\l| \frac{1}{T} \|\Y_T -\P_{\wh{\SU}_{T,r} }\Y_T\|_F^2 - \E_y\| y - \P_{\wh{\SU}_{T,r}}y \|_2^2  \r|.
\label{eq:brk down of error rem}
\end{align}
That is, we bound the population risk with the empirical risk. We control the empirical risk in the first part of the proof by  noting that 
\begin{align}
\|\Y_T-\P_{\wh{\SU}_{T,r}} \Y_T\|_F & = 
\| \P_{\wh{\SU}_{T,r}^{\perp} } \Y_T\|_F \nonumber\\
& = \| \P_{\wh{\SU}_{T,r}^{\perp}  } ( \Y_T - \wh{\Y}_{T,r})\|_F 
\qquad \mbox{(see \eqref{eq:span of output of MOSES})}
\nonumber\\
& \le  \|  \Y_T - \wh{\Y}_{T,r}\|_F, 
\end{align}
where the last line gauges how well the output of $\alg$ approximates the data matrix $\Y_T$, see \eqref{eq:near residual of moses thm}. 
We then bound $\|\Y_T-\wh{\Y}_{T,r}\|_F$ in two steps: As it is common in these types of arguments, the first step finds a deterministic upper bound for this norm, which is then evaluated for our particular stochastic setup. 
\begin{itemize}
\item The deterministic bound appears in Lemma \ref{lem:online trunc SVD} and gives an upper bound for $\|\Y_T -\wh{\Y}_{T,r}\|_F$ in terms of the overall ``innovation''. Loosely speaking, the innovation $\| \P_{\SU_{(k-1)b,r}^\perp} \y_k\|_F$ at iteration $k$ is the part of the new data block $\y_k$ that cannot be described by the leading $r$ principal components of data arrived so far, which span the subspace $\SU_{(k-1)b,r}$.   
\item The stochastic bound is given in Lemma \ref{lem:stochastic bnd} and uses a tight perturbation result. 
\end{itemize}
Our argument so far yields an upper bound on the empirical loss $\|\Y_T - \P_{\wh{\SU}_{T,r}}\Y_T\|_F$ that holds with high probability.  In light of \eqref{eq:brk down of error rem}, it remains to control
\begin{align}
\l| \frac{1}{T}\|\Y_T -\P_{\wh{\SU}_{T,r} }\Y_T\|_F^2 - \E_y\|y - \P_{\wh{\SU}_{T,r}}y \|_2^2 \r|
& =  \frac{1}{T}  \l|  \|\Y_T -\P_{\wh{\SU}_{T,r} }\Y_T\|_F^2 - \E \|\Y_T -\P_{\wh{\SU}_{T,r} }\Y_T\|_F^2  \r| \nonumber\\
& =  \frac{1}{T}  \l|  \|\P_{\wh{\SU}^\perp_{T,r} }\Y_T\|_F^2 - \E \|\P_{\wh{\SU}^\perp_{T,r} }\Y_T\|_F^2  \r| 
\end{align}
with a standard large deviation bound. 

\paragraph{Other stochastic models.} While our results were restricted to the Gaussian distribution, they extend easily and with minimal change to the larger class of subgaussian distributions. Beyond subgaussian data models, Lemma \ref{lem:online trunc SVD} is the key deterministic result, relating  the $\alg$ error to the overall innovation. One might therefore control the overall innovation, namely the right-hand side of  \eqref{eq:deterministic online svd} in Lemma \ref{lem:online trunc SVD}, for any other stochastic model at hand.

\begin{figure}
 \centering
 \iftoggle{OVERLEAF}{
 \scalebox{1.3}{\begin{tikzpicture}
    
    \node[text width=.5cm] at (0.15, 6.25) {{\boldmath $y_{1}$}};
    \node[text width=.5cm] at (0.15, 6) {$\bullet$};
    \draw[draw=gray, dashed, very thick, ->] (0.1, 6) -- (1.9, 6);
    
    \node[text width=.5cm] at (2.1, 6.25) {{\boldmath $y_{2}$}};
    \node[text width=.5cm] at (2.15, 6) {$\bullet$};
    \draw[draw=gray, dashed, very thick, ->] (2.1, 6) -- (3.9, 6);
    
    \node[text width=.5cm] at (4.1, 6.25) {{\boldmath $y_{3}$}};
    \node[text width=.5cm] at (4.15, 6) {$\bullet$};
    
    \node[text width=.7cm] at (6.05, 6) {\Large $\cdots$};
    
    \draw[draw=gray, dashed, very thick, ->] (8, 6) -- (9.9, 6);
    \node[text width=.5cm] at (10.15, 6) {$\bullet$};
    \node[text width=.7cm] at (10.2, 6.25) {{\boldmath $y_{K}$}};
    
    \node[text width=.8cm] at (0.23, 5.5) {{\boldmath $\widehat{Y}_{1,r}$}};
    
    \node[text width=.5cm] at (2.15, 5) {$\bullet$};
    \node[text width=.7cm] at (2.18, 4.5) {{\boldmath$\widehat{Y}_{2,r}$}};
    \draw[very thick, ->] (1.99, 6) -- (1.99, 5.08);
    
    \node[text width=.5cm] at (4.15, 4) {$\bullet$};
    \node[text width=.7cm] at (4.18, 3.5) {{\boldmath $\widehat{Y}_{3,r}$}};
    \draw[very thick, ->] (3.99, 6) -- (3.99, 4.08);
    
    \draw[very thick, ->] (9.99, 6) -- (9.99, 1.08);
    \node[text width=.5cm] at (10.15, 1) {$\bullet$};
    \node[text width=.8cm] at (10.24, 0.5) {{\boldmath $\widehat{Y}_{K,r}$}};
    
    
    \draw[very thick, ->] (0, 6) -- (1.92, 5.03);
    
    \draw[very thick, ->] (2, 5) -- (3.92, 4.03);
    
    \node[text width=.7cm, rotate=-27] at (6.03, 2.96) {\Large $\cdots$};
    
    \draw[very thick, ->] (8, 2) -- (9.92, 1.03);
\end{tikzpicture}}
 }{
 \scalebox{1.3}{\begin{tikzpicture}
    
    \node[text width=.5cm] at (0.15, 6.25) {{\boldmath $y_{1}$}};
    \node[text width=.5cm] at (0.15, 6) {$\bullet$};
    \draw[draw=gray, dashed, very thick, ->] (0.1, 6) -- (1.9, 6);
    
    \node[text width=.5cm] at (2.1, 6.25) {{\boldmath $y_{2}$}};
    \node[text width=.5cm] at (2.15, 6) {$\bullet$};
    \draw[draw=gray, dashed, very thick, ->] (2.1, 6) -- (3.9, 6);
    
    \node[text width=.5cm] at (4.1, 6.25) {{\boldmath $y_{3}$}};
    \node[text width=.5cm] at (4.15, 6) {$\bullet$};
    
    \node[text width=.7cm] at (6.05, 6) {\Large $\cdots$};
    
    \draw[draw=gray, dashed, very thick, ->] (8, 6) -- (9.9, 6);
    \node[text width=.5cm] at (10.15, 6) {$\bullet$};
    \node[text width=.7cm] at (10.2, 6.25) {{\boldmath $y_{K}$}};
    
    \node[text width=.8cm] at (0.23, 5.5) {{\boldmath $\widehat{Y}_{1,r}$}};
    
    \node[text width=.5cm] at (2.15, 5) {$\bullet$};
    \node[text width=.7cm] at (2.18, 4.5) {{\boldmath$\widehat{Y}_{2,r}$}};
    \draw[very thick, ->] (1.99, 6) -- (1.99, 5.08);
    
    \node[text width=.5cm] at (4.15, 4) {$\bullet$};
    \node[text width=.7cm] at (4.18, 3.5) {{\boldmath $\widehat{Y}_{3,r}$}};
    \draw[very thick, ->] (3.99, 6) -- (3.99, 4.08);
    
    \draw[very thick, ->] (9.99, 6) -- (9.99, 1.08);
    \node[text width=.5cm] at (10.15, 1) {$\bullet$};
    \node[text width=.8cm] at (10.24, 0.5) {{\boldmath $\widehat{Y}_{K,r}$}};
    
    
    \draw[very thick, ->] (0, 6) -- (1.92, 5.03);
    
    \draw[very thick, ->] (2, 5) -- (3.92, 4.03);
    
    \node[text width=.7cm, rotate=-27] at (6.03, 2.96) {\Large $\cdots$};
    
    \draw[very thick, ->] (8, 2) -- (9.92, 1.03);
\end{tikzpicture}}
 }
 \caption{Streaming problems may be interpreted as a special case of distributed computing. Each data block $\y_k$ lives on a node of the chain graph and the nodes are combined, from left to right, following the structure of the ``cone'' tree.   \label{fig:cone tree}}
\end{figure}
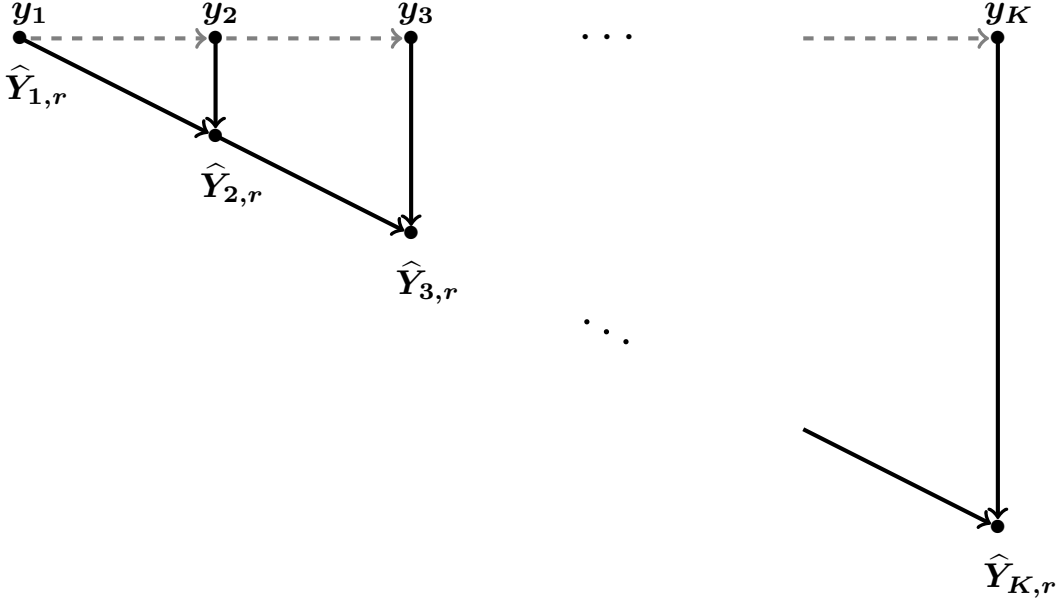

\section*{Acknowledgements}
 AE is supported by the Alan
Turing Institute under the EPSRC grant EP/N510129/1 and also by the Turing Seed Funding grant SF019. RAH is supported by EPSRC grant EP/N510129/1. AG is supported by the Alan Turing Institute under the
EPSRC grant EP/N510129/1 and TU/C/000003. AE is grateful to Chinmay Hedge, Mike Wakin, Jared Tanner, and Mark Davenport for insightful suggestions and  valuable feedback. Parts of this project were completed when AE was a Leibniz Fellow at Oberwolfach Research Institute for Mathematics and AE is extremely grateful for their hospitality.

\section{Notation and Toolbox \label{sec:notation}}

This section collects the notation  and a number of useful results in one place for the convenience of the reader. 
We will always use bold letters for matrices and calligraphic letters for subspaces, for example matrix $\B{A}$ and subspace $\mathcal{S}$. In particular, $\B{0}_{a\times b}$ denotes the $a\times b$ matrix of all zeros. For integers $a\le b$, we use the convention that $[a:b]=\{a,\cdots,b\}$. We will  also use MATLAB's matrix notation to represent rows, columns, and blocks of matrices, for example $\B{A}[1:r,:]$ is the restriction of matrix $\B{A}$ to its first $r$ rows.  Throughout, $C$ is an absolute constant, the value of which might change in every appearance.

In the appendices, $\lambda_1(\B{A})\ge \lambda_2(\B{A})\ge \cdots$ denote the eigenvalues of a symmetric matrix $\B{A}$ and  $\sigma_1(\B{B})\ge \sigma_2(\B{B})\ge \cdots$ denotes the singular values of a matrix $\B{B}$. Also  $\rho_r^2(\B{B}) = \sum_{i\ge r+1} \sigma_i^2(\B{B})$ stands for the residual of matrix $\B{B}$.

Let us also recall some of the  spectral properties of a standard random Gaussian matrix, namely a matrix populated with independent random Gaussian variables with zero-mean and unit variance. For a standard Gaussian matrix $\B{G}\in\R^{a\times b}$ with $a\ge b$  and for fixed $\alpha \ge 1$, Corollary 5.35 in \cite{vershynin2010introduction} dictates that 
\begin{equation}
\sqrt{a} - \alpha \sqrt{b} \le \sigma_{b}(\B{G}) \le \sigma_{1}(\B{G}) \le \sqrt{a}+\alpha\sqrt{b},
\label{eq:bnd on Gaussians}
\end{equation}
except with a probability of at most $e^{-C\alpha^2 b}$.  Moreover, for a matrix $\B{\Gamma}\in\R^{a'\times a}$ and $\alpha\ge 1$,  an application of the Hensen-Wright inequality \cite[Theorem 1.1]{rudelson2013hanson} yields  that 
\begin{equation}
\l| \l\| \B{\Gamma} \B{G} \r\|_F^2 - \E\| \B{\Gamma} \B{G} \|_F^2 \r|  \le \beta,
\label{eq:scalar Bernie full}
\end{equation}
for $\beta \ge0$ and except with a probability of at most 
$$
\exp\l(-\min\l( \frac{\beta^2}{b\|\B{\Gamma}\|^2 \| \B{\Gamma}\|_F^2} , \frac{\beta}{\|\B{\Gamma} \|^2}\r)   \r),
$$
where $\|\cdot \|$ stands for  spectral norm.
In particular, with the choice $\beta =\alpha^2 \|\B{\Gamma}\|_F^2 b$ above and $\alpha \ge 1$, we find that 
\begin{equation}
\|\B{\Gamma} \B{G} \|_F^2 \le (1+\alpha^2) \|\B{\Gamma}\|_F^2 b \le 2\alpha^2 \|\B{\Gamma}\|_F^2 b,
\label{eq:scalar Bernie}
\end{equation}
except with a probability of at most 
$$
\exp\l({-C{\alpha^2 b\|\B{\Gamma}\|_F^2}/\|\B{\Gamma}\|^2}\r) \le \exp(-C\alpha^2 b).
$$ 
In a different regime, with the choice of $\beta = \alpha^2 \| \B{\Gamma}\|_F^2\sqrt{b}$ in \eqref{eq:scalar Bernie full} and $\alpha^2 \le \sqrt{b}$, we arrive at  
\begin{equation}
\l| \|\B{\Gamma} \B{G} \|_F^2 - \E\|\B{\Gamma}\B{G} \|_F^2\r| = \l| \|\B{\Gamma} \B{G} \|_F^2 - b\| \B{\Gamma} \|_F^2\r| \le \alpha^2 \|\B{\Gamma}\|_F^2 \sqrt{b},
\label{eq:scalar Bernie near}
\end{equation}
except with a probability of at most 
$$
\exp\l({-C{\alpha^4 \|\B{\Gamma}\|_F^2}/\|\B{\Gamma}\|^2}\r) \le \exp(-C\alpha^4).
$$

\section{Proof of Proposition \ref{prop:first app} \label{sec:proof of first app}}

Let 
\begin{equation} 
 \B{\Xi} =\B{S} \B{\Lambda} \B{S}^* =  \B{S} \B{\Sigma}^2 \B{S}^* \in \R^{n\times n}
 \label{eq:def of Xi proof of proposition}
\end{equation} 
be the eigen-decomposition of the covariance matrix $\B{\Xi}$, where $\B{S}\in\R^{n\times n}$ is an orthonormal matrix and the diagonal matrix $\B{\Lambda} = \B{\Sigma}^2\in\R^{n\times n}$ contains the eigenvalues of $\B{\Xi}$  in nonincreasing order, namely 
\begin{equation}
\B{\Lambda} = \B{\Sigma}^2 = 
\l[
\begin{array}{cccc}
\sigma_1^2 \\
& \sigma_2^2 \\
& & \ddots \\
& & & \sigma_n^2
\end{array}
\r]
\in\R^{n\times n},
\qquad \sigma_1^2 \ge \sigma_2^2 \ge \cdots\ge \sigma_n^2.
\label{eq:Sigma 2 proposition}
\end{equation}
Throughout, we also make use of the condition number and residual, namely 
\begin{equation}
\kappa_r = \frac{\sigma_1}{\sigma_r}, \qquad 
\rho_r^2 = \rho_r^2(\B{\Xi}) = \sum_{i=r+1}^n \sigma_i^2. \qquad \mbox{(see \eqref{eq:res of gaussian dist})}
\label{eq:shorthand}
\end{equation}
Recall that $\{y_t\}_{t=1}^T\subset\R^n$ are the data vectors drawn from the  Gaussian measure $\mu$ with zero mean and covariance matrix $\B{\Xi}$,  and that $\Y_T\in\R^{n\times T}$ is obtained by concatenating $\{y_t\}_{t=1}^T$.  It follows that  
\begin{equation*}
y_t = \B{S} \B{\Sigma} g_t,\qquad  t\in [1:T],
\end{equation*}
\begin{equation}
\Y_T=\B{S} \B{\Sigma} \G_T,
\label{eq:def of YT gaussian}
\end{equation}
where $g_t\in\R^n$ and $\G_T\in\R^{n\times T}$ are standard random Gaussian vector and matrix, respectively. That is, $g_t$ and $\G_T$ are populated with independent Gaussian random variables with zero mean and unit variance. 
With these preparations, we are now ready to prove Proposition \ref{prop:first app}. For $y$ drawn from the Gaussian measure $\mu$, note that 
\begin{align}
\E_y\| y-\P_{\SU_{T,r}}y\|_2^2 
& = \E_y  \| \P_{\SU_{T,r}^\perp} y\|_2^2 \nonumber\\
& = \E_y \langle \P_{\SU_{T,r}^\perp}, yy^* \rangle  \nonumber\\
& = \langle \P_{\SU_{T,r}^\perp}, \B{\Xi} \rangle \nonumber\\
& = \l\langle \P_{\SU_{T,r}^\perp}, \B{\Xi} - \frac{\Y_T\Y_T^*}{T}\r \rangle + \frac{1}{T} \langle \P_{\SU_{T,r}^\perp}, {\Y_T\Y_T^*} \rangle \nonumber\\
& =  \l\langle \P_{\SU_{T,r}^\perp}, \B{\Xi} - \frac{\Y_T\Y_T^*}{T} \r\rangle + \frac{1}{T} \|   \P_{\SU_{T,r}^\perp} \Y_T \|_F^2  \nonumber\\
& =  \l\langle \P_{\SU_{T,r}^\perp}, \B{\Xi} - \frac{\Y_T\Y_T^*}{T} \r\rangle + \frac{\rho_r^2(\Y_T)}{T} 
\qquad \mbox{(see Program \eqref{eq:empirical})}
\nonumber\\
& = \frac{1}{T} \l( \E\| \P_{\SU_{T,r}^\perp} \Y_T\|_F^2 -\|\P_{\SU^\perp_{T,r}} \Y_T\|_F^2 \r)+ \frac{\rho_r^2(\Y_T)}{T}.  
\qquad  \mbox{(see \eqref{eq:def of YT gaussian})}
\label{eq:bias var decomp -1}
\end{align}
Let us next control the two components in the last line above. The first component above involves the deviation of random variable $\|\P_{\SU_{T,r}^\perp}\Y_T\|_F^2$ from its expectation. By invoking the Hensen-Wright inequality in Section \ref{sec:notation} and for $\widetilde{\alpha}^2\le \sqrt{T}$, we write that 
\begin{align}
{\E\| \P_{\SU_{T,r}^\perp} \Y_T\|_F^2 - \|\P_{\SU^\perp_{T,r}} \Y_T\|_F^2} & 
= {\E\| \P_{\SU_{T,r}^\perp} \B{S} \B{\Sigma} \cdot \B{G}_T\|_F^2 - \|\P_{\SU^\perp_{T,r}} \B{S} \B{\Sigma} \cdot  \B{G}_T \|_F^2}  
\qquad \text{(see \eqref{eq:def of YT gaussian})} \nonumber\\
& 
\le \widetilde{\alpha}^2 \| \P_{\SU^\perp_{T,r}}\B{S} \B{\Sigma} \|_F^2 {\sqrt{T}}
\qquad \mbox{(see \eqref{eq:scalar Bernie near})} \nonumber \\
& \le  {\widetilde{\alpha}^2 \|\P_{\SU_{T,r}^\perp} \B{S}\|^2_F \l\|  \B{\Sigma} \r\|^2} {\sqrt{T}} 
\nonumber\\
& \le {\widetilde{\alpha}^2  (n-r) \sigma_1^2 }{\sqrt{T}},
\qquad \mbox{(see (\ref{eq:Sigma 2 proposition},\ref{eq:shorthand}))}
\end{align}
except with a probability of at most $e^{-C\widetilde{\alpha}^4}$. In particular, for the choice of $\widetilde{\alpha}^2 = \alpha^2 \sqrt{\log T}$ with $\alpha^2 \le \sqrt{T/\log T}$, we find that 
\begin{equation}
{\E\| \P_{\SU_{T,r}^\perp} \Y_T\|_F^2 - \|\P_{\SU^\perp_{T,r}} \Y_T\|_F^2}
\le \alpha^2 (n-r)\sigma_1^2 \sqrt{T {\log T} }, 
\label{eq:deviation with alpha}
\end{equation}
except with a probability of $T^{-C\alpha^4}$. 
We next bound the second term in the last line of \eqref{eq:bias var decomp -1}, namely the residual of $\Y_T$. 
Note that 
\begin{align}
\rho_r^2(\Y_{T}) & = \rho_r^2(\B{S} \B{\Sigma} \B{G}_{T}) 
\qquad \mbox{(see \eqref{eq:def of YT gaussian})}
\nonumber\\
& = \rho_r^2 (\B{\Sigma} \G_{T})
\qquad \l( \B{S}^* \B{S} = \B{I}_n \r) \nonumber\\
& =  \min_{\operatorname{rank}(\X)=r} \|\B{\Sigma} \G_{T}-\X\|_F^2.
\qquad \mbox{(see \eqref{eq:shorthand})}
\label{eq:simplification of residual}
\end{align}
By substituting above the suboptimal choice of 
\begin{equation}
\X_o = 
\l[
\begin{array}{c}
\B{\Sigma}[1:r,1:r] \cdot \G_{T}[1:r,:] \\
\B{0}_{(n-r)\times T}
\end{array} \r],
\label{eq:def of Xo other}
\end{equation}
we find that 
\begin{align}
\rho_r^2(\Y_{T}) & = \min_{\text{rank}(\X)=r} \|\B{\Sigma} \G_{T}-\X\|_F^2 
\qquad \mbox{(see \eqref{eq:simplification of residual})}
 \nonumber\\
& \le \|\B{\Sigma} \G_{T}-\X_o\|_F^2\nonumber\\
& = \| \B{\Sigma}[r+1:n,r+1:n] \cdot \G_{T}[r+1:n,:] \|_F.
\qquad \mbox{(see \eqref{eq:def of Xo other})}
\label{eq:finding residual pre}
\end{align}
Note that $\G_{T}[r+1:n,:] \in \R^{(n-r)\times T}$ is a standard Gaussian matrix. For $\alpha\ge 1$, an application of the Hensen-Wright inequality in Section \ref{sec:notation} therefore implies that 
\begin{align}
\rho_r^2(\Y_{T}) 
&\le \| \B{\Sigma}[r+1:n,r+1:n] \cdot \G_{T}[r+1:n,:] \|_F^2 
\qquad \mbox{(see \eqref{eq:finding residual pre})}
\nonumber\\
& \le 2\alpha^2  \|\B{\Sigma}[r+1:n,r+1:n]\|_F^2 T
\qquad \mbox{(see \eqref{eq:scalar Bernie})}
\nonumber\\
& =2\alpha^2 \rho_r^2 T,
\qquad \mbox{(see \eqref{eq:shorthand})}
\label{eq:second norm}
\end{align}
except with a probability of at most $e^{-C\alpha^2 T}$. 
We now substitute the bounds in \eqref{eq:deviation with alpha} and  \eqref{eq:second norm} back into \eqref{eq:bias var decomp -1} to arrive at 
\begin{align}
\E\| y-\P_{\SU_{T,r}}y\|_2^2 & \le\alpha^2 (n-r)\sigma_1^2 \sqrt{T {\log T} }+2\alpha^2\rho_r^2,
\end{align}
when $\alpha^2 \le \sqrt{T/\log T}$ and except with a probability of at most 
\begin{align*}
T^{-C\alpha^4}+e^{-C\alpha^2 T} \le T^{-C\alpha^4},
\qquad 
\l( \alpha^2 \le \sqrt{T/\log T}\r)
\end{align*}
where we have used the abuse of notation in which $C$
is a universal constant that is allowed to change in every appearance. 
This completes the proof of Proposition \ref{prop:first app}.

\section{Proof of Theorem \ref{thm:stochastic result} \label{sec:theory}}

In the rest of this paper, we slightly unburden the notation by using  $\Y_{k}\in\R^{n\times kb}$ to denote $\Y_{kb}$. For example, we will use $\Y_K\in \R^{n\times T}$ instead of $\Y_T$ because $T=Kb$. We also write $\widehat{\mathcal{S}}
_{k,r}$ instead of $\widehat{\mathcal{S}}
_{kb,r}$. As with the proof of Proposition \ref{prop:first app}, we argue that 
\begin{align}
\E_y \| y - \P_{\wh{\SU}_{K,r}}y \|_2^2 & \le \frac{1}{T} \l( \E\| \P_{\wh{\SU}_{K,r}^\perp} \Y_T\|_F^2 - {\|\P_{\wh{\SU}^\perp_{K,r}} \Y_T\|_F^2} \r) +\frac{1}{T} \| \P_{\wh{\SU}_{K,r}^\perp} \Y_K\|_F^2
\qquad \mbox{(similar to \eqref{eq:bias var decomp -1})}
\nonumber\\ 
& \le \alpha^2 (n-r)\sigma_1^2 \sqrt{\frac{ {\log T} }{T}} + \frac{1}{T} \| \P_{\wh{\SU}_{K,r}^\perp} \Y_K\|_F^2  \qquad \mbox{(see \eqref{eq:deviation with alpha})} \nonumber\\
& =  \alpha^2 (n-r)\sigma_1^2 \sqrt{\frac{ {\log T} }{T}}  + \frac{1}{T} \|  \P_{\wh{\SU}_{K,r}^\perp} ( \Y_K - \wh{\Y}_{K,r}) \|_F^2
\qquad \mbox{(see \eqref{eq:span of output of MOSES})}
\nonumber\\
& \le  \alpha^2 (n-r)\sigma_1^2 \sqrt{\frac{ {\log T} }{T}}  + \frac{1}{T} \|  \Y_K - \wh{\Y}_{K,r}\|_F^2, 
\label{eq:moses err brk down}
\end{align}
except with a probability of at most $T^{-C\alpha^4}$ and provided that $\alpha^2 \le  \sqrt{T/\log T}$. It therefore remains to control the norm in the last line above. Recall that the output of $\alg$, namely $\wh{\Y}_{K,r}$, is intended to approximate a rank-$r$ truncation of $\Y_K$.  
We will therefore compare the error $\|\Y_{K}-\wh{\Y}_{K,r}\|_F$ in \eqref{eq:moses err brk down} with the true residual $\rho_r(\Y_{K})$. 
To that end, our analysis consists of a deterministic bound  and a stochastic evaluation of this bound. The deterministic bound is as follows, see Section \ref{sec:proof of deterministic lemma} for the proof. 
\begin{lem}\label{lem:online trunc SVD} For every $k\in[1:K]$, let $\Y_{k,r}=\SVD_r(\Y_{k}) \in\mathbb{R}^{n\times kb}$ be a rank-$r$ truncation of $\Y_{k}$ and set $\Ysub_{k,r} = \mbox{span}(\Y_{k,r})\in\GR(n,r)$.  For $\p>1$, we also set 
\begin{equation}
\theta_k :=  1+ \frac{\p^{\frac{1}{3}}\|\y_k\|^2}{\sigma_r(\Y_{k-1})^2}.
\label{eq:def of theta}
\end{equation}
Then the output of $\alg$, namely $\wh{\Y}_{K,r}$, satisfies   
\begin{align}
\| \Y_{K}-\wh{\Y}_{K,r} \|_F^2 
& \le \frac{\p^{\frac{1}{3}}}{\p^{\frac{1}{3}}-1}
  \sum_{k=2}^K \l( \prod_{l=k+1}^K \theta_l \r)  \| \P_{\Ysub_{k-1,r}^\perp} \y_k\|_F^2,
\label{eq:deterministic online svd}
\end{align}
where $\P_{\Ysub_{k-1,r}^\perp}\in\R^{n\times n}$ is the orthogonal projection onto the orthogonal complement of $\Ysub_{k-1,r}$. Above, we use the convention that $\prod^{K}_{l=K+1} \theta_l=1$. 

\end{lem}
In words, \eqref{eq:deterministic online svd} gives a deterministic bound on the performance of $\alg$. The term $\| \P_{\Ysub_{k-1,r}^\perp} \y_k\|_F$ 
in \eqref{eq:deterministic online svd} is in  a sense the ``innovation'' at iteration $k$, namely the part of the new data block $\y_k$ that cannot be described by the current estimate $\SU_{k-1,r}$.  The overall innovation in  \eqref{eq:deterministic online svd} clearly controls the performance of $\alg$.
In particular, if the data blocks  are drawn from the same distribution, this innovation gradually reduces as $k$ increases. For example, if $\{\y_k\}_{k=1}^K$  are drawn from a distribution with a rank-$r$ covariance matrix, then the innovation term vanishes  almost surely after finitely many iterations. In contrast, when the underlying covariance matrix is high-rank, the innovation term decays more slowly and never completely disappears even as $k\rightarrow \infty$. 
We will next evaluate the right-hand side of \eqref{eq:deterministic online svd} in a  stochastic setup, see Section \ref{sec:stochastic bnd} for the proof. 
\begin{lem}\label{lem:stochastic bnd}
Suppose that $\{y_t\}_{t=1}^T$  are drawn from a zero-mean Gaussian probability measure with the covariance matrix $\B{\Xi}\in\R^{n\times n}$. Let $\sigma_1^2\ge \sigma_2^2 \ge \cdots $ be the eigenvalues of $\B{\Xi}$ and recall the notation in (\ref{eq:shorthand}). For $\p>1$, also let 
$$
\rate_r := \kappa_r + \frac{\sqrt{2}\alpha \rho_r}{\p^{\frac{1}{6}}\sigma_r}.
$$ 
For $\alpha \ge 1$,  it then holds that   
\begin{align}
 \| \Y_{K}-\wh{\Y}_{K,r} \|_F^2 
& \le   \frac{50\p^{\frac{4}{3}}\alpha^2  }{(\p^{\frac{1}{3}}-1)^2} \cdot  \min\l (   \kappa_r^2 \rho_r^2, r \sigma_1^2 +\rho_r^2\r) \eta_r^2 b  \l( \frac{2K}{\p\eta_r^2}+2\r)^{\p\eta_r^2}  ,
\label{eq:residual lemma display}
\end{align}
except with a probability of at most $e^{-C\alpha^2 r}$ 
and provided that 
$$
b
\ge \frac{\p^{\frac{1}{3}}\alpha^2 r}{(\p^{\frac{1}{6}}-1)^2},
\qquad b \ge C\alpha^2 r.
$$
\end{lem}
Substituting the right-hand side of \eqref{eq:residual lemma display} back into \eqref{eq:moses err brk down} yields that 
\begin{align}
\E_y \| y - \P_{\wh{\SU}_{K,r}}y \|_2^2 & \le  \alpha^2 (n-r)\sigma_1^2 \sqrt{\frac{ {\log T} }{T}}  + \frac{1}{T} \|  \Y_K - \wh{\Y}_{K,r}\|_F^2, 
\qquad \mbox{(see \eqref{eq:moses err brk down})}
\nonumber\\
& \le 
 \alpha^2 (n-r)\sigma_1^2 \sqrt{\frac{ {\log T} }{T}}
+
 \frac{50\p^{\frac{4}{3}}\alpha^2  }{(\p^{\frac{1}{3}}-1)^2} \cdot  \min\l (   \kappa_r^2 \rho_r^2, r \sigma_1^2 +\rho_r^2\r) \frac{\eta_r^2}{K}  \l( \frac{2K}{\p\eta_r^2}+2\r)^{\p\eta_r^2}.  
\end{align}
In particular, if $K \ge p\eta_r^2$, we may simplify the above bound to read 
\begin{align}
\E_y \| y - \P_{\wh{\SU}_{K,r}}y \|_2^2 \le 
 \alpha^2 (n-r)\sigma_1^2 \sqrt{\frac{ {\log T} }{T}}  + \frac{50\p^{\frac{1}{3}}\alpha^2 4^{p\eta_r^2} }{(\p^{\frac{1}{3}}-1)^2} \cdot  \min\l (   \kappa_r^2 \rho_r^2, r \sigma_1^2 +\rho_r^2\r)
\l(\frac{K}{\p\eta^2_r} \r)^{p\eta_r^2-1},
\label{eq:final bnd proof}
\end{align}
which completes the proof of Theorem \ref{thm:stochastic result}.

\section{Proof of Lemma \ref{lem:online trunc SVD} \label{sec:proof of deterministic lemma}}

Recall that the output of $\alg$ is the sequence of rank-$r$ matrices $\{\wh{\Y}_k\}_{k=1}^K$.  For every $k<K$, it is more convenient in the proof of Lemma \ref{lem:online trunc SVD} to pad both $\Y_k,\wh{\Y}_{k,r}\in\mathbb{R}^{n\times kb}$ with zeros to form the $n\times K b$ matrices 
\begin{equation}
\l[ 
\begin{array}{cc}
{\Y}_{k} & \B{0}_{n\times (K-k)b}
\end{array}
\r],
\qquad 
\l[ 
\begin{array}{cc}
\wh{\Y}_{k,r} & \B{0}_{n\times (K-k)b}
\end{array}
\r].
\label{eq:overloaded not}
\end{equation}
We overload the notation $\Y_{k},\wh{\Y}_{k,r}$ to show the new $n\times Kb$ matrices in \eqref{eq:overloaded not}. Let 
$$
\widehat{\SU}_{k,r}=\mbox{span}(\widehat{\Y}_{k,r})\in \GR(n,r),
$$
\begin{equation}
\wh{\mathcal{Q}}_{k,r}=\mbox{span}(\widehat{\Y}_{k,r}^*) \in \GR(Kb,r)
\label{eq:left right subspaces of iterates}
\end{equation}
denote the ($r$-dimensional) column and row spaces of the rank-$r$ matrix $\wh{\Y}_{k,r} \in \R^{n\times Kb}$, respectively. Let also $\wh{\B{S}}_{k,r}\in\mathbb{R}^{n\times r}$ and $\wh{\Q}_{k,r}\in\mathbb{R}^{Kb\times r}$ be orthonormal bases for these subspaces. We also let  $\I_k\subset \mathbb{R}^{Kb}$ 
 denote the $b$-dimensional subspace spanned by the coordinates $[(k-1)b+1:bk]$, namely
 \begin{equation}
\I_k = \mbox{span}\l(
\l[
\begin{array}{c}
\B{0}_{(k-1)b\times b}\\
\B{I}_{b}\\
\B{0}_{(K-k)b \times b}
\end{array}
\r]
 \r) \in \GR(Kb,b),
 \label{eq:def of Is}
\end{equation}
and we use the notation 
\begin{equation}
\J_{k}:=\I_1\oplus \I_2 \cdots \oplus \I_k \in \GR(Kb,kb), \qquad k\in [1:K] ,
\label{eq:def of Js}
\end{equation}
to denote the $kb$-dimensional subspace  that spans the  first $kb$ coordinates in $\mathbb{R}^{Kb}$. 
The following technical lemma, proved in Section \ref{sec:alt exp}, gives another way of expressing the output of $\alg$, namely $\{\wh{\Y}_{k,r}\}_{k=1}^K$.
\begin{lem}\label{lem:SVD lemma} For every $k\in[1:K]$, it holds that 
\begin{equation}
\wh{\Y}_{k,r} =
{\Y}_{K} \P_{\wh{\Qsub}_{k,r}}, 
\label{eq:SVD simplified}
\end{equation}
or equivalently
\begin{equation}
\wh{\Y}_{k-1,r}+\Y_k \P_{\I_k}
=  
{\Y}_{K} \P_{\wt{\Qsub}_{k}},
\label{eq:SVD simplified 2}
\end{equation}
where 
\begin{equation}
\wt{\Qsub}_{k} := \wh{\Qsub}_{k-1,r}\oplus \I_k \subset \R^{Kb}
\label{eq:def of Qtilde}
\end{equation}
is the direct sum of the two subspaces $\wh{\Qsub}_{k-1,r}$ and $\I_k$. In particular, the update rule in Algorithm \ref{alg:MOSES} can be written as 
\begin{equation}
{\Y}_{K} \P_{\wh{\Qsub}_{k,r}} = \SVD_r \l( {\Y}_{K} \P_{\wt{\Qsub}_k} \r),
\qquad k\in[2:K]. 
\label{eq:SVD simplified 3}
\end{equation}
Lastly we have the inclusion 
\begin{equation}
\wh{\Qsub}_{k,r}\subset \wt{\Qsub}_k \subset \J_{k}\in \GR(Kb,kb).
\label{eq:long inclusion}
\end{equation} 
\end{lem}
In particular, \eqref{eq:SVD simplified} and \eqref{eq:long inclusion} together imply that 
\begin{equation*}
\wh{\Y}_{k,r} = {\Y}_{K}\P_{\J_{k}}\P_{\wh{\Qsub}_{k,r}} = \Y_{k}\P_{\wh{\Qsub}_{k,r}},
\end{equation*}
that is, only $\Y_{k}$ (containing the first $kb$ data vectors) contributes to the formation of $\wh{\Y}_{k,r}$, the output of algorithm at iteration $k$, which was to be expected of course. Recall that  $\wh{\Y}_{k,r}$ is intended to approximate $\Y_{k,r}=\SVD_r(\Y_{k})$. 
In light of Lemma \ref{lem:SVD lemma}, let us now derive a simple recursive expression for  the residual $\Y_{k}-\wh{\Y}_{k,r}$. 
For every $k\in[2:K]$, it holds that 
\begin{align}
\label{eq:iterative exp for err}
\Y_{k}-\wh{\Y}_{k,r} & 
= \Y_{K}\P_{\J_{k}}- \Y_{K}\P_{\wh{\Qsub}_{k,r}}
\qquad \mbox{(see \eqref{eq:def of Js} and \eqref{eq:SVD simplified})}
 \nonumber\\
& = \Y_{K}\P_{\J_{k-1}}+\Y_{K}\P_{\I_k}- \Y_{K}\P_{\wh{\Qsub}_{k,r}} 
\qquad \mbox{(see \eqref{eq:def of Js})}
\nonumber\\
& = \Y_{k-1}+\Y_{K}\P_{\I_k}-\Y_{K}\P_{\wh{\Qsub}_{k,r}} 
\qquad \mbox{(see \eqref{eq:def of Js})}
\nonumber\\
& = \Y_{k-1}-\wh{\Y}_{k-1,r}+ \Y_{K}\P_{\wh{\Qsub}_{k-1,r}} +\Y_{K}\P_{\I_k}- \Y_{K}\P_{\wh{\Qsub}_{k,r}} 
\qquad \mbox{(see \eqref{eq:SVD simplified})}
\nonumber\\
& = \l(\Y_{k-1}-\wh{\Y}_{k-1,r} \r)+ \Y_{K}\l( \P_{\wh{\Qsub}_{k-1,r}} +\P_{\I_k}\r)- \Y_{K}\P_{\wh{\Qsub}_{k,r}}
 \nonumber\\
&=\l( \Y_{k-1}-\wh{\Y}_{k-1,r} \r)+ \Y_{K}\l(\P_{\wt{\Qsub}_{k}}-\P_{\wh{\Qsub}_{k,r}} \r).
\qquad \mbox{(see \eqref{eq:def of Qtilde})}
\end{align}
Interestingly, the two terms in the last line of \eqref{eq:iterative exp for err} are orthogonal, as proved by induction in Section \ref{sec:orth of summands}. 
\begin{lem}\label{lem:orthogonality of summands} For every $k\in[2:K]$, it holds that 
\begin{equation}\label{eq:orthogonality of summands}
\l\langle  \Y_{k-1}-\wh{\Y}_{k-1,r},  \Y_{K}\l(\P_{\wt{\Qsub}_{k}}-\P_{\wh{\Qsub}_{k,r}} \r) \r\rangle = 0.
\end{equation}
\end{lem}
For fixed $k\in[2:K]$, Lemma \ref{lem:orthogonality of summands} immediately implies that 
\begin{align}
\| \Y_{k}- \wh{\Y}_{k,r} \|_F^2 & 
= \l\| \l(\Y_{k-1}-\wh{\Y}_{k-1,r} \r) + \Y_{K}\l( \P_{\wt{\Q}_k}-\P_{\wh{\Q}_{k,r}} \r)\r\|_F^2 
\qquad \mbox{(see \eqref{eq:iterative exp for err})}
\nonumber\\
& = \| \Y_{k-1}-\wh{\Y}_{k-1,r}  \|_F^2 + \| \Y_{K}( \P_{\wt{\Qsub}_k}-\P_{\wh{\Qsub}_{k,r}} )\|_F^2 
\qquad \mbox{(see Lemma \ref{lem:orthogonality of summands})}
\nonumber\\
& = \| \Y_{k-1}-\wh{\Y}_{k-1,r}  \|_F^2 + \rho_r\l( 
\wh{\Y}_{k-1,r} + \Y_k\P_{\I_k}
 \r).
 \qquad \mbox{(see \eqref{eq:SVD simplified 3} and \eqref{eq:SVD simplified 2})}
\end{align}
Recalling from \eqref{eq:left right subspaces of iterates} that $\wh{\SU}_{k-1,r}=\mbox{span}(\wh{\Y}_{k-1,r})$, we bound the above expression  by writing that 
\begin{align}
\| \Y_{k}- \wh{\Y}_{k,r} \|_F^2 &
= \| \Y_{k-1}-\wh{\Y}_{k-1,r}  \|_F^2 + \rho_r\l( 
\wh{\Y}_{k-1,r} + \Y_k \P_{\I_k}
 \r) \nonumber\\
& \le \| \Y_{k-1}-\wh{\Y}_{k-1,r}  \|_F^2
 + \l\|  \P_{\wh{\SU}_{k-1,r}^\perp} 
\l( \wh{\Y}_{k-1,r} + \Y_k \P_{\I_k} \r)
 \r\|_F^2 \nonumber\\
 & = \| \Y_{k-1}-\wh{\Y}_{k-1,r}  \|_F^2
 + \|  \P_{\wh{\SU}_{k-1,r}^\perp} \y_k \|_F^2,
 \qquad \text{(see \eqref{eq:left right subspaces of iterates})}
 \label{eq:getting rid of residual}
\end{align}  
where the second line follows from the sub-optimality of the choice of subspace $\wh{\SU}_{k-1,r}$. 
Let us focus on the last norm above. For every $k$, let $\Y_{k,r}=\SVD_r(\Y_{k})$ be a rank-$r$ truncation of $\Y_{k}$ with the column span $\Ysub_{k,r}=\mbox{span}(\Y_{k,r})$. We now write that 
\begin{align}
\| \P_{\wh{\SU}_{k-1,r}^\perp} \y_k \|_F & \le 
\| \P_{\wh{\SU}_{k-1,r}^\perp} \P_{\Ysub_{k-1,r}}\y_k \|_F + \| \P_{\wh{\SU}_{k-1,r}^\perp} \P_{\SU_{k-1,r}^\perp }\y_{k} \|_F 
\qquad \mbox{(triangle inequality)}
\nonumber\\
& \le  \| \P_{\wh{\SU}_{k-1,r}^\perp} \P_{\Ysub_{k-1,r}}\|_F \cdot \| \y_k\|
+ \| \P_{\Ysub_{k-1,r}^\perp }\y_k \|_F.
\label{eq:nearly an angle}
\end{align} 
The first norm in the last line above gauges the principal angles between the two $r$-dimensional subspaces $\wh{\SU}_{k-1,r}$ and $\Ysub_{k-1,r}$. We can bound this norm with a standard perturbation result, for example see   \cite[Lemma 6]{eftekhari2016snipe} or \cite{wedin1972perturbation}. More specifically,  we may imagine that $\Y_{k-1}$ is a perturbed copy of $\Y_{k-1,r}$. Then the angle between $\SU_{k-1,r}=\text{span}(\Y_{k-1,r})$ and  $\wh{\SU}_{k-1,r}=\mbox{span}(\wh{\Y}_{k-1,r})$ is controlled by the amount of perturbation, namely with the choice of $\B{A}=\wh{\Y}_{k-1,r},\B{B}=\Y_{k-1},\B{B}_r = \Y_{k-1,r}$ in \cite[Lemma 6]{eftekhari2016snipe}, we find that 
\begin{align}
\| \P_{\wh{\SU}_{k-1,r}^\perp} \P_{\Ysub_{k-1,r}}\|_F  
& \le \frac{ \|  \Y_{k-1}- \wh{\Y}_{k-1,r} \|_F}{\sigma_r\l(\Y_{k-1} \r)}.
\label{eq:pert bound in online SVD}
\end{align}
By plugging \eqref{eq:pert bound in online SVD} back into \eqref{eq:nearly an angle}, we find that 
\begin{align}
\| \P_{\wh{\SU}_{k-1,r}^\perp} \y_k \| \le \frac{\|\y_k\|}{\sigma_r\l(\Y_{k-1} \r)} \cdot \| \Y_{k-1}-\wh{\Y}_{k-1,r} \|_F +  \| \P_{\SU_{k-1,r}^\perp} \Y_k\|_F. 
\label{eq:full circle}
\end{align}
In turn, for $\p>1$, substituting the above inequality into \eqref{eq:getting rid of residual} yields that 
\begin{align}
\| \Y_{k}-\wh{\Y}_{k,r} \|_F^2  & \le \| \Y_{k-1}-\wh{\Y}_{k-1,r} \|_F^2 + 
\| \P_{\wh{\SU}_{k-1,r}^\perp} \y_k\|_F^2 
\qquad \mbox{(see \eqref{eq:getting rid of residual})}
\nonumber\\
& \le \l( 1+ \frac{\p^{\frac{1}{3}}\|\y_k\|^2 }{\sigma_r\l(\Y_{k-1} \r)^2}\r) \| \Y_{k-1}-\wh{\Y}_{k-1,r} \|_F^2 + \frac{\p^{\frac{1}{3}}}{\p^{\frac{1}{3}}-1}\| \P_{\Ysub_{k-1,r}^\perp} \y_k\|_F^2
\qquad\mbox{(see \eqref{eq:full circle})}
\nonumber\\
& =: \theta_k  \| \Y_{k-1}-\wh{\Y}_{k-1,r} \|_F^2 + \frac{\p^{\frac{1}{3}}}{\p^{\frac{1}{3}}-1}\| \P_{\Ysub_{k-1,r}^\perp} \y_k\|_F^2.
\label{eq:folded recursion}
\end{align}
where we used the inequality $(a_1+a_2)^2 \le q a_1^2 + \frac{q a_2^2}{q-1}$ for scalars $a_1,a_2$ and $q>1$, with the choice of $q=p^{\frac{1}{3}}$.  
By unfolding the recursion in \eqref{eq:folded recursion}, we arrive at 
\begin{align}
\| \Y_{K}-\wh{\Y}_{K,r} \|_F^2 & \le \frac{\p^{\frac{1}{3}}}{\p^{\frac{1}{3}}-1} \sum_{k=2}^K \l( \prod_{l=k+1}^K \theta_l \r)  \| \P_{\Ysub_{k-1,r}^\perp} \y_k\|_F^2,
\end{align}
which completes the proof of Lemma \ref{lem:online trunc SVD}.

\section{Proof of Lemma \ref{lem:stochastic bnd} \label{sec:stochastic bnd}}

Recall  that $\y_k\in\R^{n\times b},\Y_k\in\R^{n\times kb}$ denote the $k$th block and the concatenation of the first $k$ blocks of data, respectively.
Since the data vectors are independently drawn from a zero-mean Gaussian probability measure with covariance matrix $\B{\Xi}$, it follows from (\ref{eq:def of Xi proof of proposition},\ref{eq:Sigma 2 proposition}) that
\begin{equation*}
\y_k=\B{S} \B{\Sigma} \g_k,
\end{equation*}
\begin{equation}
\Y_k = \B{S}\B{\Sigma} \G_k,
\label{eq:def of Y1k gaussian}
\end{equation}
for every $k\in [1:K]$, where   $\g_k\in\R^{n\times b}$ and $\G_k\in\R^{n\times kb}$ are standard random Gaussian matrices. 
 For fixed $k\in[2:K]$, let us  now study each of the random quantities on the right-hand side of \eqref{eq:deterministic online svd}.  The following results are proved in Appendices \ref{sec:proof of bnd on yk spec norm} and \ref{sec:proof of bnd on sigmar Yk}, respectively. 
\begin{lem}\textbf{\emph{(Bound on $\|\y_k\|$)}} \label{lem:bnd on yk spec norm}
For $\alpha\ge 1$, $\p >1$,  and fixed $k\in [1:K]$, it holds that 
\begin{align}
\|\y_k\| & \le   \p^{\frac{1}{6}}( \sigma_1  +\sqrt{2}\alpha \p^{-\frac{1}{6}}\rho_r ) \sqrt{ b},\label{eq:bnd on sigma1 Yk}
\end{align}
except with a probability of at most $ e^{-C\alpha^2 b}$ and provided that 
\begin{equation}
b\ge \frac{\alpha^2 r}{(p^{\frac{1}{6}}-1)^2}.
\end{equation}
\end{lem}
\begin{lem}\textbf{\emph{(Bound on $\sigma_r(\Y_{k})$)}} \label{lem:bnd on sigmar Yk}
For $\alpha \ge 1$, $\p>1$, and fixed $k\in[ 1:K]$, it holds that 
\begin{align}
\sigma_r(\Y_{k}) & \ge \p^{-\frac{1}{6}} \sigma_r \sqrt{kb},
\label{eq:bnd on sigmar Y1k}
\end{align}
except with a probability of at most $e^{-C\alpha^2 r}$ and provided that 
\begin{equation}
b\ge \frac{\alpha^2 r}{(1-p^{\frac{-1}{6}})^2}.
\label{eq:lwr bnd on b}
\end{equation}
\end{lem}
By combining Lemmas \ref{lem:bnd on yk spec norm} and \ref{lem:bnd on sigmar Yk}, we find for fixed $k\in[2:K]$ that 
\begin{align}
\theta_k & = 1+\frac{\p^{\frac{1}{3}}\|\y_k\|^2}{\sigma_r(\Y_{k-1})^2}
\qquad \mbox{(see \eqref{eq:def of theta})}
 \nonumber\\
 & \le 1+ \frac{\p ( \sigma_1  +\sqrt{2} \alpha \p^{-\frac{1}{6}}\rho_r )^2 b}{\sigma_r^2 (k-1)b} 
 \qquad \mbox{(see Lemmas \ref{lem:bnd on yk spec norm} and \ref{lem:bnd on sigmar Yk})}
 \nonumber\\
 & =:  1+ \frac{ \p \rate_r^2  }{k-1 },
 \label{eq:bnd on theta k}
\end{align}
except with a probability of at most $e^{-C\alpha^2 r}$ and provided that \eqref{eq:lwr bnd on b} holds. 
In particular, it follows that 
\begin{align}
\prod_{l=k+1}^K \theta_l & \le  \prod_{l=k+1}^K 
\l( 1+  \frac{ \p\rate_r^2 }{l-1 } \r) 
\qquad \mbox{(see \eqref{eq:bnd on theta k})}
\nonumber\\
& \le \frac{(K-1+p\eta_r^2)^{K-1+p \eta_r^2}}{(K-1)^{K-1}} \cdot \frac{(k-1)^{k-1}}{(k-1+p \eta_r^2)^{k-1+p \eta_r^2}}
\qquad \text{(see below)}
\nonumber\\
& = \l(1+ \frac{\p \eta_r^2}{K-1} \r)^{K-1}  \l(1+\frac{\p \eta_r^2}{k-1} \r)^{-k+1} \l(\frac{K-1+\p \eta_r^2}{k-1+\p \eta_r^2} \r)^{\p \eta_r^2},
\label{eq:bnd on prod -1}
\end{align}
holds for every $k\in [2:K]$ and except with a probability of at most $Ke^{-C\alpha r}$, where the failure probability follows from an application of the union bound. The second line above is obtained by bounding the logarithm of the product in that line with the corresponding integral. More specifically, it holds that 
\begin{align}
& \log \l( \prod_{l=k+1}^K  
 \l( 1+ \frac{\p \eta_r^2}{l-1 } \r)  \r) 
\nonumber\\ & = \sum_{l=k}^{K-1} 
\log \l(1+ \frac{\p \eta_r^2}{l }  \r) \nonumber\\
& \le \int_{k-1}^{K-1} 
\log \l(1+ \frac{\p \eta_r^2}{x}  \r) \, dx \nonumber\\
& = (K-1+\p \eta_r^2)\log(K-1+\p \eta_r^2) - (K-1)\log(K-1) \nonumber\\
& \qquad 
-(k-1+\p \eta_r^2)\log(k-1+\p \eta_r^2) +(k-1)\log (k-1),
\end{align}
where the third line above follows because the integrand is decreasing in $x$. 
Let us further simplify \eqref{eq:bnd on prod -1}. Note that $K\ge k\ge 2$ and that $\p \eta_r^2 \ge 1$ by its definition in \eqref{eq:bnd on theta k}. Consequently, using the relation $2\le (1+1/x)^x \le e$ for $x\ge 1$, we can write that 
\begin{equation}
2 \le  \l(1+\frac{\p \eta_r^2}{k-1} \r)^{\frac{k-1}{\p \eta_r^2}} \le e,
\qquad 
2 \le \l(1+ \frac{\p \eta_r^2}{K-1} \r)^{\frac{K-1}{\p \eta_r^2}}  \le e. 
\label{eq:natural log ineq}
\end{equation}
In turn,  (\ref{eq:natural log ineq}) allows us to simplify \eqref{eq:bnd on prod -1} as follows:
\begin{align}
\prod_{l=k+1}^K \theta_l &  \le \l(1+ \frac{\p \eta_r^2}{K-1} \r)^{K-1}  \l(1+\frac{\p \eta_r^2}{k-1} \r)^{-k+1} \l(\frac{K-1+\p \eta_r^2\rate_r}{k-1+\p \eta_r^2} \r)^{\p \eta_r^2}
\qquad \mbox{(see \eqref{eq:bnd on prod -1})}
 \nonumber\\
& \le \l(\frac{e}{2}\r)^{\p \eta_r^2} \l(\frac{K-1+\p \eta_r^2}{k-1+\p \eta_r^2} \r)^{\p \eta_r^2}.
\qquad \mbox{(see \eqref{eq:natural log ineq})}
\label{eq:bnd on prod}
\end{align}
Next we control the random variable $\|\P_{\Ysub^\perp_{k-1}} \y_k\|_F$ in \eqref{eq:deterministic online svd} with the following result, proved in Section \ref{sec:proof of innovation}. 
\begin{lem}\textbf{\emph{(Bound on the Innovation)}}\label{lem:innovation}
For $\alpha\ge 1$ and fixed $k\in [2:K]$, it holds that 
\begin{equation}
\|\P_{\Ysub^\perp_{k-1,r}} \y_k\|_F \le  5 \alpha \min\l (   \kappa_r \rho_r, \sqrt{r} \sigma_1+\rho_r\r) \sqrt{b}, 
\end{equation}
except with a probability of at most $e^{-C\alpha^2 r}$ and provided that $b\ge C\alpha^2 r$. 
\end{lem}
By combining Lemma \ref{lem:innovation} and \eqref{eq:bnd on prod}, we finally find a stochastic bound for the right-hand side of \eqref{eq:deterministic online svd}. More specifically, it holds that 
\begin{align}
& \| \Y_{K}-\wh{\Y}_{K,r} \|_F^2  \nonumber\\
& \le \frac{\p^{\frac{1}{3}}}{\p^{\frac{1}{3}}-1}
  \sum_{k=2}^K \l( \prod_{l=k+1}^K \theta_l \r)  \| P_{\Ysub_{k-1,r}^\perp} \y_k\|_F^2
\qquad \mbox{(see \eqref{eq:deterministic online svd})}  
  \nonumber\\
  & \le \frac{50\p^{\frac{1}{3}}\alpha^2}{\p^{\frac{1}{3}}-1} \min\l (   \kappa_r^2 \rho_r^2, r \sigma_1^2 +\rho_r^2\r) b \cdot  \l(\frac{e}{2}\r)^{\p\eta_r^2} \l( K-1+\p\eta_r^2\r)^{\p\eta_r^2} \sum_{k=2}^K  \l(k-1+\p\eta_r^2\r)^{-\p\eta_r^2}
\qquad \mbox{(see \eqref{eq:bnd on prod} and Lemma \ref{lem:innovation})}  
   \nonumber\\
  & \le \frac{50\p^{\frac{1}{3}}\alpha^2}{\p^{\frac{1}{3}}-1} \min\l (   \kappa_r^2 \rho_r^2, r \sigma_1^2 +\rho_r^2\r) b\cdot  \l(\frac{e}{2}\r)^{\p\eta_r^2} \l( K-1+\p\eta_r^2\r)^{\p\eta_r^2} \int_{\p\eta_r^2}^{\infty} x^{-\p\eta_r^2}\, dx \nonumber\\
  & = \frac{50\p^{\frac{1}{3}}\alpha^2}{\p^{\frac{1}{3}}-1}\min\l (   \kappa_r^2 \rho_r^2, r \sigma_1^2 +\rho_r^2\r) b \cdot  \l(\frac{e}{2}\r)^{\p\eta_r^2} \l( K-1+\p\eta_r^2\r)^{\p\eta_r^2} \cdot \frac{(\p\eta_r^2)^{-\p\eta_r^2+1}}{\p\eta_r^2-1} \nonumber\\
  & \le \frac{50\p^{\frac{1}{3}}\alpha^2}{\p^{\frac{1}{3}}-1}\min\l (   \kappa_r^2 \rho_r^2, r \sigma_1^2 +\rho_r^2\r) b  \l( \frac{2K}{\p\eta_r^2}+2\r)^{\p\eta_r^2}  \frac{\p\eta_r^2}{\p\eta_r^2-1} \nonumber\\
  & \le 
  \frac{50\p^{\frac{4}{3}}\alpha^2  }{(\p^{\frac{1}{3}}-1)^2} \cdot \min\l (   \kappa_r^2 \rho_r^2, r \sigma_1^2 +\rho_r^2\r) \eta_r^2  b  \l( \frac{2K}{\p\eta_r^2}+2\r)^{\p\eta_r^2},
  \qquad \l(p,\eta_r\ge 1 \r)
\end{align}
except with a probability of at most $e^{-C\alpha^2 r}$ and provided that 
$$
b
\ge \frac{\p^{\frac{1}{3}}\alpha^2 r}{(\p^{\frac{1}{6}}-1)^2},
\qquad 
b\ge C\alpha^2 r.
$$
This completes the proof of Lemma \ref{lem:stochastic bnd}.

\section{Proof of Lemma \ref{lem:SVD lemma} \label{sec:alt exp}}
The proof is by induction. For $k=1$, it holds that 
\begin{align}
\wh{\Y}_{1,r}& =\SVD_r(\Y_1)
\qquad \text{(see Algorithm \ref{alg:MOSES})}
\nonumber\\
& = \Y_1P_{\wh{\Qsub}_{1,r}} 
\qquad \text{(see \eqref{eq:left right subspaces of iterates})}
\nonumber\\
& = \Y_{K} \P_{\I_1}\P_{\wh{\Qsub}_{1,r}} \nonumber\\
& = \Y_{K} \P_{\wh{\Qsub}_{1,r}},
\qquad \l( \wh{\Qsub}_{1,r} \subseteq \I_1\r)
\end{align}
which proves the base of induction. 
Next suppose that (\ref{eq:SVD simplified}-\ref{eq:long inclusion}) hold for $[2:k]$ with $k< K$.  We now show that   (\ref{eq:SVD simplified}-\ref{eq:long inclusion}) hold also for $k+1$. We can then write that 
\begin{align}
\wh{\Y}_{k+1,r} & = 
\SVD_r\l( \wh{\Y}_{k,r}+
\l[
\begin{array}{ccc}
\B{0}_{n\times kb} & \y_{k+1} & \B{0}_{n\times (K-k-1)b}
\end{array}
\r]\r) 
\qquad \text{(see Algorithm \ref{alg:MOSES})}
\nonumber\\
& = 
\SVD_r\l( 
\Y_{K} \P_{\wh{\Qsub}_{k,r}} + \Y_{K}\P_{\I_{k+1}}
\r) 
\qquad \text{(assumption of induction)}
\nonumber\\
& = \SVD_r\l( 
\Y_{K} \P_{\wt{\Qsub}_{k+1}}\r),
\qquad \text{(see \eqref{eq:def of Qtilde})}
\end{align}  
which completes the proof of Lemma \ref{lem:SVD lemma}.

\section{Proof of Lemma \ref{lem:orthogonality of summands} \label{sec:orth of summands} }
In this proof only, it is convenient to use the notation $\mbox{rowspan}(\B{A})$ to denote the row span of a matrix $\B{A}$, namely $\mbox{rowspan}(\B{A})=\mbox{span}(\B{A}^*)$. 
For every $k\in[1:K]$, recall from \eqref{eq:SVD simplified 3} that $\Y_{K}(\P_{\wt{\Qsub}_k}-\P_{\wh{\Qsub}_{k,r}})$ is the residual of rank-$r$ truncation of $\Y_{K}\P_{\wt{\Qsub}_k}$.  Consequently, 
\begin{equation}
\Y_{K} (\P_{\wt{\Qsub}_k}-\P_{\wh{\Qsub}_{k,r}}) =
\Y_{K} \P_{\wh{\Qsub}_{k,r}^C},
\qquad k\in[1:K],
\label{eq:residual written}
\end{equation}
where $\wh{\Qsub}_{k,r}^C$ is the orthogonal complement of $\wh{\Qsub}_{k,r}$ with respect to $\wt{\Qsub}_k$, namely 
\begin{equation}
\label{eq:decomposition in k iteration}
\wt{\Qsub}_{k} = \wh{\Qsub}_{k,r} \oplus \wh{\Qsub}_{k,r}^C, \qquad \wh{\Qsub}_{k,r} \perp \wh{\Qsub}_{k,r}^C \qquad k\in [1:K],
\end{equation}
in which we conveniently set $\wt{\Qsub}_1 = \I_1$, see \eqref{eq:def of Is}. 
Using \eqref{eq:residual written}, we can rewrite \eqref{eq:iterative exp for err} as 
\begin{align}
\Y_{k}-\wh{\Y}_{k,r} & = ( \Y_{k-1}-\wh{\Y}_{k-1,r} ) +\Y_k(\P_{\wt{\Qsub}_k} - \P_{\wh{\Qsub}_{k,r}} )
\qquad \mbox{(see \eqref{eq:iterative exp for err})}
 \nonumber\\
&
 = (\Y_{k-1}-\wh{\Y}_{k-1,r}) + \Y_{K}\P_{\wh{\Qsub}_{k,r}^C}, \qquad k\in[2:K]. 
 \label{eq:recursive rewritten}
\end{align}
With the preliminaries out of the way, let us rewrite the claim of Lemma \ref{lem:orthogonality of summands} as 
\begin{equation}
\l\langle \Y_{k-1}-\wh{\Y}_{k-1,r},  \Y_{K}\P_{\wh{\Qsub}_{k,r}^C}\r\rangle = 0,
\qquad k\in [2:K],
\end{equation}
see \eqref{eq:orthogonality of summands} and \eqref{eq:residual written}. Because $\wh{\Qsub}_{k,r}^C\subset \wt{\Qsub}_k$ by \eqref{eq:decomposition in k iteration}, it suffices to instead prove  the stronger claim that 
\begin{equation}
\mbox{rowspan}(\Y_{k-1}-\wh{\Y}_{k-1,r} ) \perp \wt{\Qsub}_{k},
\qquad k\in [2:K].
\label{eq:orthogonality general}
\end{equation}
We next prove \eqref{eq:orthogonality general} by induction. 
 The base of induction, namely $k=2$,  is trivial. Suppose now that  \eqref{eq:orthogonality general} holds for $[2:k]$ with $k<K$. We next show that \eqref{eq:orthogonality general} holds for $k+1$ as well.  
Note that 
\begin{align}
\label{eq:induction leg 0}
\mbox{rowspan}(\Y_{k}-\wh{\Y}_{k,r} ) & 
 = \mbox{rowspan}\l( ( \Y_{k-1}-\wh{\Y}_{k-1,r} )+ \Y_{K} \P_{\wh{\Qsub}_{k,r}^C} \r) 
\qquad \mbox{(see \eqref{eq:recursive rewritten})} 
 \nonumber\\
& \subseteq \mbox{rowspan} ( \Y_{k-1}-\wh{\Y}_{k-1,r} ) \oplus \wh{\Qsub}_{k,r}^C. 
\end{align}  
As we next show, both subspaces in the last line above are orthogonal to $\wt{\Qsub}_{k+1}$. Indeed, on the one hand, 
\begin{align}
& \begin{cases}
\mbox{rowspan} ( \Y_{k-1}-\wh{\Y}_{k-1,r} ) \perp \wt{\Qsub}_{k} \supseteq \wh{\Qsub}_{k,r}, & \mbox{(induction hypothesis and \eqref{eq:long inclusion})}\\
\mbox{rowspan} ( \Y_{k-1}-\wh{\Y}_{k-1,r} ) \subset \J_{k-1} \perp \I_{k+1},
& \mbox{(see \eqref{eq:long inclusion} and \eqref{eq:def of Js})}
\end{cases}
\nonumber
\\
& \Longrightarrow
\mbox{rowspan} ( \Y_{k-1}-\wh{\Y}_{k-1,r} ) \perp ( \wh{\Qsub}_{k,r} \oplus \I_{k+1}) = \wt{\Qsub}_{k+1}. 
\qquad \text{(see \eqref{eq:def of Qtilde})}
\label{eq:induction arg leg 1}
\end{align}
On the other hand, 
\begin{align}
& \begin{cases}
\wh{\Qsub}_{k,r}^C \perp \wh{\Qsub}_{k,r}, \\
\wh{\Qsub}_{k,r}^C \subset \wt{\Qsub}_k \subset \J_{k} \perp \I_{k+1}, 
& \mbox{(see \eqref{eq:long inclusion} and \eqref{eq:def of Js})}
\end{cases} \nonumber\\
& \Longrightarrow 
\wh{\Qsub}_{k,r}^C \perp ( \wh{\Qsub}_{k,r} \oplus \I_{k+1}) = \wt{\Qsub}_{k+1}. 
\qquad \text{(see \eqref{eq:def of Qtilde})}
\label{eq:induction arg leg 2}
\end{align}
By combining \eqref{eq:induction arg leg 1} and \eqref{eq:induction arg leg 2}, we conclude that 
\begin{align}
\mbox{rowspan}(\Y_{k}-\wh{\Y}_{k,r} ) 
& \subseteq
\mbox{rowspan} ( \Y_{k-1}-\wh{\Y}_{k-1,r} ) \oplus \wh{\Qsub}_{k,r}^C 
\qquad \mbox{(see \eqref{eq:induction leg 0})}
\nonumber\\
&  \perp \wt{\Qsub}_{k+1}. 
\qquad \text{(see (\ref{eq:induction arg leg 1},\ref{eq:induction arg leg 2}))}
\end{align}
Therefore,  \eqref{eq:orthogonality general} holds for every $k\in[2:K]$ by induction. In particular, this proves Lemma \ref{lem:orthogonality of summands}.

\section{Proof of Lemma \ref{lem:bnd on yk spec norm} \label{sec:proof of bnd on yk spec norm}}

Note that 
\begin{align}
\|\y_k\| & = \| \B{S}\B{\Sigma} \g_k\| 
\qquad \mbox{(see \eqref{eq:def of Y1k gaussian})}
 \nonumber\\
& = \|\B{\Sigma} \B{g}_k\| \qquad \l( \B{S}^*\B{S} = \B{I}_n\r)
\nonumber\\
& \le \| \B{\Sigma}[1:r,1:r]\cdot  \g_k[1:r,:]\| + \| \B{\Sigma}[r+1:n,r+1:n]\cdot  \g_k[r+1:n,:]\| 
\qquad \mbox{(triangle inequality)} \nonumber\\
& \le \sigma_1 \cdot \l\|\g_k[1:r,:]\r\| + \| \B{\Sigma}[r+1:n,r+1:n]\cdot  \g_k[r+1:n,:]\| 
\nonumber\\
& \le \sigma_1 \cdot \l\|\g_k[1:r,:]\r\| + \| \B{\Sigma}[r+1:n,r+1:n]\cdot  \g_k[r+1:n,:]\|_F,
\label{eq:bnd on Yk pre}
\end{align}
where we used MATLAB's matrix notation as usual. 
Note that both $\g_k[1:r,:]\in\R^{r\times b}$ and $\g_k[r+1:n,:]\in\R^{(n-r)\times b}$  in \eqref{eq:bnd on Yk pre} are  standard Gausssian random matrices. For $\alpha\ge 1$ and $\p>1$,  invoking the results about the spectrum of Gaussian random matrices in Section \ref{sec:notation}  yields that 
\begin{align}
\|\y_k\| & \le \sigma_1 \cdot \|\g_k[1:r,:]\| +  \|
\B{\Sigma}[r+1:n,r+1:n] \cdot \g_k[r+1:n,:]\|_F \qquad \mbox{(see \eqref{eq:bnd on Yk pre})} \nonumber\\
& \le  \sigma_1(\sqrt{b}+\alpha\sqrt{r}) + \sqrt{2} \alpha \|\B{\Sigma}[r+1:n,r+1:n]\|_F \sqrt{b}
\qquad \l(\mbox{see (\ref{eq:bnd on Gaussians},\ref{eq:scalar Bernie}) and }   b\ge r \r)
 \nonumber\\
 & =  \sigma_1(\sqrt{b}+\alpha\sqrt{r}) + \alpha \rho_r\sqrt{  2b}
\qquad \mbox{(see (\ref{eq:Sigma 2 proposition},\ref{eq:shorthand}))}
 \nonumber\\
& \le  \p^{\frac{1}{6}}\sigma_1 \sqrt{b} +\alpha \rho_r \sqrt{2 b},
\qquad \l( \mbox{if }  b\ge \frac{\alpha^2 r}{(p^{\frac{1}{6}}-1)^2} \r)
\end{align}
except with a probability of at most $e^{-C\alpha^2 r }+e^{-C\alpha^2 b} \le e^{-C\alpha^2 r}$, where this final inequality follows from the assumption that $b\ge r$. 
This completes the proof of Lemma \ref{lem:bnd on yk spec norm}. 
We remark that a slightly stronger bound can be obtained by using Slepian's inequality for comparing Gaussian processes, see \cite[Section 5.3.1]{vershynin2010introduction} and~\cite[Section 3.1]{ledoux2013probability}.

\section{Proof of Lemma \ref{lem:bnd on sigmar Yk} \label{sec:proof of bnd on sigmar Yk}}
For a matrix $\B{A}\in\R^{n\times kb}$, it follows from the Fisher-Courant representation of the singular values that 
\begin{equation}
\sigma_r(\B{A})\ge \sigma_r(\B{A}[1:r,:]).
\label{eq:fisher}
\end{equation}
Alternatively, \eqref{eq:fisher} might be verified using Cauchy's interlacing theorem applied to $\B{A}\B{A}^*$. For a vector $\gamma\in\R^{r\times r}$ and matrix $\B{A}\in\R^{r\times r}$, we also have the useful inequality 
\begin{equation}
\sigma_r(\mbox{diag}(\gamma) \B{A}) \ge \min_{i\in [r]} |\gamma[i]| \cdot \sigma_r(\B{A}),
\label{eq:diagonal sigma}
\end{equation}
where $\mbox{diag}(\gamma)\in\R^{r\times r}$ is the diagonal matrix formed from the entries of $\gamma$. 
Using the above inequalities, we may write that 
\begin{align}
\sigma_r(\Y_{k}) & = \sigma_r( \B{S}\B{\Sigma} \G_{k}) 
\qquad \mbox{(see \eqref{eq:def of Y1k gaussian})}
\nonumber\\
& = \sigma_r(\B{\Sigma} \G_{k}) \qquad \l(\B{S}^* \B{S} = \B{I}_n \r) \nonumber\\
& \ge \sigma_r\l( \B{\Sigma}[1:r,1:r]\cdot \B{G}_{k}[1:r,:] \r) 
\qquad \mbox{(see \eqref{eq:fisher})}
\nonumber\\
& \ge \sigma_r \cdot \sigma_{r}\l(\B{G}_{k}[1:r,:]\r).
\qquad \l(\mbox{see (\ref{eq:diagonal sigma},\ref{eq:Sigma 2 proposition})}\r)
\label{eq:bnd on Yk 0}
\end{align}
Note also that $\G_{k}[1:r,:]\in\R^{r\times kb}$ above is a standard Gaussian  random matrix. Using the spectral properties listed in Section \ref{sec:notation}, we can therefore write that 
\begin{align}
\sigma_r(\Y_{k}) & \ge \sigma_r\cdot  \sigma_r\l(\G_{k}[1:r,:]\r)
\qquad \mbox{(see \eqref{eq:bnd on Yk 0})}
 \nonumber\\
& \ge \sigma_r  \cdot (\sqrt{kb}-\alpha\sqrt{r} )
\qquad \l(\mbox{see \eqref{eq:bnd on Gaussians} and  } b\ge r \r)
\nonumber\\
& \ge \sigma_r \cdot p^{-\frac{1}{6}}\sqrt{kb},
\qquad \l(\mbox{if } b\ge \frac{\alpha^2 r}{(1-p^{-\frac{1}{6}})^2} \r)
\end{align}
except with a probability of at most $e^{-C\alpha^2 r}$.  This completes the proof of Lemma \ref{lem:bnd on sigmar Yk}. 

\section{Proof of Lemma \ref{lem:innovation} \label{sec:proof of innovation}}

Without loss of generality, we set $\B{S}=\B{I}_n$ in \eqref{eq:def of Xi proof of proposition} to simplify the presentation, as this renders the contribution of the bottom rows
of $\B{y}_k$ to the innovation typically small.
We first separate this term via the inequality 
\begin{align}
\| \P_{\Ysub_{k-1,r}^\perp} \y_k\|_F & =
\l\| \P_{\Ysub_{k-1,r}^\perp}
\l[
\begin{array}{c}
\y_k[1:r,:] \\
\y_k[r+1:n,:]
\end{array}
\r]
\r\|_F \nonumber\\
& \le 
\l\| 
\P_{\Ysub_{k-1,r}^\perp} 
\l[
\begin{array}{c}
\y_k[1:r,:] \\
\B{0}_{(n-r)\times b}
\end{array}
\r]
\r\|_F + 
\l\| 
\y_k[r+1:n,:]
\r\|_F. 
\qquad \mbox{(triangle inequality)}
\label{eq:main term of innovation}
\end{align}
To control the last norm above, we simply write that 
\begin{align}
\l\| 
\y_k[r+1:n,:]
\r\|_F & = \l\| \B{\Sigma}[r+1:n,r+1:n] \cdot \g_k[r+1:n,:] \r\|_F 
\qquad \mbox{(see \eqref{eq:def of Y1k gaussian})}
\nonumber\\
& \le  \alpha \| \B{\Sigma}[r+1:n,r+1:n]\|_F \sqrt{2b} \qquad \mbox{(see \eqref{eq:scalar Bernie})} \nonumber\\
& = \alpha\rho_r \sqrt{2b},
\qquad \mbox{(see \eqref{eq:shorthand})}
\label{eq:residual left of out innovation}
\end{align}
except with a probability of at most $e^{-C\alpha^2 b}$. In the second line above, we used the fact that $\g_k$ is a standard Gaussian random matrix. It therefore remains to control the first norm in the last line of \eqref{eq:main term of innovation}. Note that 
\begin{align}
\l\| 
\P_{\Ysub_{k-1,r}^\perp} 
\l[
\begin{array}{c}
\y_k[1:r,:] \\
\B{0}_{(n-r)\times b}
\end{array}
\r]
\r\|_F & = 
\l\| 
\P_{\Ysub_{k-1,r}^\perp} 
\l[
\begin{array}{cc}
\B{I}_{r} \\
& \B{0}_{n-r}
\end{array}
\r]
\cdot 
\l[
\begin{array}{c}
\y_k[1:r,:] \\
\B{0}_{(n-r)\times b}
\end{array}
\r]
\r\|_F \nonumber\\
& =: 
\l\| 
\P_{\Ysub_{k-1,r}^\perp} 
\B{J}_r
\cdot 
\l[
\begin{array}{c}
\y_k[1:r,:] \\
\B{0}_{(n-r)\times b}
\end{array}
\r]
\r\|_F \nonumber\\
& \le  
\| 
\P_{\Ysub_{k-1,r}^\perp} 
\B{J}_r
\|_F
\cdot 
\l\|
\y_k[1:r,:] 
\r\| \nonumber\\
& \le 
\| 
\P_{\Ysub_{k-1,r}^\perp} 
\B{J}_r
\|_F
\cdot
\| 
\B{\Sigma}[1:r,1:r]  \| \cdot 
\l\| \g_k[1:r,:]
\r\| 
\qquad \mbox{(see \eqref{eq:def of Y1k gaussian})}
\nonumber\\
& \le \| 
\P_{\Ysub_{k-1,r}^\perp} 
\B{J}_r
\|_F
\cdot \sigma_1 \cdot (\sqrt{b}+ \alpha\sqrt{r} ) \qquad
\mbox{(see (\ref{eq:Sigma 2 proposition},\ref{eq:bnd on Gaussians}))} \nonumber\\
& \le  \| 
\P_{\Ysub_{k-1,r}^\perp} 
\B{J}_r
\|_F
\cdot \sigma_1 \sqrt{2b}, 
\qquad \l( \mbox{if } b\ge C\alpha^2 r\r)
\label{eq:main term 1st step}
\end{align}
except with a probability of at most $e^{-C\alpha^2 r}$ and provided that $b\ge C\alpha^2 r$.  The fifth line above again uses the fact that $\g_k$ is a standard Gaussian random matrix. Let us now estimate the norm in the last line above. Recall that $\P_{\Ysub_{k-1,r}}\in\R^{n\times n}$ projects onto the span of $\Y_{k-1,r}=\mbox{SVD}_r(\Y_{k-1})$, namely $\P_{\Ysub_{k-1,r}}$ projects onto the span of leading $r$ left singular vectors of $\Y_{k-1}=\B{\Sigma} \G_{k-1}$, see \eqref{eq:def of Y1k gaussian}. Because the diagonal entries of $\B{\Sigma}\in\R^{n\times n}$ are in nonincreasing order, it is natural to expect that $ \P_{\Ysub_{k-1,r}} \approx \B{J}_{r} $. We now  formalise this notion using standard  results from the perturbation theory. Note that one  might think of $\Y_{k-1,r}=\mbox{SVD}_r(\Y_{k-1})$ as a perturbed copy of $\Y_{k-1}$. Note also that $\B{J}_r $ is the orthogonal projection onto the subspace
$$
\mbox{span}
\l( 
\l[
\begin{array}{c}
\Y_{k-1}[1:r,:]\\
\B{0}_{(n-r)\times (k-1)b}
\end{array}
\r]
\r),
$$
because $\Y_{k-1}[1:r,:]$ is almost surely full-rank. 
An application of Lemma 6 in \cite{eftekhari2016snipe} with $\B{A}$ as specified inside the parenthesis above and $\B{B}=\Y_{k-1}$   yields that 
\begin{align}
\| \P_{\Ysub_{k-1,r}^\perp} \B{J}_r  \|_F & \le \frac{\l\| \Y_{k-1} -\l[
\begin{array}{c}
\Y_{k-1}[1:r,:]\\
\B{0}_{(n-r)\times (k-1)b}
\end{array}
\r] \r\|_F}{\sigma_r(\Y_{k-1})} \nonumber\\
& = \frac{\l\|\Y_{k-1}[r+1:n,:] \r\|_F}{\sigma_r(\Y_{k-1})} \nonumber\\
& = \frac{\l\| \B{\Sigma}[r+1:n,r+1:n] \cdot \G_{k-1}[r+1:n,:] \r\|_F}{\sigma_r(\Y_{k-1})} 
\qquad \mbox{(see \eqref{eq:def of Y1k gaussian})}
\nonumber\\
& \le \frac{\alpha \|\B{\Sigma}[r+1:n,r+1:n] \|_F \sqrt{2(k-1)b}}{  \sigma_r \sqrt{(k-1)b/2} }
\qquad \l( \mbox{see \eqref{eq:scalar Bernie} and Lemma \ref{lem:bnd on sigmar Yk} with } p=8 \r) \nonumber\\
& = \frac{2\alpha \rho_r}{\sigma_r},
\qquad \mbox{(see \eqref{eq:shorthand})}
\end{align}
provided that $b\ge C\alpha^2 r$ and except with a probability of at most $e^{-C\alpha^2 b }+e^{-C\alpha^2 r}\le e^{-C\alpha^2 r}$, where this last inequality follows from the assumption that $b\ge r$.  It also trivially holds that $$
\| \P_{\Ysub_{k-1,r}^\perp}\B{J}_r\|_F\le 
\| \P_{\Ysub_{k-1,r}^\perp}\| \cdot \| \B{J}_r\|_F\le  \|\B{J}_r\|_F = \|\B{I}_r\|_F 
= \sqrt{r},
$$
where we used above the definition of $\B{J}_r$ in \eqref{eq:main term 1st step}. 
 Therefore, overall we find that 
\begin{equation}
\| \P_{\Ysub_{k-1,r}^\perp} \B{J}_r  \|_F  \le \min\l(\frac{2\alpha \rho_r}{\sigma_r} ,\sqrt{r}\r). 
\label{eq:overall bnd}
\end{equation}
Substituting the above bound back into \eqref{eq:main term 1st step} yields that 
\begin{align}
\l\| 
\P_{\Ysub_{k-1,r}^\perp} 
\l[
\begin{array}{c}
\y_k[1:r,:] \\
\B{0}_{(n-r)\times b}
\end{array}
\r]
\r\|_F & \le  \| 
\P_{\Ysub_{k-1,r}^\perp} 
\B{J}_r
\|_F
\cdot \sigma_1 \sqrt{2b}
\qquad \mbox{(see (\ref{eq:main term 1st step}))} \nonumber\\
& \le  \min\l( {\alpha \kappa_r \rho_r } ,\sigma_1 \sqrt{r} \r) \sqrt{8b},
\qquad \mbox{(see (\ref{eq:overall bnd},\ref{eq:shorthand}))}
\label{eq:main innovation final}
\end{align}
except with a probability of at most $e^{-C\alpha^2 r}$.
Combining \eqref{eq:residual left of out innovation} and \eqref{eq:main innovation final} finally controls the innovation as 
\begin{align}
\| \P_{\Ysub_{k-1,r}^\perp} \y_k\|_F 
& \le 
\l\| 
\P_{\Ysub_{k-1,r}^\perp} 
\l[
\begin{array}{c}
\y_k[1:r,:] \\
\B{0}_{(n-r)\times b}
\end{array}
\r]
\r\|_F + 
\l\| 
\y_k[r+1:n,:]
\r\|_F
\qquad \mbox{(see \eqref{eq:main term of innovation})}
 \nonumber\\
 & \le \min\l( \alpha \kappa_r \rho_r, \sigma_1 \sqrt{r}  \r) \sqrt{8b} + \alpha \rho_r \sqrt{2b}  
 \qquad \text{(see (\ref{eq:main innovation final},\ref{eq:residual left of out innovation}))}
 \nonumber\\
 & \le  5\alpha \min\l (   \kappa_r \rho_r, \sigma_1\sqrt{r}+\rho_r\r) \sqrt{b}, 
 \qquad \l( \alpha, \kappa_r \ge 1 \r)
\end{align}
except with a probability of at most $e^{-C\alpha^2 r}$ and provided that $b\ge C\alpha^2 r$. This completes the proof of Lemma \ref{lem:innovation}. 


\end{document}